\documentclass{jaa}
\usepackage[T1]{fontenc}
\usepackage{ae,aecompl}
\usepackage[normalem]{ulem}
\usepackage{graphicx}	% Including figure files
\usepackage{amsmath, mathtools, xfrac}	% Advanced maths commands
\usepackage{amssymb, enumitem}	% Extra maths symbols
\usepackage{longtable, multicol}
\usepackage{rotate, setspace}
\usepackage{lscape}
\usepackage{natbib, hyperref}
\usepackage{latexsym,verbatim,xcolor}
\newcommand{\apj}[1]{ApJ, }
\newcommand{\jcp}[1]{J. Chem. Phys., }
\newcommand{\mnras}[1]{MNRAS, }
\newcommand{\aj}[1]{AJ, }
\newcommand{\apjs}[1]{ApJS, }
\newcommand{\apjl}[1]{ApJ Letter, }
\newcommand{\aap}[1]{A\&A, }
\newcommand{\aaps}[1]{A\&A Suppl. Series, }
\newcommand{\araa}[1]{Annu. Rev. A\&A, }
\newcommand{\aaas}[1]{A\&AS, }
\newcommand{\apss}[1]{Ap\&SS }
\newcommand{\bain}[1]{Bul. of the Astron. Inst. of the Netherlands,}
\newcommand{\planss}[1]{Planetary and Space Science,}
\newcommand{\nat}[1]{Nature,}
\newcommand{\actaa}[1]{Acta Astronomica,}
\newcommand{\aapr}[1]{The Astronomy and Astrophysics Review,}
\newcommand{\memsai}[1]{Memorie della Societa Astronomica Italiana,}
\begin{document}\sloppy
%\color{red}
%%paper title
%%For line breaks \\ can be used within title
\title{A comprehensive photometric and kinematical characteristic of the newly discovered QCs clusters with Gaia EDR3} %\zmat{v3}}
%Title of the paper goes here:\\ Second line}

%%author names are separated by comma (,)
%%use \and before the last author name
%%use a * along with the number separated by comma
%% for the  author for correspondence
%%\textsuperscript{number} is used for affiliation
%%\affilOne, \affilTwo etc., upto \affilTwentyfive is possible
%%Please note the first letter after \affil is capitalised in the command
\begin{singlespace}
\author{W. H. Elsanhoury\textsuperscript{1,2,*}, Magdy Y. Amin\textsuperscript{3}, A. A. Haroon\textsuperscript{4} and Z. Awad\textsuperscript{3}}
\affilOne{\textsuperscript{1}Astronomy Department, National Research Institute of Astronomy and Geophysics (NRIAG), 11421, Helwan, Cairo, 
Egypt (Affiliation ID: 60030681).\\}
\affilTwo{\textsuperscript{2}Physics Department, Faculty of Science and Arts, Northern Border University, Turaif Branch, Saudi Arabia.\\}
\affilThree{\textsuperscript{3}Department of Astronomy, Space Science, and Meteorology, Faculty of Science, Cairo University, Giza 11326, Egypt. Emails: mamin@sci.cu.edu.eg $\&$ zma@sci.cu.edu.eg\\}
\affilFour{\textsuperscript{4}Astronomy and Space Science Department, Faculty of Science, King Abdul Aziz University, Jeddah, Saudi Arabia. Email: aaharoon@kau.edu.sa}

%%escape two column mode for title, affiliation and abstract
%%by giving \twocolumn command as shown

\twocolumn[{

\maketitle

%%include \corres to print the corresponding author Email id
\corres{elsanhoury@nbu.edu.sa; welsanhoury@gmail.com; elsanhoury@nriag.sci.eg}

%%include \msinfo for
%%manuscript information such as
%%received, revised and accepted dates
%%
\msinfo{1 January 2021}{1 January 2021}

\begin{abstract}\\
%{\bf NOTE:}\zmat{All modifications are in this colour} while those added to the referee are\zma{in this colour}\\

%\zma{\bf I will work on the abstract shortly - not ready yet; today is 31 Oct 2021 \\}
%\zmat{
This study reports the first comprehensive astrometric, photometric and kinematical analysis of four newly discovered open clusters; namely QC1, QC2, QC3, and QC4, using astrometric and photometric data from the most recent Gaia EDR3 for G $<$ 17 mag. Utilizing the ASteCA code, we identified the most probable (P $\ge$ 50\%) star candidates and found the numbers of star members (N) to be 118 (QC1), 142 (QC2), 210 (QC3), and 110 (QC4). By fitting King's density profile to the cluster's RDPs, we found the internal structural parameters of each cluster such as the cluster radii that are in the range 7.00 to 11.00 arcmin. For each cluster we constructed the CMD and by fitting them with suitable isochrones we found that the metallicity range is (0.0152 -- 0.0199) which is in line with the Solar value, the logarithmic age (in yrs) range between 6.987 and 8.858. The distances derived from CMD are 1674$\pm$41, 1927$\pm$44, 1889$\pm$43, and  1611$\pm$40 (pc) for QC1, QC2, QC3, and QC4, respectively, and they are in good agreement up to 85\% with the values obtained from the astrometric data. In addition, from the MLR of the clusters, we obtained a total mass, M$_C$ in Solar units, of 158, 177, 232, and 182 and an absolute magnitude M$_G$ (mag) of 4.33, 3.80, 4.25, and 4.10 for QC1, QC2, QC3, and QC4, respectively. %The structural parameters have been computed, showing the clusters radii are ranged from 7.00 to 11.00 (arcmin). Using the extracted member candidates and the isochrone fitting with metallicities (Z) ranges from 0.01520 to 0.01987 for (G, G$_{\text{BP}}$, G$_{\text{RP}}$) passbands and color-magnitude diagrams CMDs, we have estimated log (age/yr):6.987$\pm$0.022  (QC1),8.709$\pm$0.039  (QC2),  8.858$\pm$0.114(QC3), and 8.367$\pm$0.043 (QC4), and the clusters are located at a distances 1674$\pm$41, 1927$\pm$44, 1889$\pm$43, and  1611$\pm$40 (pc) for QC1, QC2, QC3, and QC4, respectively. We have estimated the cumulative cluster masses adopting mass-luminosity relation MLR of initial mass function and synthetic CMDs, therefore the estimated masses are 158 (QC1), 177 (QC2), 232 (QC3), and 182 (QC4) in the units of Solar mass. 
The dynamical analysis and evolution parameters of the cluster members indicated that all the four clusters are dynamically relaxed; except QC1 which has an evolution parameter $\tau \sim$ 0.82 that indicates a dynamical activity within the cluster. From the kinematical analysis of the cluster data, we computed the space velocity, the %For the kinematical approaches and velocity ellipsoidal motion of member candidates, we computed the 
coordinates of the apex point (A, D) using the AD -- diagram method, as well as the Solar elements (S$_{\odot}$, l$_{A}$, b$_{A}$, $\alpha_{A}$, $\delta_{A}$).%}
\end{abstract}

%%insert keywords separated by 3 hyphens using \keywords{words}

}]
\end{singlespace}
%%close the twocolumn escape here
%%include \doinum{number}for the DOI number in the header
%%include \volnum{number} for the volume number in the header
%%include \year{yyyy} for  year of publication in the header
%%include \pgrange{num--num} page range of article in the header
%%include \artcitid{num} for the article citation id
%%include \lp to print last page of the article
%%include \setcounter{page}{pagenum} for the exact starting page of the article

\doinum{12.3456/s78910-011-012-3}
\artcitid{\#\#\#\#}
\volnum{000}
\year{2021}
\pgrange{1--}
\setcounter{page}{1}
\lp{\#}
%\rfoot{Page \thepage \hspace{1pt} of \pageref{LastPage}}

\section{Introduction}
Open clusters (OCs) are formed within giant molecular clouds (GMCs) that are located in the disk of the Milky Way Galaxy \citep{Lada03, Portegies10}. OCs that possess simple populations with relatively easy determined ages are known as associations. They are among the best tracers of the spiral arm structure and the evolution of the Galactic disk (e.g. \citealt{Trumpler30, Moffat73,Janes82,Friel95,Moitinho10, Moraux16}). Open clusters are excellent laboratories to examine stellar physical and dynamical evolution from which we explore the mechanisms of star creation and its recent history (e.g. \citealt{Vandenberg83,Barnes07,Bertelli17,Marino18}). OCs are important in the calibration of the distance scale because of the accurate determination of their distances \citep{Perren15} and in constraining both the initial luminosity and mass functions in aggregations of stars. Moreover, radial velocities of OCs are used to trace the local kinematics such as the Velocity Ellipsoid Parameters (VEPs), % S}olar motion, 
Oort's constants (A \& B), and the rotation curve. Old distant clusters are used to define disk abundance gradients, construct the age-metallicity relation, understand the complex history of chemical enrichment, and the mixing processes in the disk \citep{Friel95}.

The constellation Cygnus is located in the Galactic plane within the spiral arm that hosts our %Milky Way 
Galaxy (e.g. \citealt{Bochkarev85}). Cygnus is a place of star formation where a number of young stars, associations and open clusters have been detected with a wide range of masses, ages and uncertainties in their distances. One of the open clusters that lies in the region of Cygnus is NGC 7062 which is classified as II2p \citep{Trumpler30}. NGC 7062 is located at a heliocentric distance of about 1600 pc with an estimated age 6.9 $\times$ 10$^8$ years \citep{Kharchenko13}. %log (age/yr) ~ 8.84 
The cluster's equatorial coordinates are ($\alpha$ = 21$^h$ 23$^m$ 27$^s$.00, $\delta$= 46$^o$ 23' 24''.00) while its Galactic coordinates are ($l$ = 89$^o$.967, $b$ = -2$^o$.740) \citep{Carrera19}.

\citet{Qin21} analyzed the region surrounding NGC 7062 and the eastern part of Cygnus, (77$^{o} \le l \le$ 90$^{o}$  and -3$^{o} \le b \le$ 4$^{o}$), using data from Gaia DR2 \citep{Gaia16, Gaia18}. The authors reported the discovery of 
%detection of the open cluster NGC 7062 in addition to other 
four open clusters known as: QC1 ($l$ = 77$^o$.635, $b$ = 1$^o$.932), QC2 ($l$ = 84$^o$.802, $b$ = 3$^o$.278), QC3 ($l$ = 84$^o$.976, $b$ = 2$^o$.513), and QC4 ($l$ = 85$^o$.036, $b$ = 3 $^o$.654). Among the 2443 star cluster candidates observed and listed in the online catalog\footnote{https://vizier.u-strasbg.fr/viz-bin/VizieR?-source=J/ApJS/245/32} by \citet{Liu19}, the location of the two clusters with IDs 506 ($l$ = 84$^o$.959, $b$ = 2$^o$.529) and 600 ($l$ = 84$^o$.81, $b$ = 3$^o$.258) is very close to that of QC3 and QC2, respectively, discovered by \citet{Qin21}. 
\citet{Qin21} retrieved astrometric and photometric data for the four newly discover clusters from the Gaia DR2 and computed their structure and photometric parameters. The authors concluded that future observations as well as investigations are required to characterize the properties of the four clusters.

%\zma{$\quad$
Most recently, the Gaia mission collaborations released the early data release 3 (Gaia EDR3; \citealt{Gaia20}). 
Similar to DR2, the EDR3 provides five astrometric parameters; Galactic position ($l, b$), proper motion ($\mu_{\alpha}\cos\delta, \mu_{\delta}$), parallax ($\varpi$) and the photometric magnitude parameters in 3 filters (G, G$_{BP}$ and G$_{RP}$) for a larger number of sources up to 1.8 billion sources with brightness larger than 21 (i.e. numerically < 21). The EDR3 is complemented with data of the radial velocity (V$_r$) for about 7 million stars from DR2 \citep{Gaia20}. The source list has a slight change to DR2 with some notable changes. The significant advance of EDR3 over DR2 is the large improvement in the accuracy of the astrometric parameters; a factor 2 in proper motion accuracy and a factor of about 1.5 in the parallax accuracy. Astrometric errors were suppressed by 30 - 40\% for the parallax and by a factor of 2.5 for the proper motion.

%\zma{$\quad$ 
In addition, advances in the photometry exists with a better homogeneity for the magnitude and color due to the significant improvement in several aspects related to the photometer preprocessing, the photometric calibration process and modeling that would probably lead to a reduction in the error of the photometric magnitudes estimates (e.g. \citealt{Gaia20, Torra21, Riello21}). All of these improvements lead to a more accurate measurement of the blue (G$_{BP}$) and red (G$_{RP}$) photometric magnitudes. As a consequences for all of these improvements, reflections on the astrophysical results are expected and astronomers are optimistic to obtain a better characterization of different observed objects from different perspectives. %Fig. \ref{fig:error} illustrates both the photometric (left panels) and the astrometric (right panels) errors in Gaia EDR3 with respect to the photometric G magnitude.

%\zma{$\quad$ 
Motivated by the findings of \citet{Qin21} and the huge improvement granted by EDR3, we conducted the present comprehensive astrometric, photometric, and kinematic study in an attempt to fully characterize the four newly discovered clusters; QC1, QC2, QC3 and QC4.

The structure of the article is as follows: \S~\ref{Gaia} describes the raw data from Gaia EDR3 and the adopted methodology in this study. Results of the analysis of the EDR3 data for the new open clusters in which the basic astrometric and photometric %(\S~\ref{astr}) 
as well as the dynamical and kinematical %(\S~\ref{kin}) 
parameters are derived, are discussed in \S~\ref{res} Finally, our main conclusions are summarized in \S~\ref{conc}

\section{Data from Gaia EDR3 and analysis tools}
\label{Gaia}
\begin{figure}
\begin{center}
%% trim left bottom right top
\includegraphics[trim= 0.20cm 0.0cm 0.0cm 0.20cm,clip=true,width=6cm]{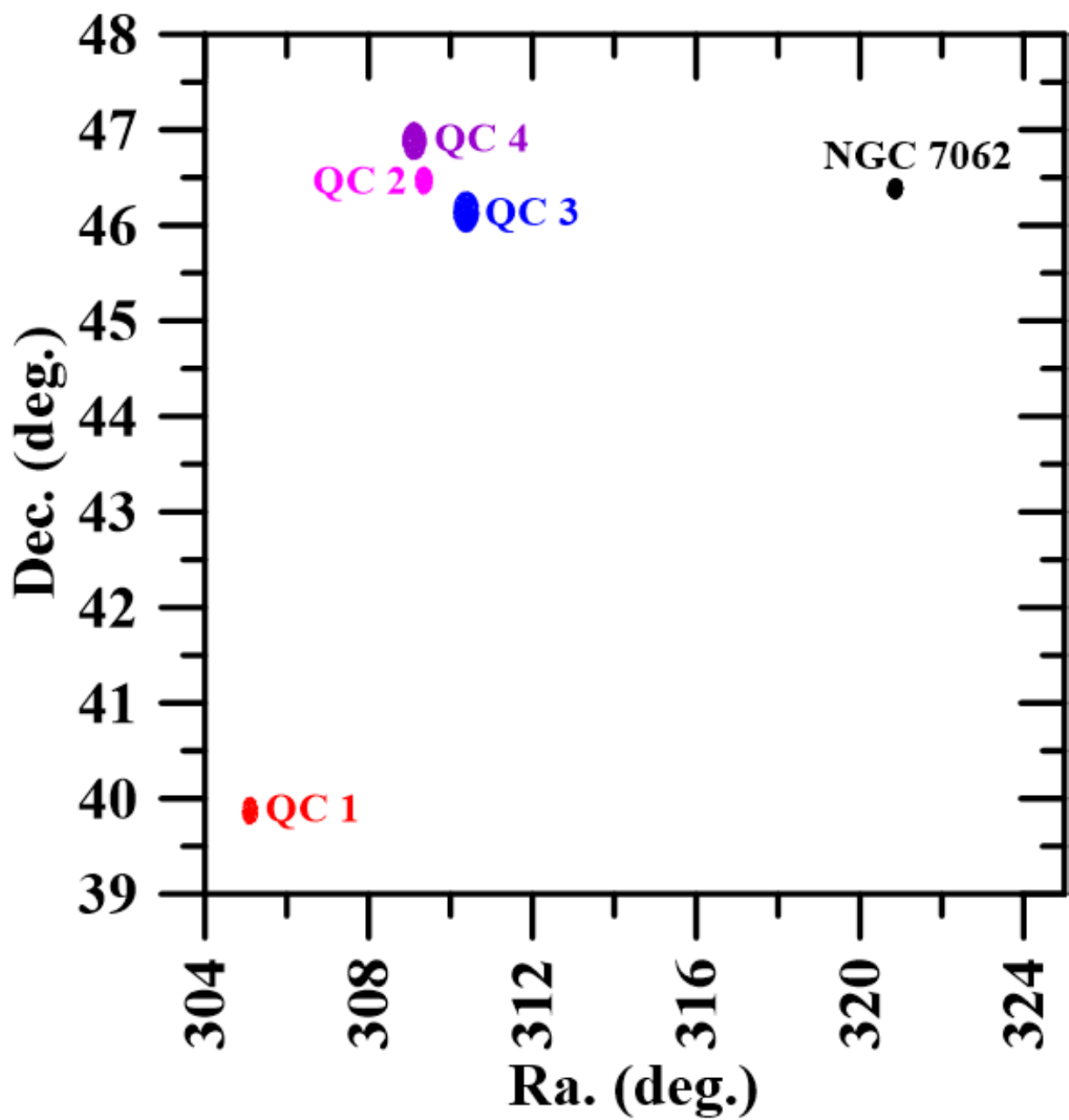} 
\caption {The spatial distribution of the four newly discovered open clusters together with the position of the well-known open cluster NGC 7062, for comparison. The five cluster positions are taken from \citet{Qin21}.}
\label{fig:1}
\end{center}
\end{figure}

\begin{figure}
\begin{center}
% trim left bottom right top
%\includegraphics[trim= 1.0cm 0.5cm 1.0cm 0.5cm,clip=true,width=5cm]{fig.pdf} 
%   \includegraphics[scale = 0.6]{plots/fig2/fig2l.pdf}width=6cm
%    \hfill
  \includegraphics[scale = 0.55]{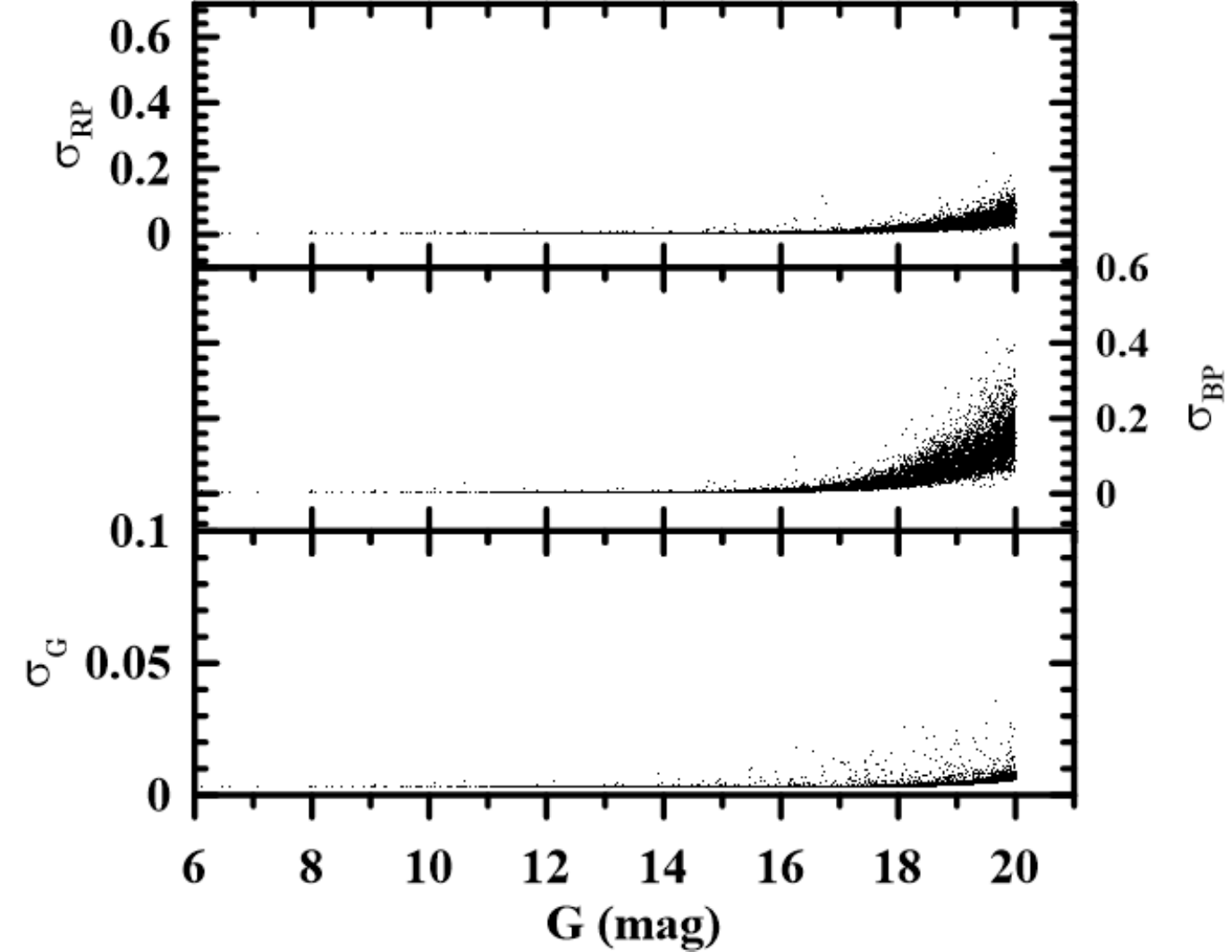}
  \includegraphics[scale = 0.55]{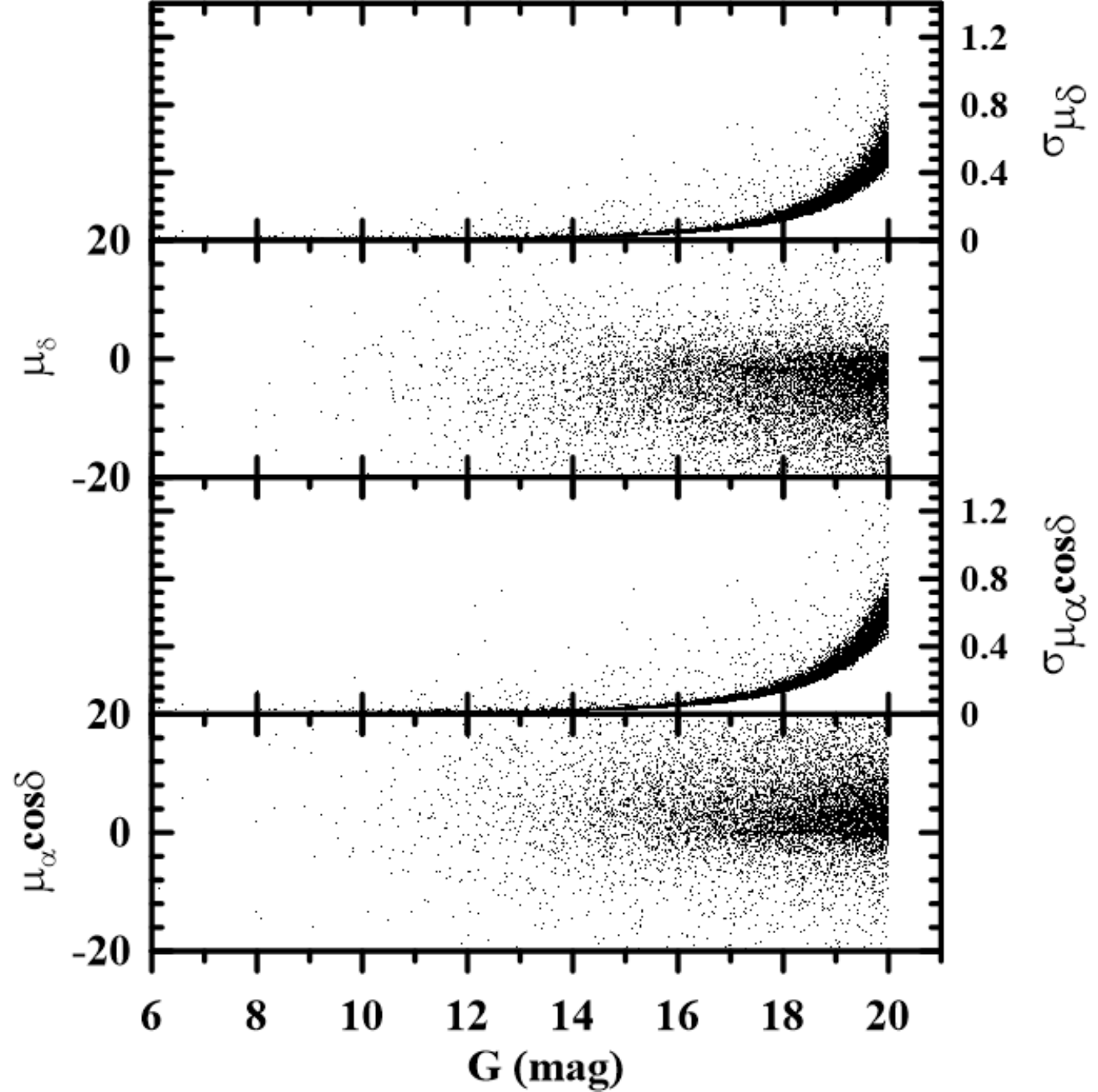}
%\vspace{0.6cm}
\caption {Uncertinities in the three photometric magnitude bands; $\sigma_G$, $\sigma_{BP}$, and $\sigma_{RP}$ (top panel) and the proper motion and its uncertinities ($\mu, \sigma_{\mu}$) in both directions (bottom panel), for the data used in this study, with respect to the photometric G magnitude.}
\label{fig:error}
\end{center}
\end{figure}

The data retrieved for the four clusters from EDR3 includes the astrometric parameters ($\alpha, \delta, l, b, \mu, \varpi$), the three photometric magnitudes (G, G$_{BP}$, G$_{RP}$) and the radial velocity (V$_r$) with their uncertainties. For each cluster, we downloaded data sheets of a surrounding space centered at the position of the cluster center ($\alpha$,$\delta$) computed by \citet{Qin21} and has an arbitrary chosen radius that is more than twice the cluster radius, r$_{cl}$, obtained by \citet{Qin21}. In this way, we initiate the calculations with a more realistic guess that would reduce the output uncertainities. The radius of QC1 and QC2 was adapted to be 20 arcmin while that for QC3 and QC4 was 25 arcmin. Table \ref{tab:fund} lists the initial boundary conditions adopted in this study taken from \citet{Qin21}. 
The spatial distribution of the four OCs with the well-known cluster NGC 7062, for comparison, is illustrated in Fig. \ref{fig:1} while errors in both the photometric G-magnitude (upper panels) and the astrometric proper motion (lower panels) in Gaia EDR3 with respect to the photometric G--magnitude for the retrived data of this study are illustrated in Fig. \ref{fig:error}. The figure shows that the maximum error in the proper motion data components is 1.6 mas/yr for G $\le$ 20 mag and 0.71 mas/yr for G $\le$ 17 mag. %the data used as initial boundary conditions for this study, taken from \citet{Qin21}, are summarised in Table \ref{tab:fund}. 
%%%%%%%%%%%
% Fig. 1                                                   %
%%%%%%%%%%%%%%%%%%%%%%%%%%%%%%%%%%%%%%%%%%%%%%%%%%%%%%%%%%%%%%%%%%%%%%%%%
%%%%%%%%%%%%%%%%%%%%%%%%%%%%%%%%%
%%%%%%%%%%
% Fig. 2                                                   %
%%%%%%%%%%%%%%%%%%%%%%%%%%%%%%%%%%%%%%%%%%%%%%%%%%%%%%%%%%%%%%%%%%%%%%%%%
%%%%%%%%%%%%%%%%%%%%%%%%%%%%%%%%%

This study is intended to provide a comprehensive characterization of the four new OCs; QC1 through QC4. Therefore, the %that have been recently discovered. The 
Automated Stellar Cluster Analysis (ASteCA) code \citep{Perren15} is a suitable tool to achieve our goal because it is designed with many functions that use positional and photometric data to get the basic parameters of a cluster with a minimal user intervention. In addition and unlike other codes in literature, ASteCA is available online as an open source for public with full documentations. %A full description of the code is given in %\citet{Perren15}. 

Briefly, the code enables us to automatically compute most of the characteristic parametrs of OCs such as center coordinates, radius, and both the mass and the luminosity functions. The code is intergrated with Bayesian field star decontamination algorithm to assign membership probabilities using photometric data alone. Moreover, the presence of an isochrone fitting process, allows ASteCA to provide accurate estimates for a cluster's metallicity, age, extinction, and distance values, as well as their uncertainties, by generating synthetic clusters from theoretical isochrones and selecting the best fit using a genetic algorithm. A full description of the code is available in \citet{Perren15} and on the code website\footnote{ASteCA website: http://asteca.github.io/}.
%Computations of the structural (e.g., radial density profiles, clusters radii, and membership probabilities), as well as the fundamental (age, distance modulus, reddening, and mass) parameters are carried out using the The ASteCA package consists of three main independent analysis blocks. The first looks into the structure study which includes the determination of a cluster region identified primarily by an over-density, the second block is concerned with individual membership probability estimation for stars inside the over-density while the third block searches for the best-fit parameters.
%
%%%%%%%%%%%
% Table-1 %
%%%%%%%%%%%
\begin{table*}
\caption{The initial conditions adopted in this work to obtain data from Gaia EDR3 as taken from \citet{Qin21}.}
%\caption{The fundamental parameters of the four open clusters QCs obtained from the data analysis of Gaia DR2 by \citet{Qin21}.}
\label{tab:fund}
\centering
\begin{tabular}{lllll} \hline 
{\bf Parameters}& {\bf QC1} & {\bf QC2}	& {\bf QC3} & {\bf QC4} \\\hline								
$\alpha$	& 20$^h$ 20$^m$ 25$^s$.40 & 20$^h$ 37$^m$ 23$^s$.70 & 20$^h$ 41$^m$ 31$^s$.40 & 20$^h$ 36$^m$ 28$^s$.00	\\[0.8 ex]
$\delta$	& 39$^o$ 52' 15''.60 & 46$^o$ 28' 08''.40 & 46$^o$ 08' 31''.20 & 46$^o$ 52' 55''.20 \\[0.8 ex]
$l$	        & 77$^o$.635 & 84$^o$.802 & 84$^o$.976 & 85$^o$.036 \\[0.8 ex]
$b$	        & 1$^o$.932  & 3$^o$.278 & 2$^o$.513 & 3$^o$.654 \\[0.8 ex]
%$\mu_{\alpha}\cos\delta$ (mas yr$^{-1}$) & -2.30 $\pm$ 0.09 & -4.05 $\pm$ 0.06 & -2.25 $\pm$ 0.10 &	-2.51 $\pm$ 0.24 \\[0.8 ex]
%$\mu_{\delta}$ (mas yr$^{-1}$) & -3.48 $\pm$ 0.12 & -4.32 $\pm$ 0.06 & -4.46 $\pm$ 0.13 &	-5.43 $\pm$ 0.17 \\[0.8 ex]
%$\varpi$ (mas)	& 0.766 $\pm$ 0.02 & 0.424 $\pm$ 0.03	&	0.400 $\pm$ 0.03 & 0.424 $\pm$ 0.04 \\[0.8 ex]
%d (pc)	        & 1261	& 2223	& 2340	& 2230	\\[0.8 ex]
%V$_r$ (km s$^{-1}$)& From Gaia	&	---	&	---	&	---	\\[0.8 ex]
r$ _{cl}$ (arcmin)& 8.76 &	8.75	&	12.82	&	11.79	\\[0.8 ex]
%r$_c$ (arcmin)	& 3.42	&	3.57	&	2.23	&	3.12	\\[0.8 ex]
%r$_t$ (arcmin)	& 18.54	&	23.91	&	-	&	16.80	\\[0.8 ex]
%log (age/yr)	& 7.00	&	8.55	&	8.60	&	8.40	\\[0.8 ex]
%	        & ---	&	$^{\dag}$6.70	&	6.60	& ---\\[0.8 ex]
%Members (P)	& 72	&	114	&	124	&	130	\\[0.8 ex]
\hline
\end{tabular}
%\flushleft
%$^{\dag}$ These values were obtained by \citet{Liu19} from the analysis of Gaia DR2 for QC2 and QC3.
\end{table*}
%%%%%%%%%%%%%%%%%%%%%%%%%%%%%%%%%%%%%%%%%%%%%%%%%%%%%%
%%%%%%%%%%%%%%%%%%%%%%%%%%%%%%%%%%%%%%%%%%%%%%%%%%%%%
\section{Results and Discussion}
\label{res}
%This section presents and discusses the main results we obtained 
\subsection{\bf Astrometric structural analysis}
%\label{astr}
\subsubsection{\bf Re-determination of the cluster's centers\\}
%\label{cent}
% In this section we put Fig. 3 and Table 2

In order to determine the position of the new centers of the four open clusters, it is important to know the stellar distribution within the space of these clusters by finding the number of stars (the star count) in this space. % of the cluster. 
%The first step to find the distribution is to find 
 %This was obtained by extracting all the data points from EDR3 within the space of the clusters using a previously computed values for both the cluster center ($\alpha$,$\delta$) and radius (r$_{cl}$). In this work, we adopted data obtained by \citet{Qin21} using DR2, and listed in Table \ref{tab:fund}, but the radii we used in this work, to feed EDR3, was chosen arbitrary to be $\ge$ 2r$_{cl}$. 
Since the cluster diameters we used for our sample are greater than 10 arcmin (see \S~\ref{Gaia}), then we can divide the space along the right ascension (RA) and declination (Dec.) into equal bins each of size 1.00 arcmin (0.017 degrees) following \citet{Maciejewski07} and \citet{Maciejewski09}. The star counts we obtained %within the space of the four clusters 
are: 17,208 for QC1, 28,765 for QC2, 52,319 for QC3 and 37,117 stars in QC4. 

\begin{figure*}
\begin{center}
  \includegraphics[scale = 0.45]{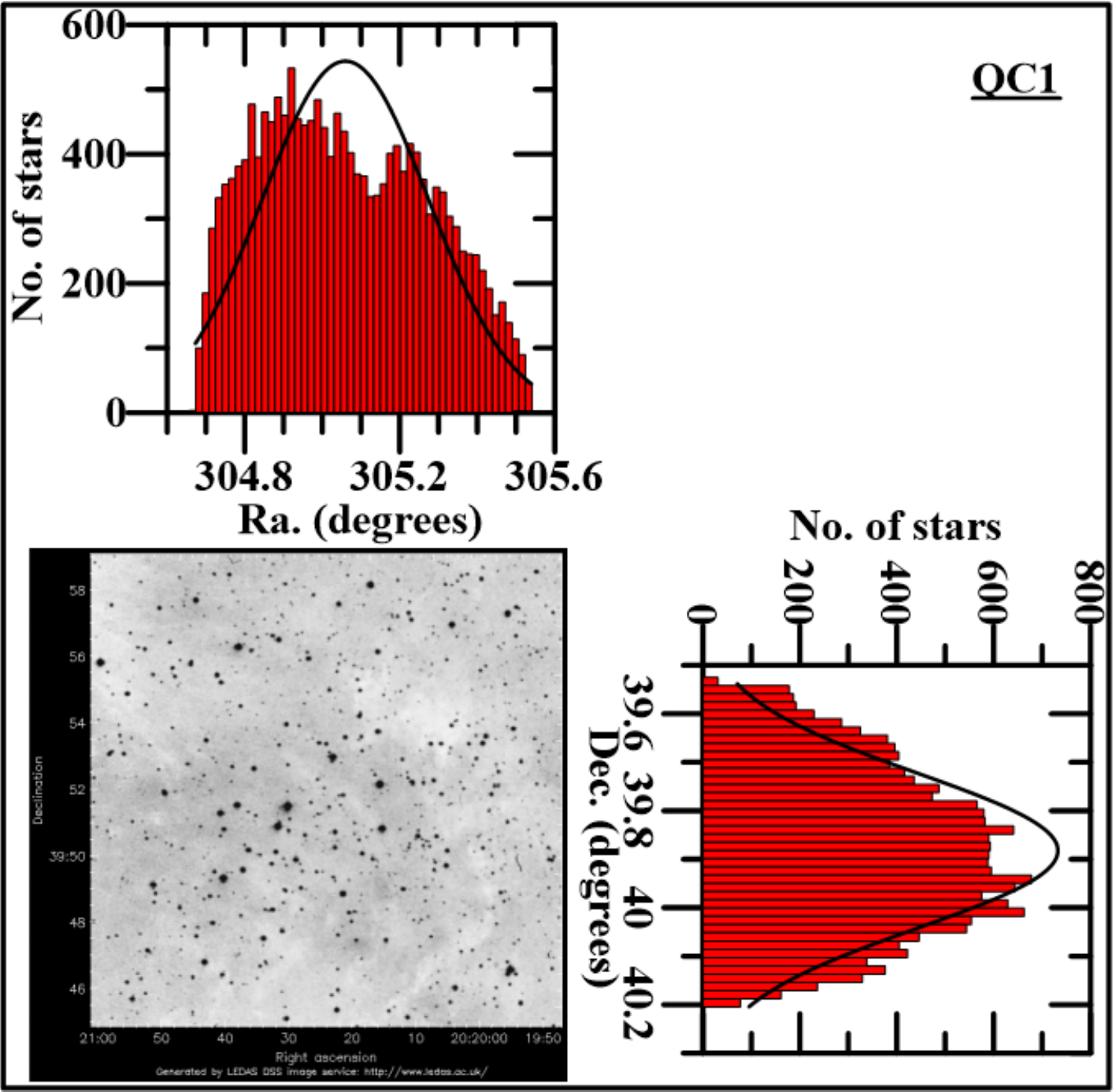}
  \includegraphics[scale = 0.45]{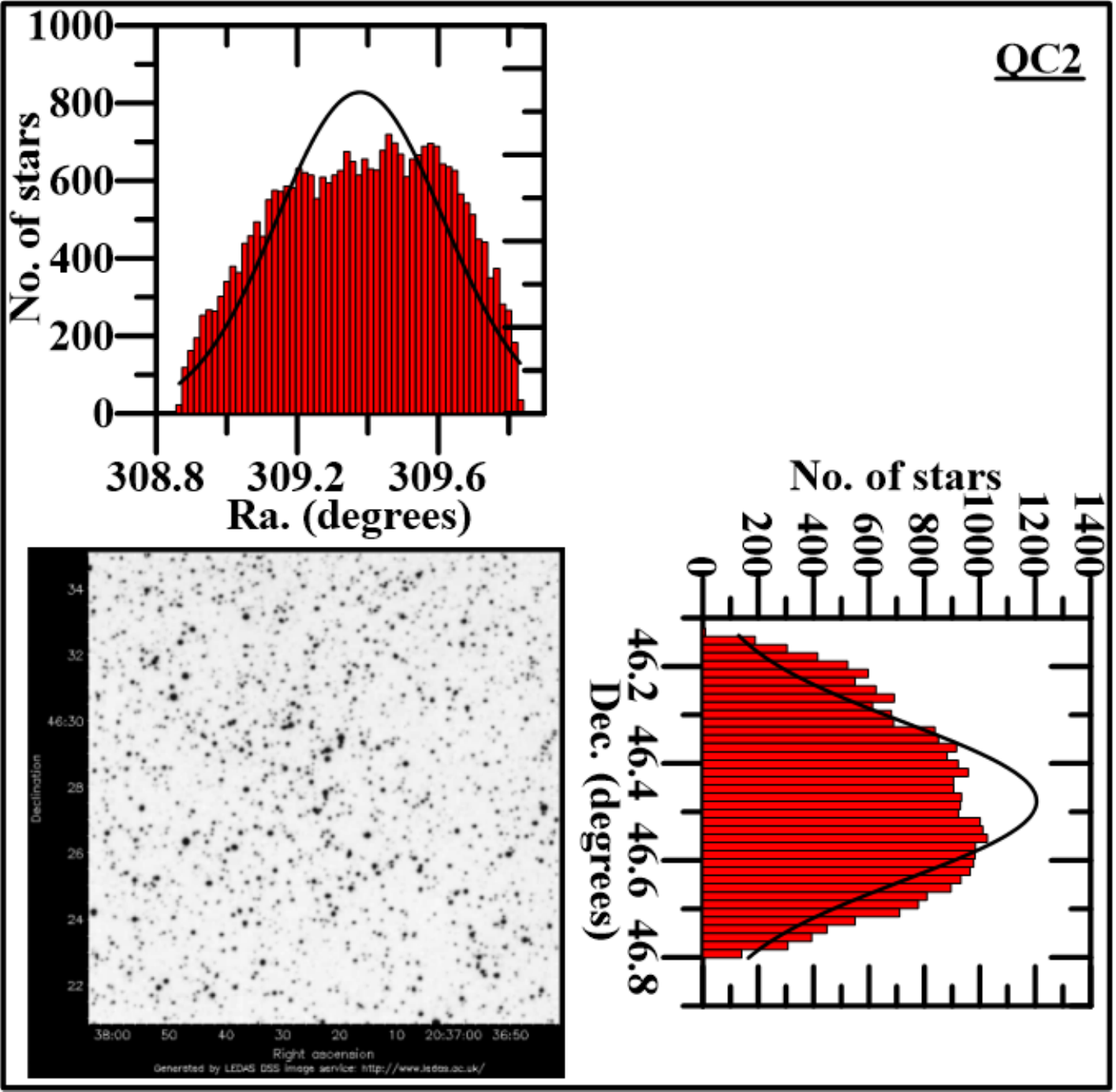}
%    \vspace{0.3cm}
  \includegraphics[scale = 0.45]{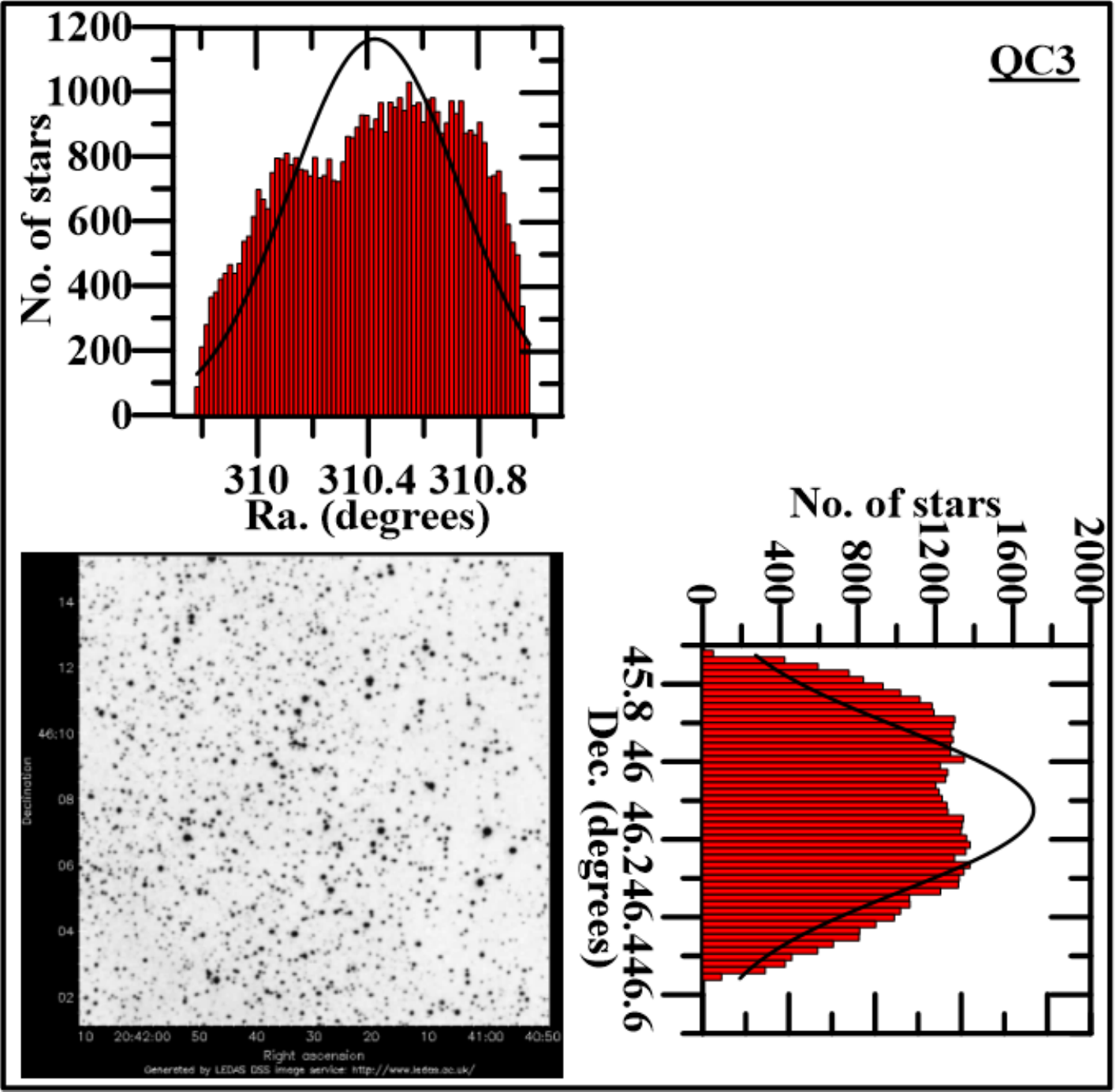}
  \includegraphics[scale = 0.45]{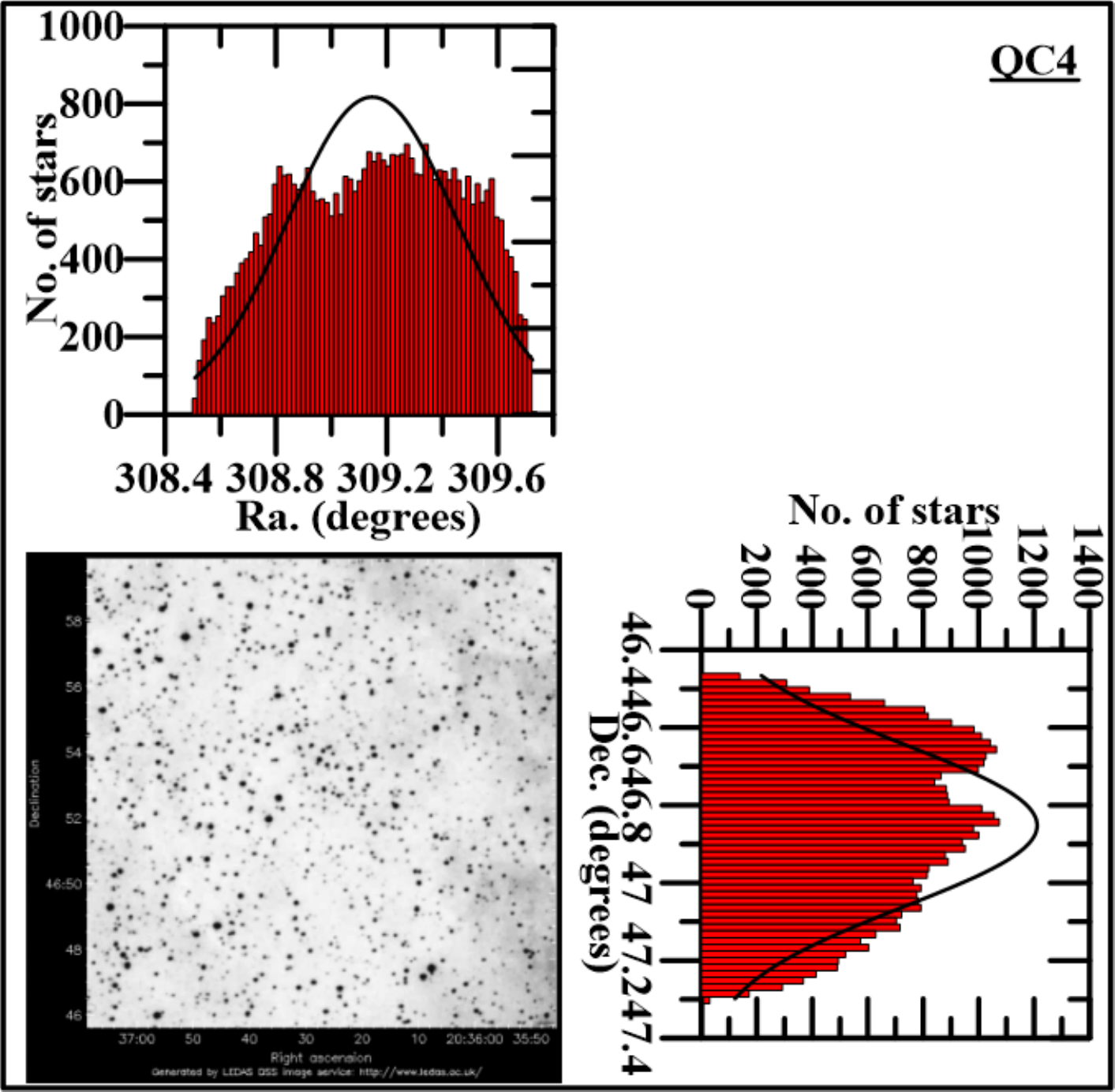}
\caption{The Gaussian fitting (solid lines) for the stars occupied within the space of each open cluster. The coordinates where the density peaks %(i.e. areas with the highest density) 
along the right ascension (RA) and the declination (Dec.) refers to the position of the new center of the cluster. The identification maps of the four QCs are taken from LEicester Database and Archive Service (LEDAS) Digitized Sky Survey (DSS) available at (https://www.ledas.ac.uk/DSSimage).}
\label{fig:3}
\end{center}
\end{figure*}

Fig. \ref{fig:3} represents the Gaussian distribution for the stars within the space of each cluster along RA and Dec. directions. %the right ascention and declination directions. 
The peak of the distribution in each direction, where star counts concentrate, marks the new position of the center of the cluster. The results showed that the positions of the new centers of the four clusters are in good agreement with those obtained by \citet{Qin21} in the RA direction while they show a minimal change in the Dec. direction ($\Delta\delta \le$ 1' 7''). 
The new estimated centers of the clusters QC1, QC2 , QC3 and QC4 in the equatorial ($\alpha$, $\delta$) and Galactic ($l$, $b$) coordinate systems are listed in Table \ref{tab:center}. 
%are (20$^h$ 20$^m$ 14$^s$.31, 39$^o$ 52' 56''.41), (20$^h$ 37$^m$ 30$^s$.79, 46$^o$ 28' 40''.34), (20$^h$ 41$^m$ 41$^s$.68, 46$^o$ 07' 24''.51) and (20$^h$ 36$^m$ 35$^s$.37, 46$^o$ 51' 07''.36), respectively. 
%The new centres we obtained for QC1, QC2 and QC3 are shifted towards larger values of right ascentions while for QC4 the position is shifted towards a lower value. For the declination component, QC1 and QC2 moved to higher locations while QC3 and QC4 moved to a lower positions. Table \ref{tab:center} lists the new\zmat{center positions of the OCs} %ly estimated values of the centers of the four clusters; both in the equatorial ($\alpha$, $\delta$) and Galactic ($l$, $b$) coordinates. % (see Table  \ref{tab:center}). 

%%%%%%%%%%%
% Table-2 %
%%%%%%%%%%%
\begin{table*}
\caption{The calculated coordinated of the new positions of the four clusters' centers in both equatorial ($\alpha$, $\delta$) and Galactic ($l$, $b$) systems.}
\label{tab:center}
\centering
\begin{tabular}{llllll} \hline 
%{\bf } & \multicolumn{4}{c}{\bf Cluster Center Coordinates$^{\text a}$}\\ 
{\bf  Cluster} & \multicolumn{2}{c}{\bf Equatorial Coordinates} &&\multicolumn{2}{c}{\bf Galactic Coordinates} \\[0.8 ex]
{\bf}& \multicolumn{1}{c}{\bf $\alpha$}& \multicolumn{1}{c}{\bf $\delta$} && \multicolumn{1}{c}{\bf $l$} & \multicolumn{1}{c}{\bf $b$} \\ \hline
{\bf QC 1}& 20$^h$ 20$^m$ 14$^s$.31 & 39$^o$ 52' 56''.41 && 77$^o$.6240 & 1$^o$.9678 \\[0.8 ex]
{\bf QC 2}& 20$^h$ 37$^m$ 30$^s$.79 & 46$^o$ 28' 40''.34 && 84$^o$.8011 & 3$^o$.2673 \\[0.8 ex]
{\bf QC 3}& 20$^h$ 41$^m$ 41$^s$.68 & 46$^o$ 07' 24''.51 && 84$^o$.9793 & 2$^o$.4785\\[0.8 ex]
{\bf QC 4}& 20$^h$ 36$^m$ 35$^s$.37 & 46$^o$ 51' 07''.36 && 85$^o$.0249 & 3$^o$.6190 \\[0.8 ex]
\hline 
\end{tabular}
%\flushleft
%References: (a) \citealt{asp09} , (b) \citealt{awad16}
\end{table*}
%%%%%%%%%%%%%%%%%%%%%%%%%%%%%%%%%%%%%%%%%%%%%%%%%%%%%%
%%%%%%%%%%%
% Fig. 3                                                   %
%%%%%%%%%%%%%%%%%%%%%%%%%%%%%%%%%%%%%%%%%%%%%%%%%%%%%%%%%%%%%%%%%%%%%%%%%

%%%%%%%%%%%%%%%%%%%%%%%%%%%%%%%%%
\subsubsection{\bf The radial density profiles (RDP)\\}
\label{rdp}

\begin{figure*}
\begin{center}
%% trim left bottom right top
    \includegraphics[trim= 0.0cm 0.0cm 0.0cm 0.0cm,clip=true,width=12cm]{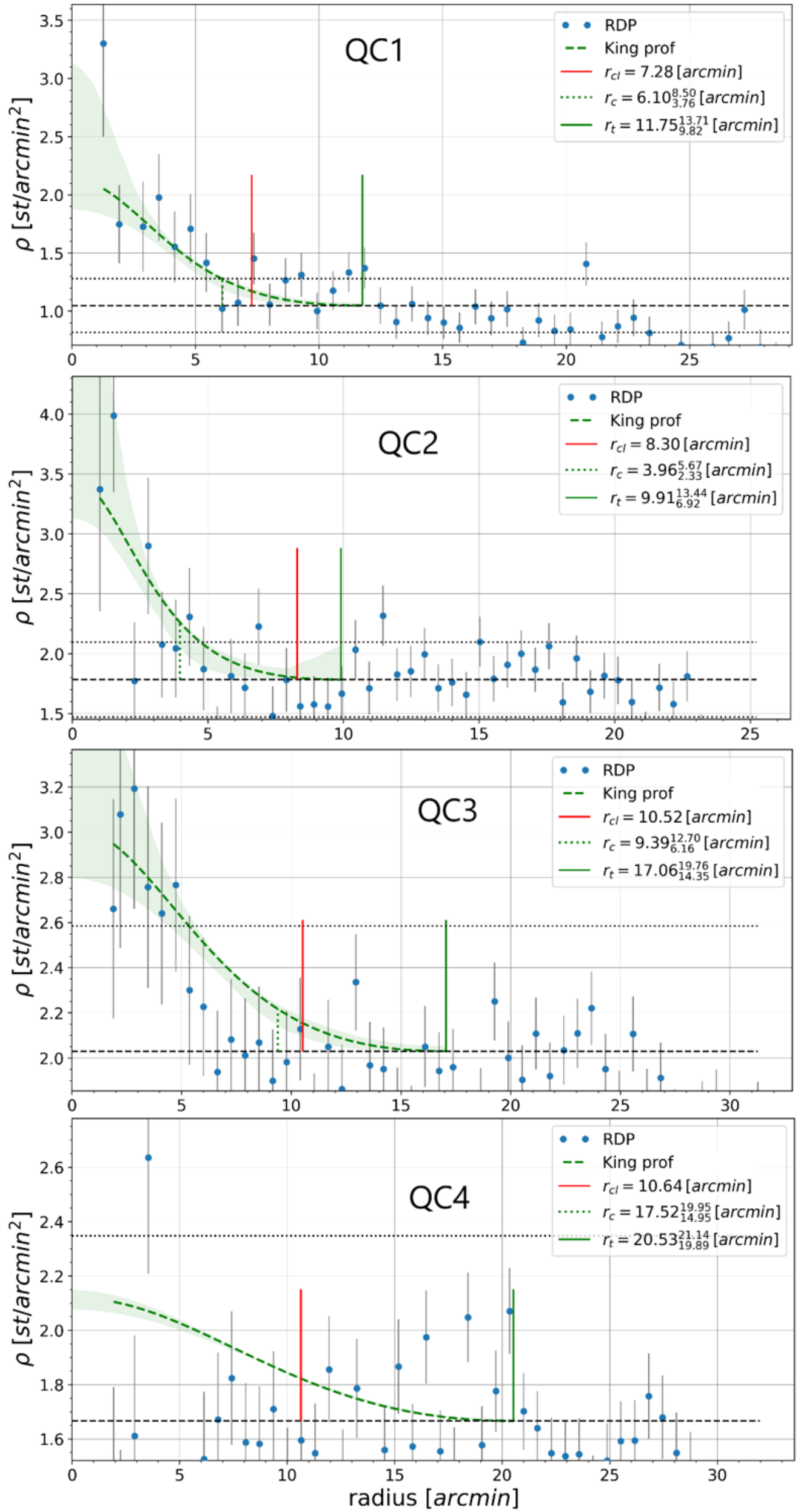}
%    \includegraphics[trim= 1.0cm 2.6cm 0.0cm 0.0cm,clip=true,[width=11cm]{plots/fig4/Fig4-1.pdf}
%    \hfill
%    \includegraphics%[trim= 1.0cm 2.5cm 0.5cm 0.0cm,clip=true,[width=11cm]{plots/fig4/Fig4-2.pdf}
%   \vspace{0.3cm}
%   \includegraphics%[trim= 1.0cm 3.2cm 0.5cm 0.0cm,clip=true,[width=11cm]{plots/fig4/Fig4-3.pdf}
%    \hfill
%    \includegraphics%[trim= 1.0cm 0.4cm 0.3cm 0.3cm,clip=true,[width=11cm]{plots/fig4/Fig4-4.pdf}
\caption {The radial density profile (RDP) for the 4 open clusters (QC1, QC2, QC3, and QC4) with the aid of the ASTeCA code (blue dots). The green dashed line and shaded area represent the King's density profile while the black dashed and dotted lines denote the background field density ($\rho_{bg}$) and the central surface density ($\rho_{o}$), respectively. Vertical lines indicates the structural parameters (r, r$_{cl}$, \& r$_t$) of each cluster, see the figure key.}
\label{fig:4}
\end{center}
\end{figure*}

The slight change in the position of a cluster center may lead to a change in the star count of the cluster. Therefore, we extracted a new data sheet for each of the four clusters to re-estimate its star count, within %the cluster space, with 
the same radii used in identifying the new center (in \S~\ref{Gaia}), but centered at the new position of these centers. The new counts are: 17,604 for QC1, 29,044 for QC2, 53,389 for QC3 and 38,041 for QC4. From these new star counts, we constructed the Radial Density Profile (RDP) from which we computed the cluster radii by obtaining the best fitting to King's density profile given in Eq. \ref{eq:1} \citep{King62}. Fig. \ref{fig:4} illustrates the four RDPs of the four clusters fitted by King's profile (green dash line and shaded area).
%({\it hereafter}  , for each cluster, which represent the density variation along the line-of-sight. The RDP will enable us to explore the structure of the cluster. A first step towards constructing the RDP is to 

To know the radial density distribution we divided the cluster into a number of cocentric zones, centered at the new center coordinates, with equal sizes and computed the star density within each of these zones. The zone widths are 0'.64, 0'.51, 0'.63 and 0'.65 %0.31', 0.38', 0.26' and 0.05' 
for QC1, QC2, QC3 and QC4, respectively. % (See Table \ref{tab:3}). %count within that divide the total area of the cluster into zones. These zones are centered at the new defined center of the cluster under study. 
For each zone, the stellar number density ($\rho$) is the number of stars (N) per zone area (A), and hence $\rho =  N / A $. %Hence, for the i$^{th}$ zone, the number density of stars is $\rho_i =  N_i / A_i $, where N$_i$ is the number of stars and A$_i$ is the area of the i$^{th}$ zone, respectively. 
The variation of the stellar number density along the cluster radius defines the radial density profile of the cluster. 

With the aid of ASteCA code, we generated four RDP, one for each cluster, and fitted it with King's profile in Eq. \ref{eq:1}, to estimate the internal cluster structural parameters such as the cluster core radius (r$_c$), limiting radius (r$_{cl}$), tidal radius (r$_t$), central surface density ($\rho_o$), and background surface density ($\rho_{bg}$). All radii are measured in arcminutes (arcmin) while all densities are in stars/arcmin$^2$.
\begin{equation}
\label{eq:1}
\rho(r)~=~\rho_{bg}+\frac{\rho_{o}}{1+\left(\sfrac{\text r}{\text{r}_{c}}\right)^2},
\end{equation}

The core radius r$_{\text{c}}$ is the distance at which the value of $\rho(r)$ becomes half that of the central density $\rho_o$. 
We estimated core radii: 6.10$^{+8.50}_{-3.76}$ (QC1), 3.96$^{+5.67}_{-2.33}$ (QC2), 9.39$^{+12.70}_{-6.16}$ (QC3) and 17.52$^{+19.95}_{-14.95}$ (QC4). Our results for all clusters, apart from QC2 (despite the uncertainty), are not in agreement with results of \citet{Qin21}. 

The cluster limiting radius r$_{cl}$ is the radius at which the line represents the value of the background density (dash black lines in Fig. \ref{fig:4}) intersects the King profile model fitting curve. At this point, the background star density $\rho_{b}$ is given by ($\rho_{bg} + 3\sigma_{bg}$) where $\sigma_{bg}$ is the uncertainty of $\rho_{bg}$. \citet{Bukowiecki11} derived an expression for the limiting radius to be
\begin{equation}
\label{eq:2}
r_{cl} = r_{c}~\sqrt{\frac{\rho_{o}}{3~\sigma_{bg}} - 1}
\end{equation}

%limit after which stars are no longer gravitationally bound to the cluster. Thus, r$_{cl}$ may reflect the boundary of the cluster and hence its radius. This radius is calculated by comparing $\rho(r)$, from Eq. \ref{eq:1}, to a background density level $\rho_{bg}$ (black dash lines in Fig. \ref{fig:4}) given by ($\rho_{bg} + 3\sigma_{bg}$) where $\sigma_{bg}$ is the uncertainty of $\rho_{bg}$.
\
Our calculations showed that the cluster radii are 7'.28, 8'.30, 10'.52 and 10'.64 for QC1, QC2, QC3 and QC4, respectively. These values are in good agreement (83--95\%) with those obtained by \citet{Qin21}. 
The large uncertainties in the computed values of the cluster radii may imply a non-spherical geometry of these open clusters. Therefore, the clusters might be extended along their radii. 

Other parameters may be used to characterize the structure of open clusters such as the density contrast parameter ($\delta_c$) and the concentration parameter (C) (e.g. \citealt{King66, Peterson75, Bonatto09, Santos12, Maurya21}). So far, the present study is the first to compute these two parameters for the four open clusters QC1--QC4. 
The $\delta_c$ is the stellar density contrast of these clusters against the background population. It is a measure of the compactness of the cluster \citep{Bonatto09} and can be expressed mathematically as 
$$\delta_c = 1 + \frac{\rho_{o}}{\rho_{bg}}$$ 
For QC1 through QC4, the estimated $\delta_c$ is 2.60, 3.91, 1.87 and 2.68 which indicates that the four clusters are scattered with respect to their background density. This finding is supported by \citet{Bonatto09} who found that for compact clusters (nearly spherical in geometry) the contrast parameter lies in the range (7 $\le ~ \delta_c ~\le$  23). In addition, this result may give support to our explanation of the large obtained uncertainty in the estimated radii of the clusters. 

%\citet{Peterson75} 
\citet{King66} introduced the concentration parameter (C) to be the ratio between the cluster limiting, r$_{cl}$, and core, r$_{c}$, radii % to characterize stellar clusters and their structures by  which may  and is defined} as  
$$C = r_{cl} / r_c $$
which may reflect the central concentration of the cluster.
%The range of the C parameter in \citet{King66} is 0.5 -- 2.5 for their observed sample. %The parameter may reflect the central concentration of the cluster \citep{King66}. 
On the other hand, \citet{Santos12} defined the C parameter to be the reciprocal of \citet{King66} definition. \citet{Santos12} concluded that young clusters are expected to have small C values because most of their members are still concentrated around the center and did not have enough time to spread away towards the edge of the cluster. The authors obtained C range (0.04 -- 1.03) which is equivalent to (0.97 -- 25) using King's definition (adopted in this study). Our obtained C range is (0.60 -- 2.18) which is in fair agreement with \citet{Santos12} and about 2 to 4 times lower than the range (2.45 -- 5.75) obtained by \citet{Qin21}; derived from their Table 2.
%In the present work, the defination of the concentration parameter is the receprocal of that of our estimated values of C are ranged between 0.6 and 2.18 that are in fair agreement with the range (0.04 -- 0.4 25 - 0.9) obtained by \citet{Santos12} for the selection of young clusters they studied. However, our results are in disagrees with the values (2.4 -- 5.7) derived from Table 2 in \citet{Qin21}. 

Moreover, \citet{Maciejewski07} found that the limiting radius is (2 -- 7) times the value of the core radius (see their Table 3). These values are in line with our estimation of only QC2 and larger than the values of the rest of the clusters, however this range is in good agreement with \citet{Qin21} results. 

%\zma{$\quad$ 
The results for QC4 are irregular and further analysis is needed to be able to drew a solid conclusion about it. %These results are in good agreement with the results of \citet{Maciejewski07} in which they found that the angular size of the coronal region \zma{surrounding a cluster} has the range (2--7)r$_c$. %\citet{Nilakshi02} found that the angular size of the coronal region \zma{surrounding a cluster} is almost 6r$_c$ and \citet{Maciejewski07} showed that r$_{cl}$ is in the range (2--7)r$_c$. The results of our analysis revealed that the concentration parameter C for the four studied QCs is ranged between 0.60 and 2.18. 
However, given the small value of the C parameter (0.60) due to the core radius (17'.52) is greater than the cluster radius (10'.64) which is coming from the fact that both the cluster density and the field density for QC4 have nearly the same level, then we may conclude that QC4 might not be a genuine cluster.
The numerical results of both parameters for the four studied QCs are given in Table \ref{tab:3}.
%%%%%%%%%%%%%%%%%%%%%%%%%%%%%
% Table 3 
%%%%%%%%%%%%%%
\begin{table*}
\caption{Our obtained structural properties for the 4 OCs as computed from the RDP fitted by King's density profile, using ASteCA code, in comparison with the results of \citet{Qin21} indicated as Q21 in the last column of the table.}%other published ones -- TBmodified}
\centering
%\resizebox{\textwidth}{!}{
\begin{tabular}{llllll}
%\topline\hline 
\hline
%Table 3: Our obtained structural properties of the QCs cluster with other published ones.
%\hline
{\bf Parameters}&{\bf QC1}&{\bf QC2}&{\bf QC3}&	{\bf QC4} &{\bf Ref.}\\   \hline
%-------------------
%RDP-bin width (arcmin)	& 0.31 & 0.38 & 0.26 & 0.05 &   \\[0.8 ex]
%-----------------
r$_c$ (arcmin)&	6.1$^{+8.50}_{-3.76}$ &3.96$^{+5.67}_{-2.33}$& 9.39$^{+12.7}_{-6.16}$&$17.52^{+19.95}_{-14.95}$& \\[0.8 ex]
	      
              & 3.42 $\pm$ 0.67 & 3.57 $\pm$ 0.25&2.23 $\pm$ 0.15&3.12 $\pm$ 0.25& Q21\\[0.8 ex]
%-------------
r$_{cl}$ (arcmin)&7.28&8.30&10.52&10.64& 	\\[0.8 ex]
	
                &8.76 $\pm$ 0.76&8.75 $\pm$ 0.63&12.82 $\pm$ 0.83&11.79 $\pm$ 0.81& Q21	\\[0.8 ex]
%-------------------------
r$_t$ (arcmin) & 11.75$^{+13.7}_{-9.8}$ & 9.91$^{+13.4}_{-6.9}$  & 17.1$^{+19.7}_{-14.3}$  & 20.53$^{+21.1}_{-19.8}$  \\ [0.8ex]
%5.83 $\pm$ 1.21 & 5.26 $\pm$ 1.15 & 9.51 $\pm$ 1.54 & 9.66 $\pm$ 1.55& \\[0.8 ex] %(ASteCA)
%-----------------------
%r$_t$  (pc) (dynamical)&7.42 $\pm$ 1.72&7.71 $\pm$ 1.78	&8.44 $\pm$ 1.90&7.78 $\pm$ 1.79& \\[0.8 ex]
%	                       &6.80 $\pm$ 1.30	&15.46 $\pm$ 1.97&---&10.90 $\pm$ 1.65& Q21\\[0.8 ex]
                       & 18.54 & 23.91 & --- & 16.80 & Q21\\[0.8 ex]
% ------------------------------
$\rho_{bg}$ (stars arcmin$^{-2}$)&1.047 $\pm$ 0.23&1.783 $\pm$ 0.32&2.029 $\pm$ 0.26&1.666 $\pm$ 0.68& \\[0.8 ex]
%---------------
$\rho_o$ (stars arcmin$^{-2}$)&1.673 $\pm$ 0.08&5.178 $\pm$ 0.05&1.759 $\pm$ 0.08&2.793 $\pm$ 0.06& \\[0.8 ex]
%-------------
$\delta_c$& 2.60 & 3.91 & 1.87 & 2.68	& 	\\[0.8 ex]
%----------------------
C	& 1.18	& 2.18 & 1.07& 0.60	& 	\\[0.8 ex]
%----------------------

%-------
\hline
\label{tab:3}
\end{tabular}
%}\flushleft  {\bf References:} (1) \citealt{Qin21}\\ \zmaa{A 2nd look to all tables is mandatory}%\sc dr what is the eq. for $r_t$ dynamical - this should not be in this table. Also The values reported to Qin 2021, are not the ones reported in his table 1 - please correct these numbers}
%}
\end{table*}
%%%%%%%%%%%%%%%%%%%%%%%%%%%%%%%%%%
%%%%%%%%%%%
% Fig. 4                                                   %
%%%%%%%%%%%%%%%%%%%%%%%%%%%%%%%%%%%%%%%%%%%%%%%%%%%%%%%%%%%%%%%%%%%%%%%%%
%%%%%%%%%%%%%%%%%%%%%%%%%%%%%%%%%
 
\subsubsection{\bf Astrometric parameters and distance determination\\}
\label{par}
In order to proceed with the analysis and parameter investigations, it is very important to identify true (most probable) stellar membership of each of the clusters in question. Thus, we performed a membership analysis using the available astrometric parameters (proper motion and parallax) from Gaia EDR3 database. 
In order to identify the cluster members, we applied the selection criteria of \citet{Qin21} to retrieve stellar data from Gaia EDR3. We obtained a full data sheet for all stars that have a photometric magnitude (G $<$ 17 mag) which corresponds to an uncertainty of 0.20 mas/yr and 0.10 mas in the proper motion ($\sigma_{\mu}$) and the parallax ($\sigma_{\varpi}$), respectively. All the data points were entered into the ASteCA code %\citep{Cabrera90} 
which we used to assign membership probability by searching for a meaningful stellar over-densities and compare them to the surrounding stellar field. For this study, only stars with probabilities (P $\ge$ 50\%) are assigned as cluster most probable members. The results of the ASteCA code revealed that the number of the most probable members for the four clusters are 118 (QC1), 142 (QC2), 210 (QC3), and 110 (QC4). This result is what we are relying on throughout the rest of this study in computing the other parameters that require a knowledge of the true membership of the cluster.

\begin{figure*}
\begin{center}
%% trim left bottom right top
\begin{minipage}{0.6\textwidth}
  \includegraphics[width = 5.3cm]{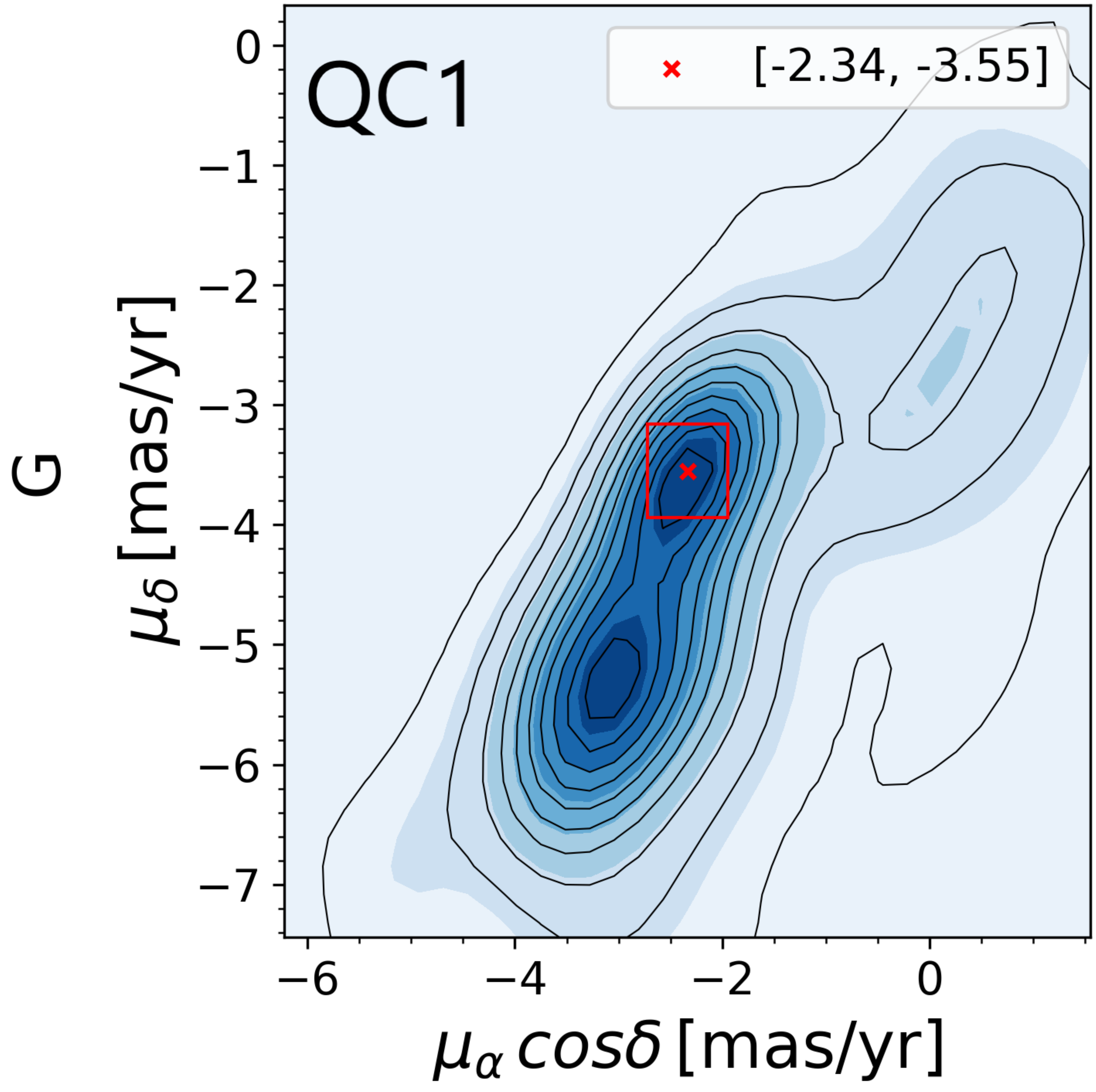}
  \includegraphics[trim= 0.2cm 0cm 0.0cm 0.0cm,clip=true,width = 5.3cm]{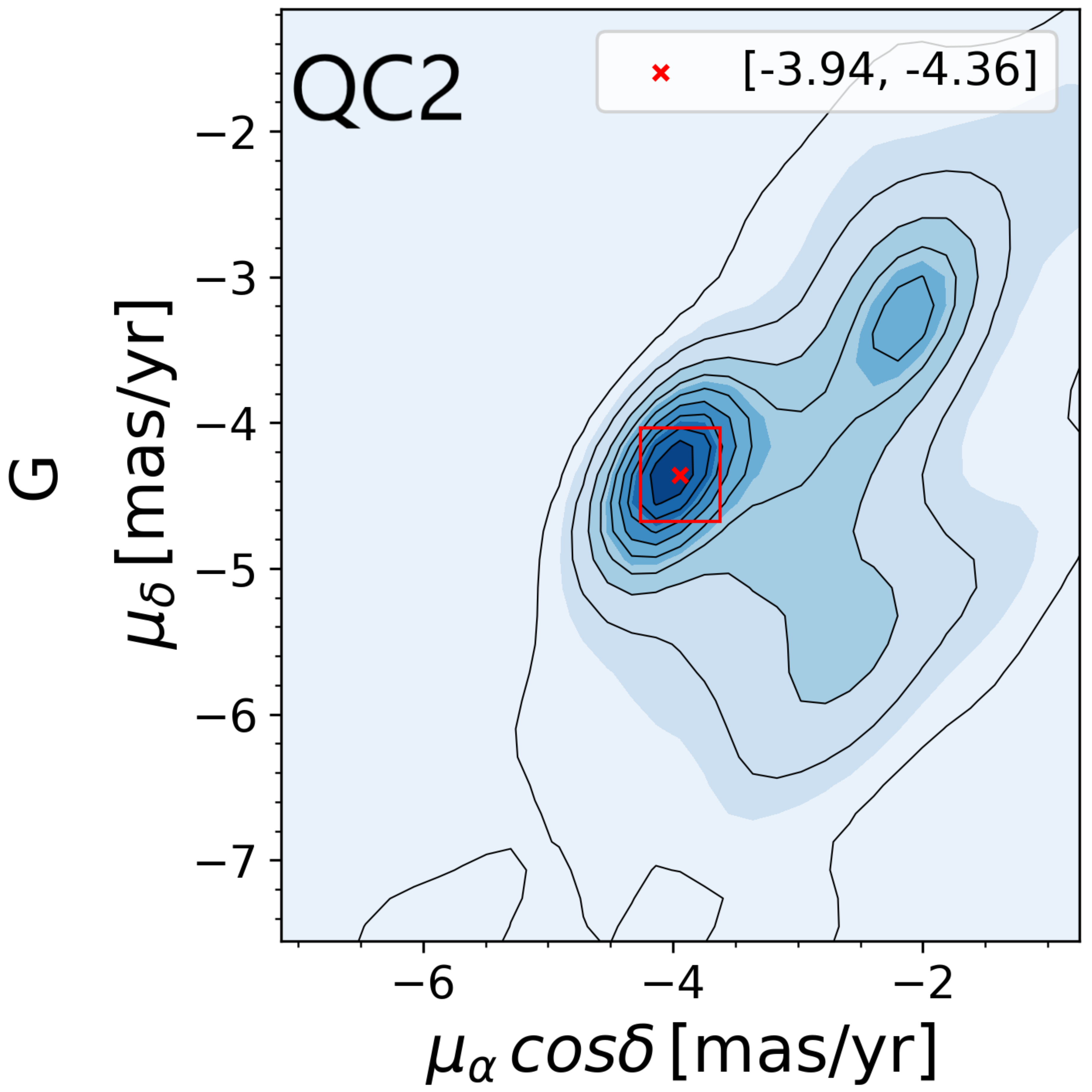}
\vspace{0.5cm}%\hfill
  \includegraphics[width = 5.3cm]{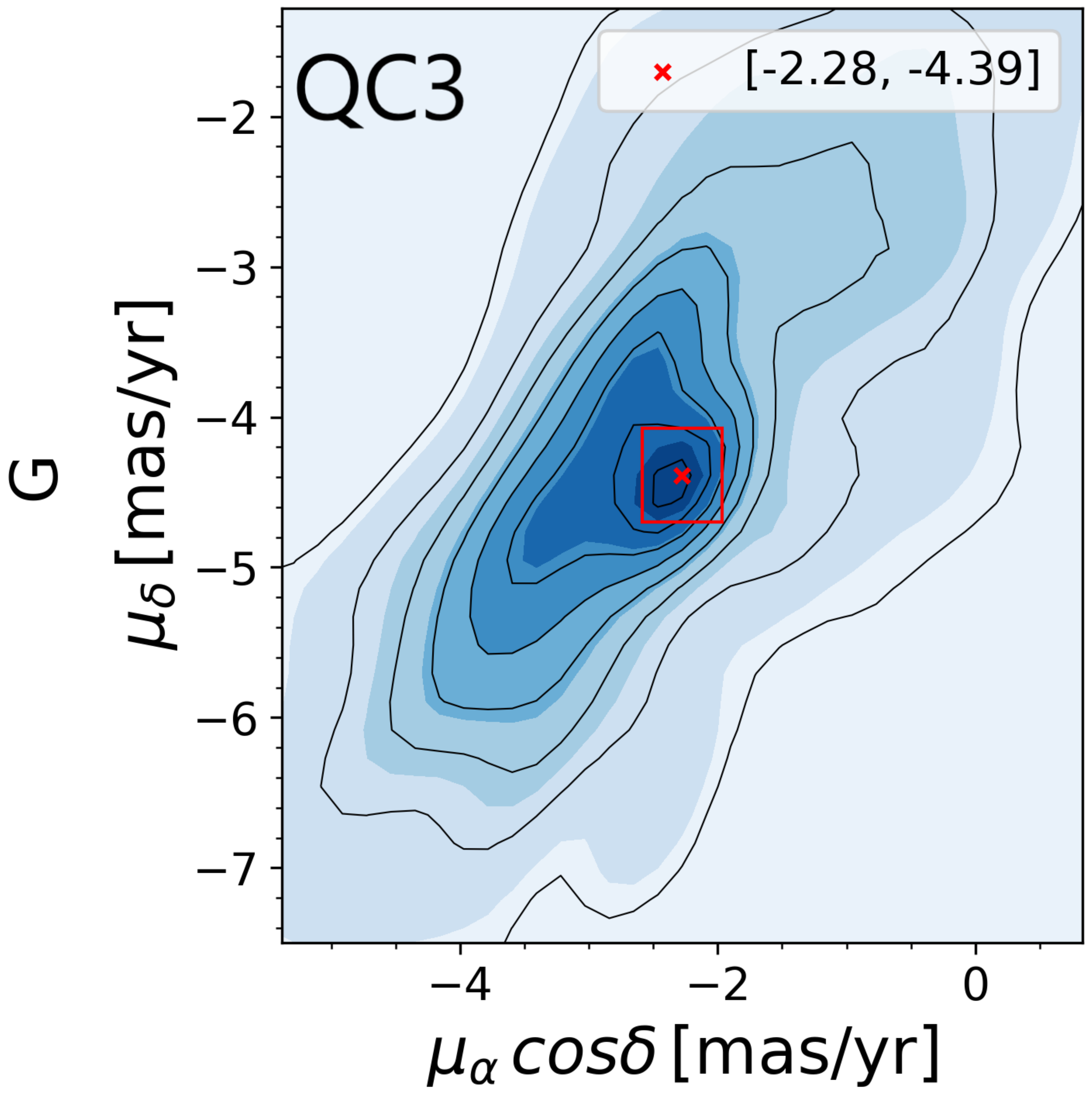}
  \includegraphics[width = 5.3cm]{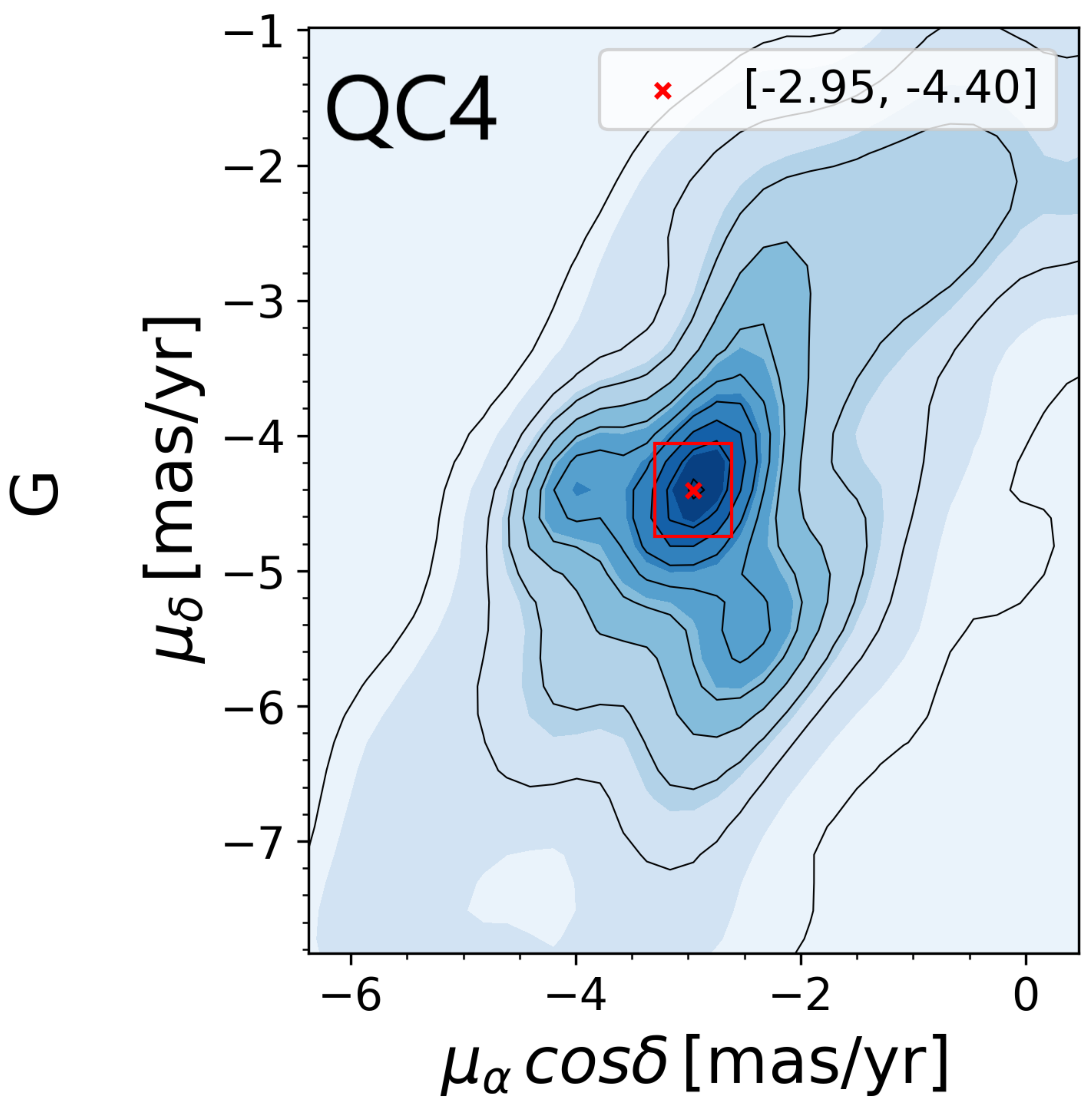}
\end{minipage}
%\vspace{1cm}
\begin{minipage}{0.6\textwidth}
  \includegraphics[width = 5.3cm]{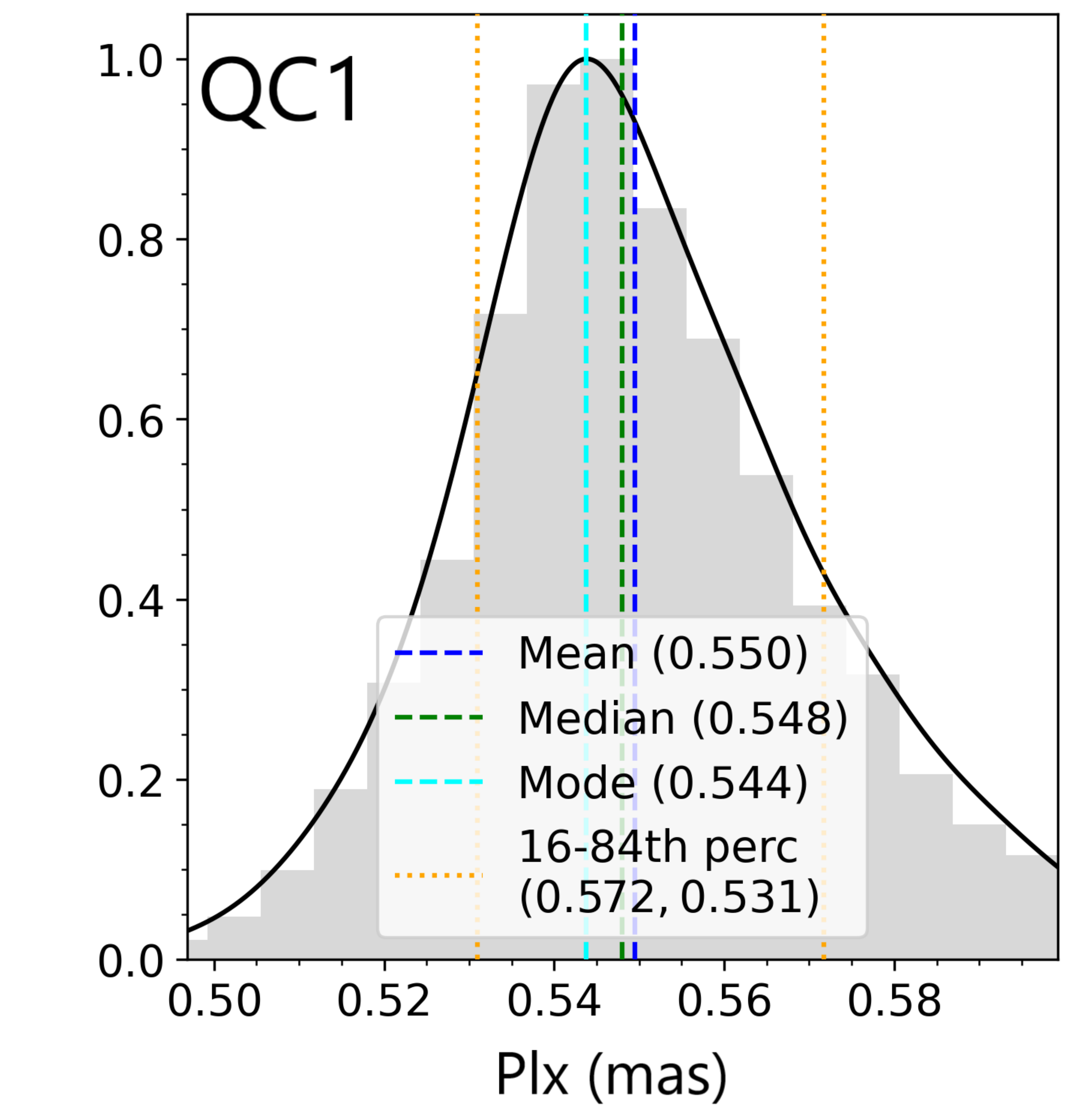}
  \includegraphics[width = 5.0cm]{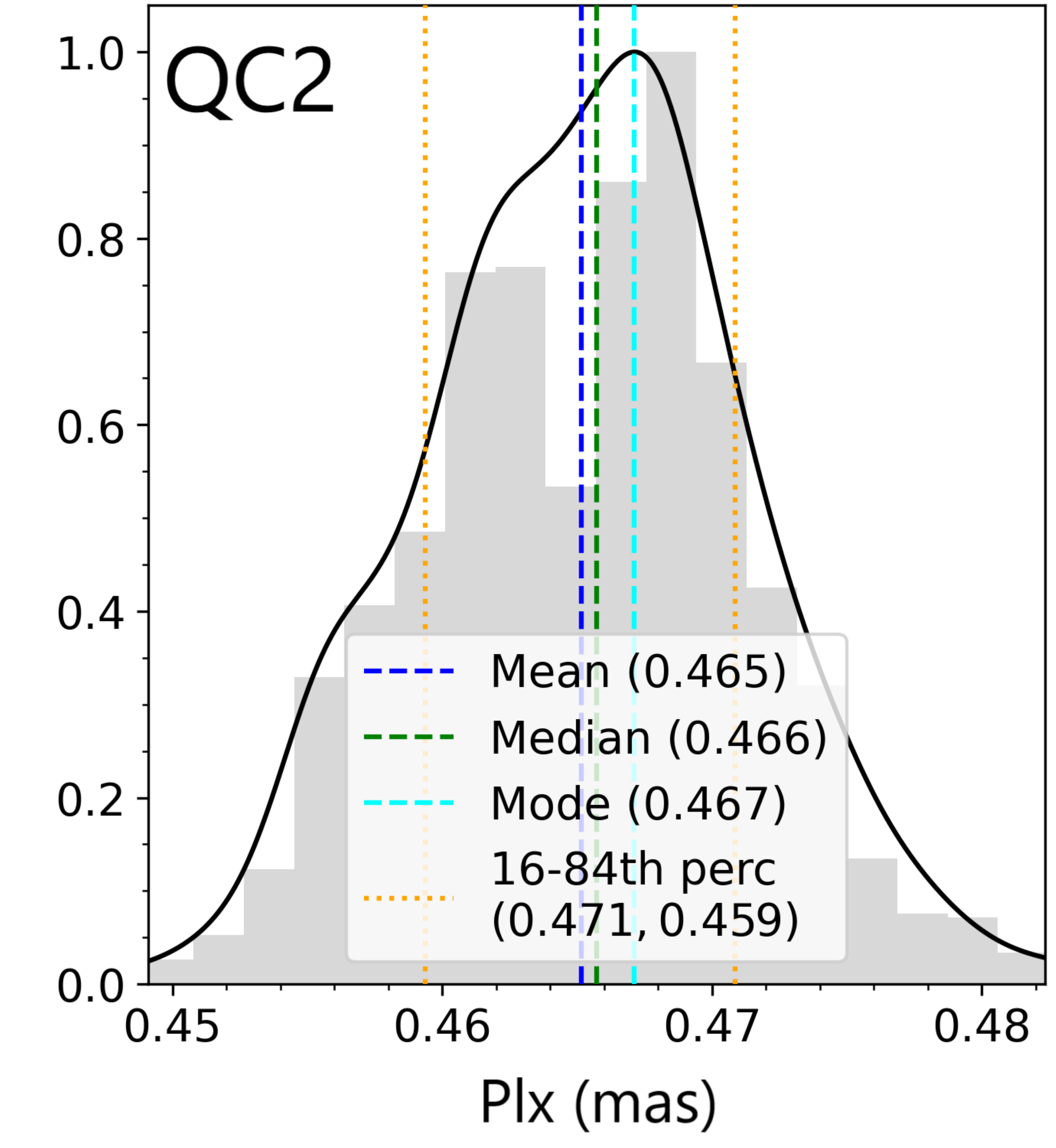}
  \includegraphics[width = 5.3cm]{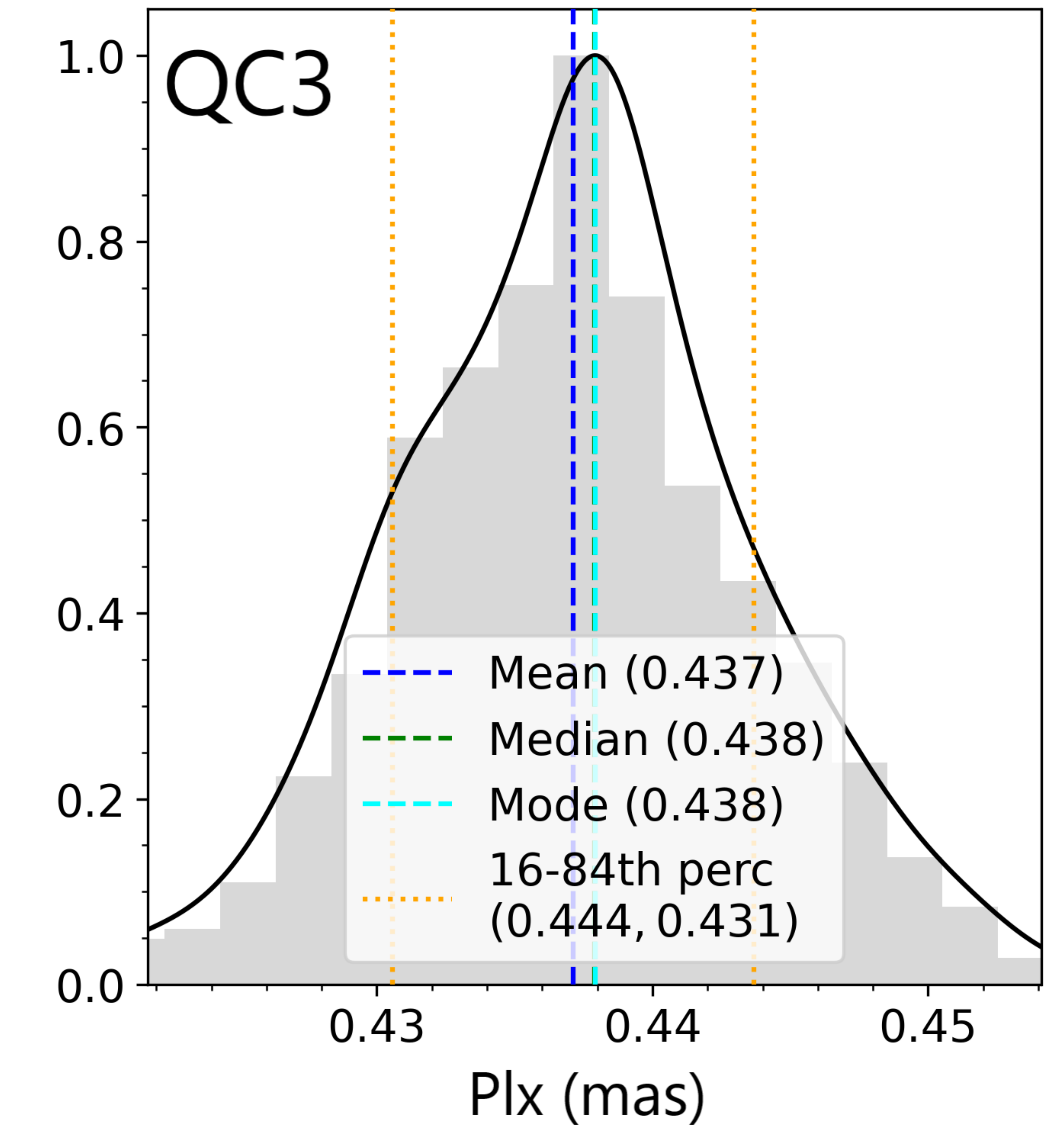}
  \includegraphics[width = 5.3cm]{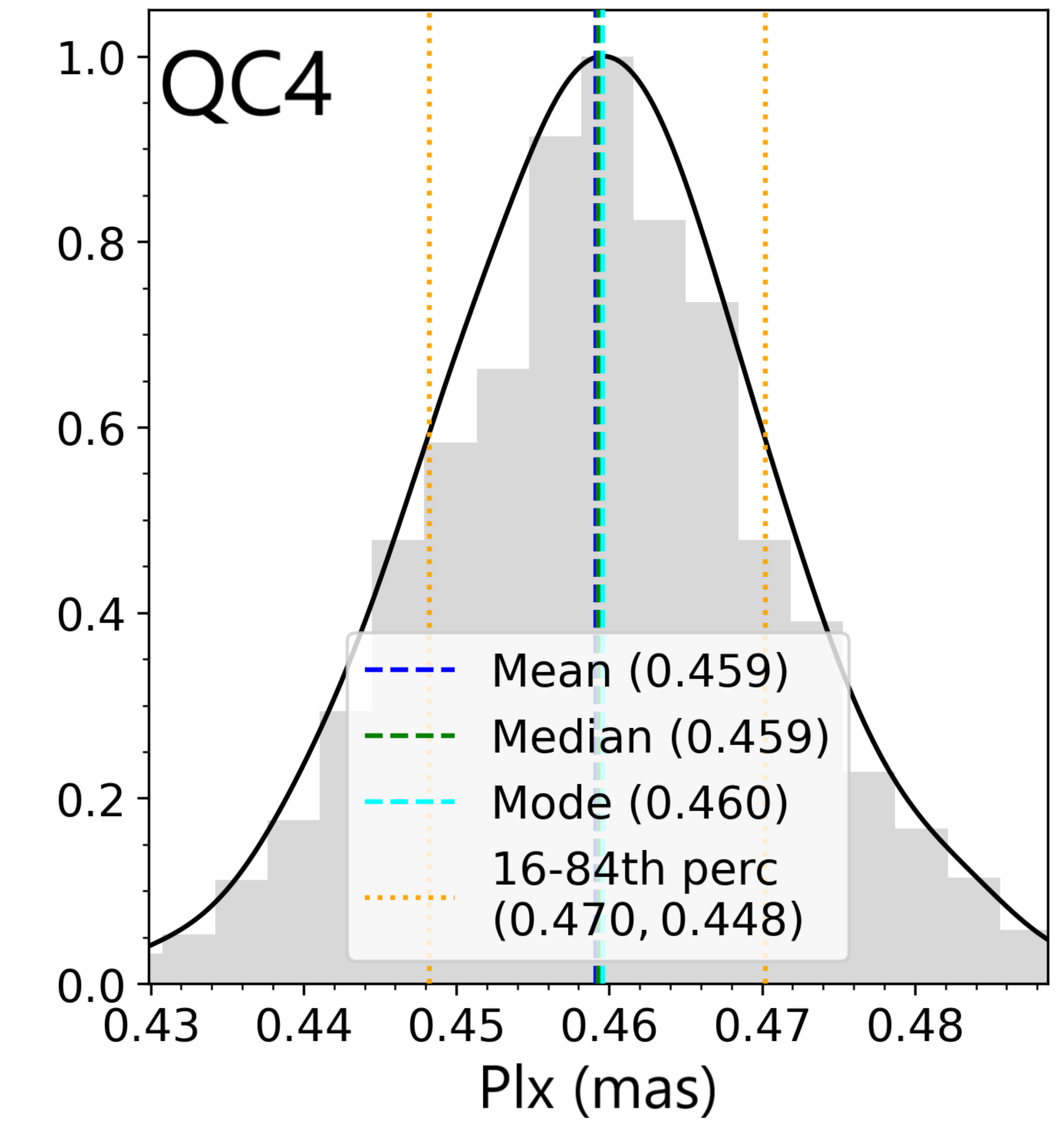}
\end{minipage}
\caption {Upper panel: The distribution of the mean proper motion ($\mu$; mas/yr) in both directions of right ascension and declination. Lower panel: The normalized parallax distribution for all stellar member candidates in the cluster space. }
\label{fig:mu}
\end{center}
\end{figure*}

For each cluster, we plotted its stellar distribution in the proper motion space ($\mu_{\alpha}\cos{\delta}$, $\mu_{\delta}$), see upper panels in Fig. \ref{fig:mu}, from which we determined the mean proper motion of the cluster by acquiring a Gaussian fitting along the corresponding directions. %In this way, we found mean PMs in the right ascension and declination directions for all clusters and are listed in Table 4. 
The lower panel of Fig. \ref{fig:mu} are four histograms that represent the parallax distribution of the candidate members of the four clusters with bin sizes (in mas) of 0.0060 (QC1), 0.0017 (QC2), 0.0020 (QC3), and 0.0034 (QC4). The black line on the histograms is the Gaussian fitting from which the mean parallax of each cluster was estimated. 

At this stage of the analysis, we are capable of computing the distance to each cluster knowing its parallax. The calculated distances (d$_{plx}$, in pc) are: 1819 $\pm$ 43, 2151 $\pm$ 64, 2288 $\pm$ 48 and 2179 $\pm$ 47, for QC1, QC2, QC3 and QC4, respectively. These results are in good agreement with those obtained by \citet{Qin21}.
%%%%%%%%%%%
% Fig. 5                                                   %
%%%%%%%%%%%%%%%%%%%%%%%%%%%%%%%%%%%%%%%%%%%%%%%%%%%%%%%%%%%%%%%%%%%%%%%%%
%%%%%%%%%%%%%%%%%%%%%%%%%%%%%%%%%
%\subsubsection{\bf Photometric parameters\\}
%\label{age} 

\subsection{\bf The photometric analysis}
\subsubsection{\bf Age, reddening, and the distance modulus\\}
\label{phot}
\begin{figure*}
\begin{center}
%% trim left bottom right top
%\begin{minipage}{0.99\textwidth}
  \includegraphics[width = 6.3cm]{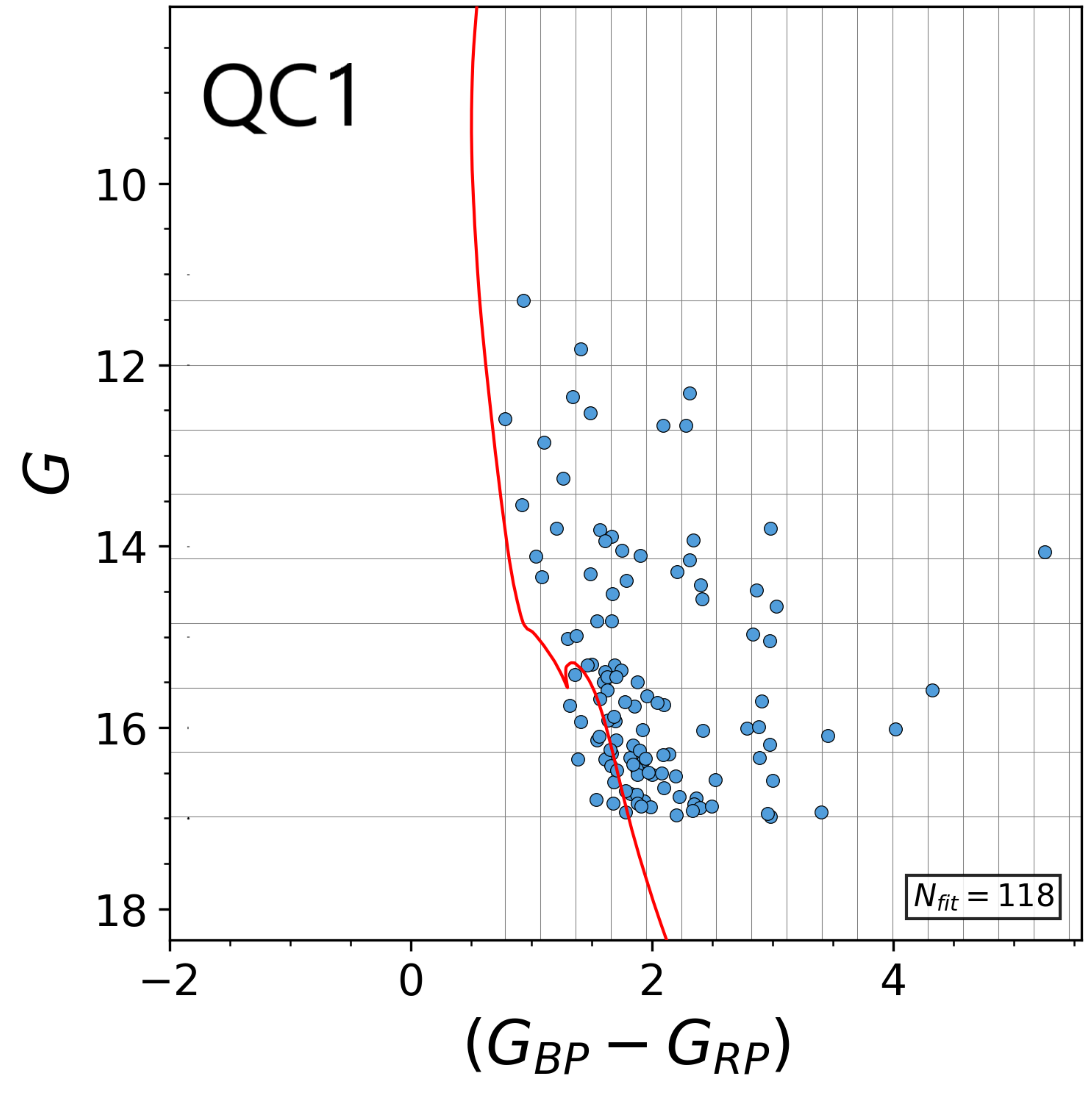}
  \includegraphics[width = 6.3cm]{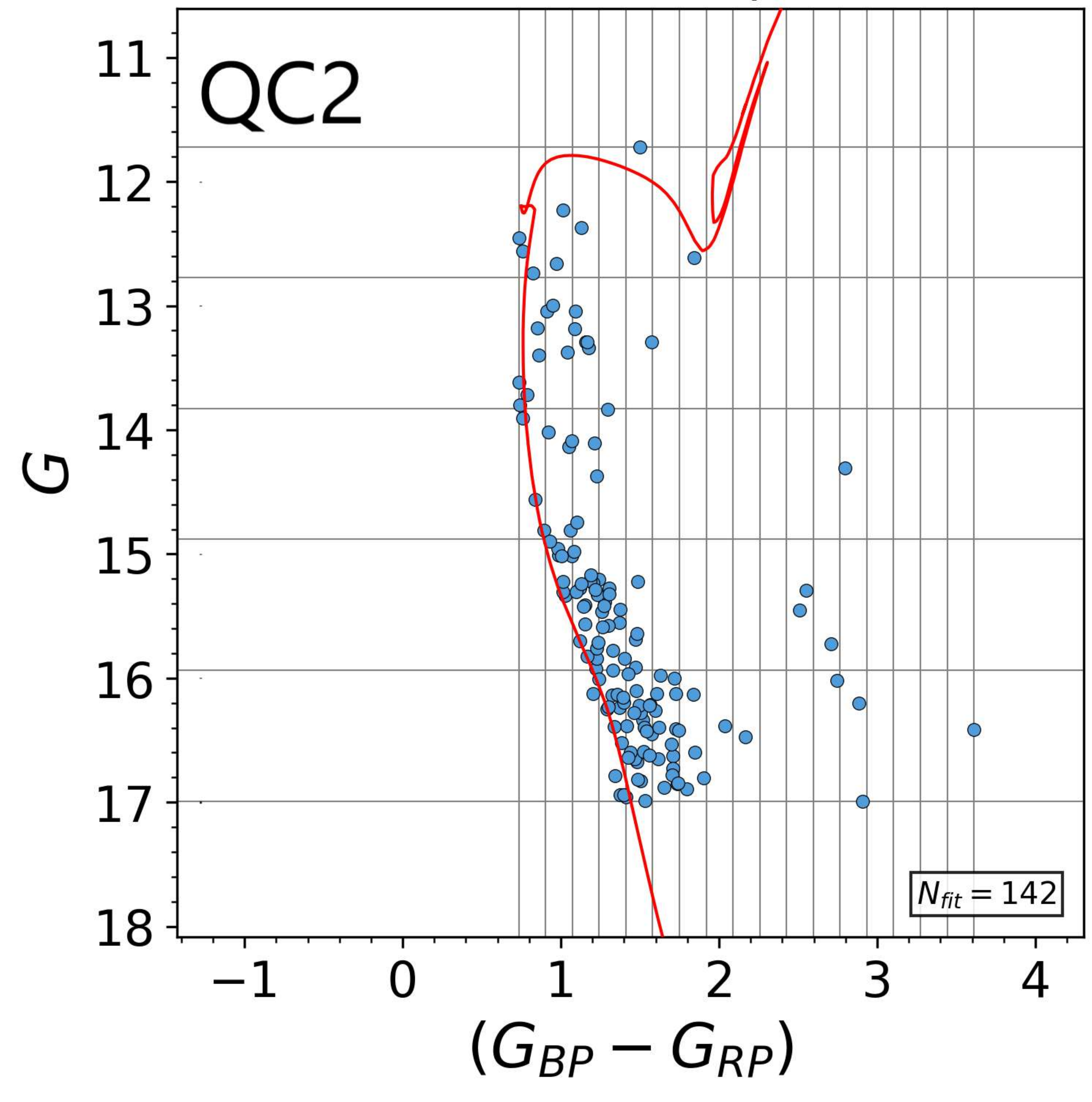}
  \includegraphics[width = 6.3cm]{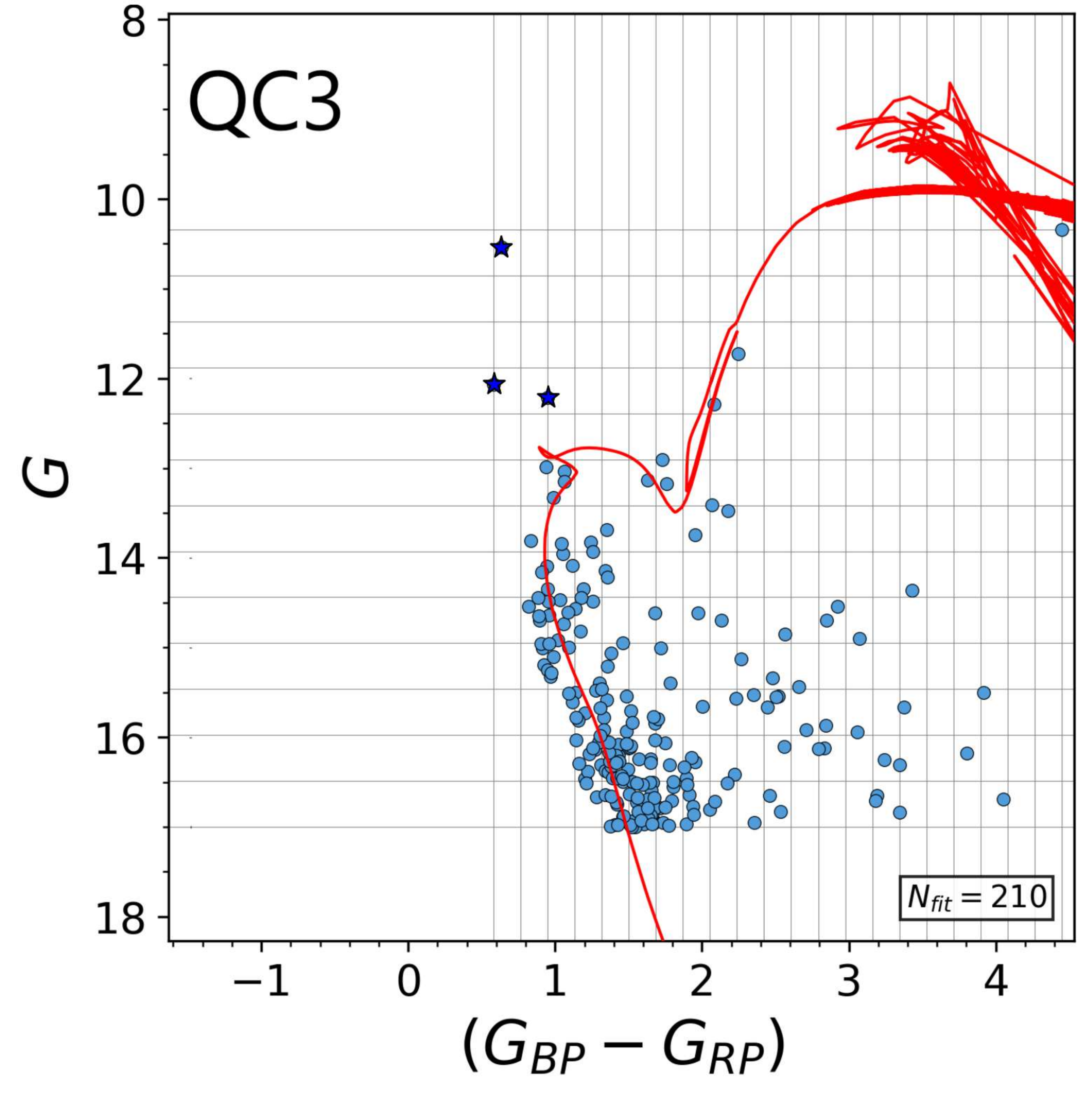}
  \includegraphics[width = 6.3cm]{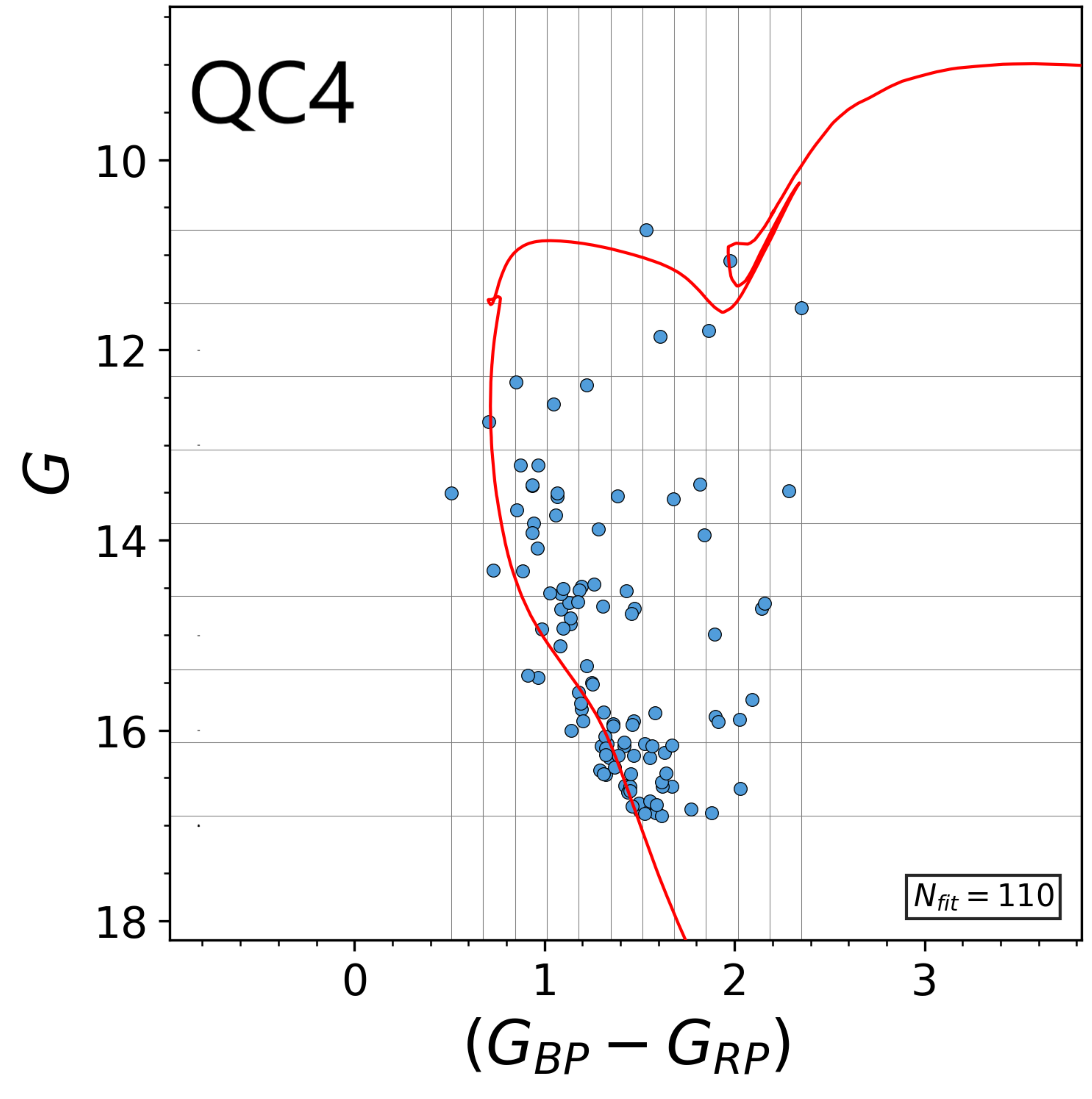}
%\end{minipage}
\caption {The CMDs for star members of the open clusters QC1, QC2, QC3 (the three BSS are pointed by Asterisk), and QC4 fitted with the extinction isochrone corrected by \citet{Bressan12} indicated by the red line.}%) isochrones over G vs. (GBP-GRP) are shown (red solid curves).}
\label{fig:cmd}
\end{center}
\end{figure*}

The Color-Magnitude Diagram (CMD) is a valuable tool in photometric analysis because we can use it to infer the cluster reddening, distance modulus and age parameters knowing the number of true members. For each cluster, we drew the corresponding CMD using EDR3 photometric magnitudes (G, G$_{BP}$, G$_{RP}$) for the member stars (the blue dots in Fig. \ref{fig:cmd}). For each CMD of a cluster, we used the ASteCA code and applied the PARSEC v1.2S \citep{Bressan12} theoretical isochrones\footnote{http://stev.oapd.inaf.it/cgi-bin/cmd} to obtain the cluster metallicity and age. %with a solar metalicity Z$_{\odot}$ = 0.0152 \citep{Caffau09} to compute the age parameter. 
The best fit values revealed a metallicity (Z) of 0.0152 $\pm$ 0.0010, 0.0150 $\pm$ 0.0004, 0.01520 $\pm$ 0.0004 and 0.0199 $\pm$ 0.0018 for QC1, QC2, QC3 and QC4, respectively. These values are in line with the standard Solar metallicity adopted by \citet{Qin21}. The logarithmic age, log (age/yr), of the four clusters QC1 through QC4 was estimated to be 6.987 $\pm$ 0.022, 8.524 $\pm$ 0.046, 8.858 $\pm$ 0.114  and 8.367 $\pm$ 0.043, respectively. For QC3 cluster, we found that there are three member stars above the turn-off point of its CMD, so we suggest that, those members are Blue Stragglers Stars (BSS) according to the criteria of \citet{Rain21}. These ages are in agreement with those previously calculated using Gaia DR2 \citep{Qin21}.

The reddening parameter, $E(G_{BP} - G_ {RP})$, was determined from the CMDs by using the formula 
$$E(G_{BP} - G_ {RP}) = \text{1.289}~ E(B - V)$$ 
The observed data have been corrected for the reddining with a line-of-sight extinction coefficient ($A_{G}$) computed by 
$$A_{G} = \text{2.74} \times E(B - V)$$ \citep{Casagrande18, Zhong19}. The reddening values for all star members, of each cluster, can be checked with Stilism\footnote{https://stilism.obspm.fr/} 3D dustmaps \citep{Capitanio17}.

Another important astrophysical parameter determined from the fitted CMDs is the distance modulus (m $-$ M) from which the distance to the clusters can be estimated. The determined moduli for QC1, QC2, QC3 and QC4 are 13.24 $\pm$ 0.27, 13.28 $\pm$ 0.28, 13.18 $\pm$ 0.28 and 12.93 $\pm$ 0.28 mag, respectively, that are in agreement with those of \citet{Qin21}. However, we think that our data are more accurate than the previously estimated values due to the improvements occurred in the EDR3 measurements for the different photometric parameters. The corresponding distances (in pc) for these distance moduli are 1674 $\pm$ 41, 1927 $\pm$ 44, 1889 $\pm$ 43 and 1611 $\pm$ 40. These distances are in agreement with those we obtained from astrometric measurements; i.e. using the cluster parallax (d$_{plx}$). All obtained results of the astrophysical and photometric parameters appeared in Table \ref{tab:4}.
 %and the Kroupa (2002) form the initial mass function.

From the estimated photometric distances to the cluster, we can infer the distance to the Galactic center (R$_{gc}$) using 
$$R_{gc} = \sqrt {R_{o}^{2} + (d~cosb)^{2} - 2~R_{o}~d~cosb~cosl}$$
where $R_{o}$ = 8.20 $\pm$ 0.10 kpc \citep{Bland19}. %awthorn et al. 2019\zmaa{\sc Dr this reference is missing, please add it to the bib file and rename as `Bland19' similar to the other references in the file.}).%\citep{Bland19}. 
After that, the projected distances towards the Galactic plane (X$_{\odot}$, Y$_{\odot}$) and the distance above the Galactic plane (Z$_{\odot}$) can be computed by using the following relationships
%$$ 
\begin{equation}
\label{eq:Xsun}
\begin{split}
X_{\odot} &= d~cosb~cosl, \\%\quad  
Y_{\odot} &= d~cosb~sinl, \\%\quad  
Z_{\odot} &= d~sinb. 
\end{split}%$$
\end{equation}

%%%%%%%%%%%
% Fig. 5                                                   %
%%%%%%%%%%%%%%%%%%%%%%%%%%%%%%%%%%%%%%%%%%%%%%%%%%%%%%%%%%%%%%%%%%%%%%%%%
%%%%%%%%%%%%%%%%%%%%%%%%%%%%%%%%%
%X} rsub {⨀} = d cos b cos l , {Y} rsub {⨀} =  , ∧ {Z} rsub {⨀} =  . left (4 right )$$s and the results are listed in Table 4.
%\sc{start here}
%%%%%%%%%%%%%%%%%%%%%%%%%%%%%
% Table 4 
%%%%%%%%%%%%%%
%\begin{landscape}
\begin{table*}
\begin{center}
\caption{The calculated astrophysical and photometric parameters of the four open clusters in comparison with data taken from \citet{Qin21} and indicated as Q21 in the last column of the table.}%\zmaa{Dr. please double check the calculations of the distances d$_{plx}$ and the one from m-M. I put them in blue to ease spotting them}}
%\centering
%\resizebox{\textwidth}{!}{
\begin{tabular}{llllll}
\topline
%Parameters & Koposov 12 & Koposov 43 & References \\
%& (FSR 802) & (FSR 848) &\\\midline
%f$_{bg}$ (stars arcmin$^{-2}$)& 0.069$\pm$0.016& 0.093$\pm$0.006&      \\\hline
{\bf Parameters}  &  {\bf QC1}  &  {\bf QC2}  &  {\bf QC3}  &  {\bf QC4}  &  {\bf Ref.}\\   \hline
Number of members  &  118  &  142  &  210  &  110  &   \\[0.8 ex]
	           &  72  &  114  &  124  &  130  &  Q21\\[0.8 ex] %\hline

$\mu_{\alpha}\cos\delta$ (mas/yr)  &  -2.34  &  -3.94  &  -2.28  &  -2.95  &   \\[0.8 ex]
	  &  -2.30 $\pm$ 0.09  &  -4.05 $\pm$ 0.06  &  -2.25 $\pm$ 0.10  &  -2.51 $\pm$ 0.17  &  Q21\\[0.8 ex]%\hline

$\mu_{\delta}$ (mas/yr)  &  -3.55  &  -4.36  &  -4.39  &  -4.40  &   \\[0.8 ex]
	  &  -3.48 $\pm$ 0.12  &  -4.32 $\pm$ 0.06  &  -4.46 $\pm$ 0.13  &  -5.43 $\pm$ 0.17  &  Q21\\[0.8 ex]%\hline

$\varpi$ (mas) & 0.548$^{+0.570}_{-0.529}$ & 0.424$^{+0.471}_{-0.459}$ & 0.437$^{+0.443}_{-0.430}$ & 0.455$^{+0.467}_{-0.444}$ & \\[0.8 ex]
	  &  0.766 $\pm$ 0.02  &  0.424 $\pm$ 0.03  &  0.400 $\pm$ 0.03  &  0.424 $\pm$ 0.04  &  Q21\\[0.8 ex]

d$_{plx}$ (pc) &  1819 $\pm$ 43 & 2151 $\pm$ 64& 2288 $\pm$ 48& 2179 $\pm$ 47 &   	\\[0.8 ex]

Z  &  0.0152 $\pm$ 0.0010  &  0.0150 $\pm$ 0.0004  &  0.0152 $\pm$ 0.0004  &   0.0199 $\pm$ 0.0018	  &   	\\[0.8 ex]
%   &  0.0152	  &  0.0152	  &  0.0152	  &  0.0152	  &  Q21\\[0.8 ex]

log(age/yr)  &  6.987 $\pm$ 0.022  &  8.524 $\pm$ 0.046  &  8.858 $\pm$ 0.114  &  8.367 $\pm$ 0.043  &  \\[0.8 ex]
	  &  7.00  &  8.55  &  8.60	  &  8.40  &  Q21\\[0.8 ex]

E(G$_{BP}$-G$_{RP}$)  &  0.995 $\pm$ 0.031  &  0.875 $\pm$ 0.017  &  0.847 $\pm$ 0.031  &  0.889 $\pm$ 0.027  &  \\[0.8 ex]

E(B-V)	&  0.772 $\pm$ 0.024  &  0.679 $\pm$ 0.013  &  0.657 $\pm$ 0.024  &  	0.690 $\pm$ 0.021  &  \\[0.8 ex]
	  &  0.756 $\pm$ 0.280  &  0.552 $\pm$ 0.010  &  0.577 $\pm$ 0.019  &   0.556 $\pm$ 0.049  &  Q21	\\[0.8 ex]

A$_G$	  &  2.12  &  1.86  &  1.80	  &  1.89  &  \\[0.8 ex]
          &  2.07  &  	1.51  &  1.58  &  1.52  &  Q21\\[0.8 ex]

(m - M) & 13.24 $\pm$ 0.27 & 13.28 $\pm$ 0.28	& 13.18 $\pm$ 0.28 & 12.93 $\pm$ 0.28 &  \\ [0.8 ex]
	   &	12.55	&	13.24	&	13.42	& 13.26	& Q21	\\

d (pc)	  &  1674 $\pm$ 41  &  1927 $\pm$ 44  &  1889 $\pm$ 43  &  1611 $\pm$ 40  &  \\[0.8 ex]
	  &  1261  &  2223  &  2340  &  2230  &  Q21\\[0.8 ex]

R$_{gc}$ (kpc) & 8.010 $\pm$ 0.089 & 8.251 $\pm$ 0.091	& 8.252 $\pm$ 0.091 & 8.218 $\pm$ 0.091	& \\  [0.8 ex]

X$_{\odot}$ (kpc) & 0.359 $\pm$ 0.018 & 0.174 $\pm$ 0.013 & 0.165 $\pm$ 0.013 & 0.139 $\pm$ 0.012 &	\\ [0.8 ex]

Y$_{\odot}$ (kpc)	& 1.634 $\pm$ 0.040 & 1.916 $\pm$0.044	& 1.880 $\pm$ 0.043 & 1.602 $\pm$ 0.040 & \\ [0.8 ex]

Z$_{\odot}$ (kpc)	& 0.057 $\pm$ 0.008 & 0.110 $\pm$ 0.010	& 0.082 $\pm$ 0.009 & 0.102 $\pm$ 0.010	& \\ [0.8 ex]
\hline 
%\flushleft
\label{tab:4}
\end{tabular}
\end{center}
%\flushleft{\bf References:} Q21 \citealt{Qin21}
%}
\end{table*}
%\end{landscape}
%%%%%%%%%%%%%%%%%%%%%%%%%%%%%%%%%%
\subsubsection{\bf The luminosity and mass functions\\}
\label{mlf}
At this stage of calculations, we had estimated for each cluster the new center positions and photometric parameters. From these data, we can derive both the luminosity and the mass functions (LF and MF) because each cluster members are formed under similar physical conditions from the same molecular cloud at the same time. This process makes star clusters ideal objects to study the initial mass function (IMF; e.g. \citealt{Scalo98, Phelps93, Yadav04, Bisht19}). 
The IMF is an empirical relationship that refers to initial stellar mass distribution in the cluster. The present time IMF can be inferred from the cluster LF and MF by using the Mass-Luminosity Relation (MLR). 

\begin{figure*}
%\begin{frame}%{Panel figure 2 * 2!}
\begin{center}
\begin{minipage}{1\textwidth}
%% trim left bottom right top
  \includegraphics[width = 4.25cm]{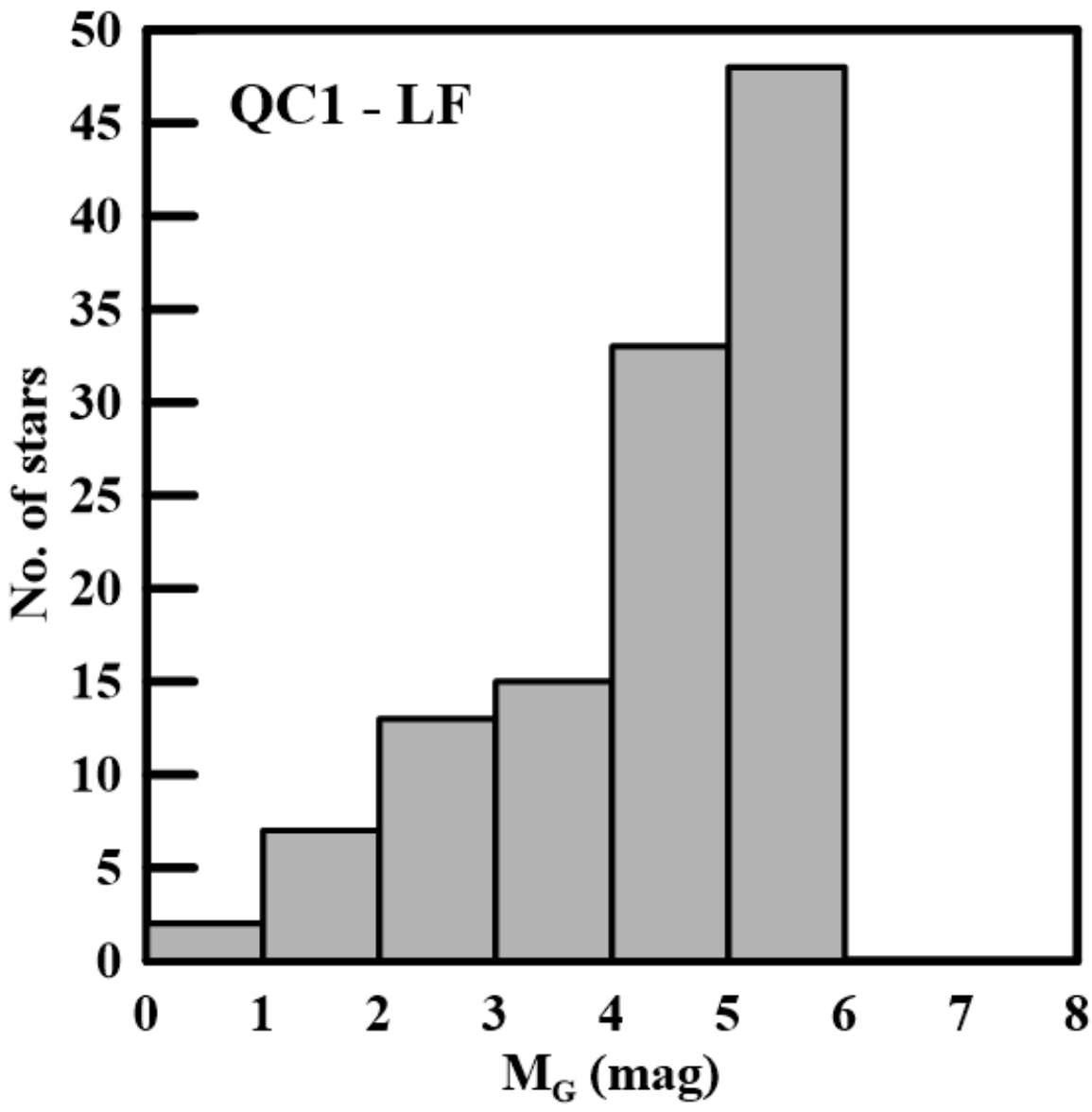}
%\hfill
  \includegraphics[width = 4.25cm]{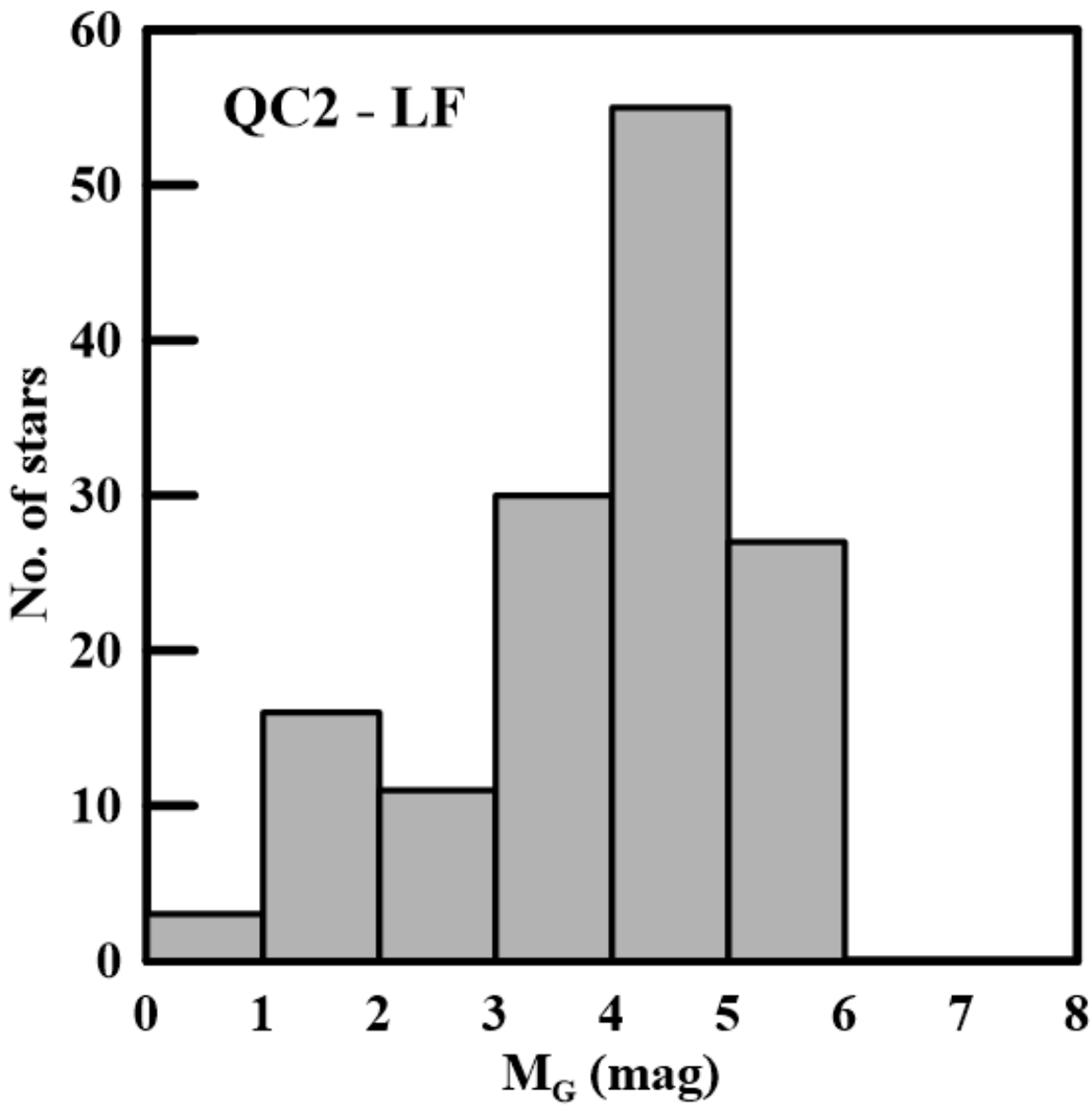}
%\vspace{0.5cm}
  \includegraphics[width = 4.25cm]{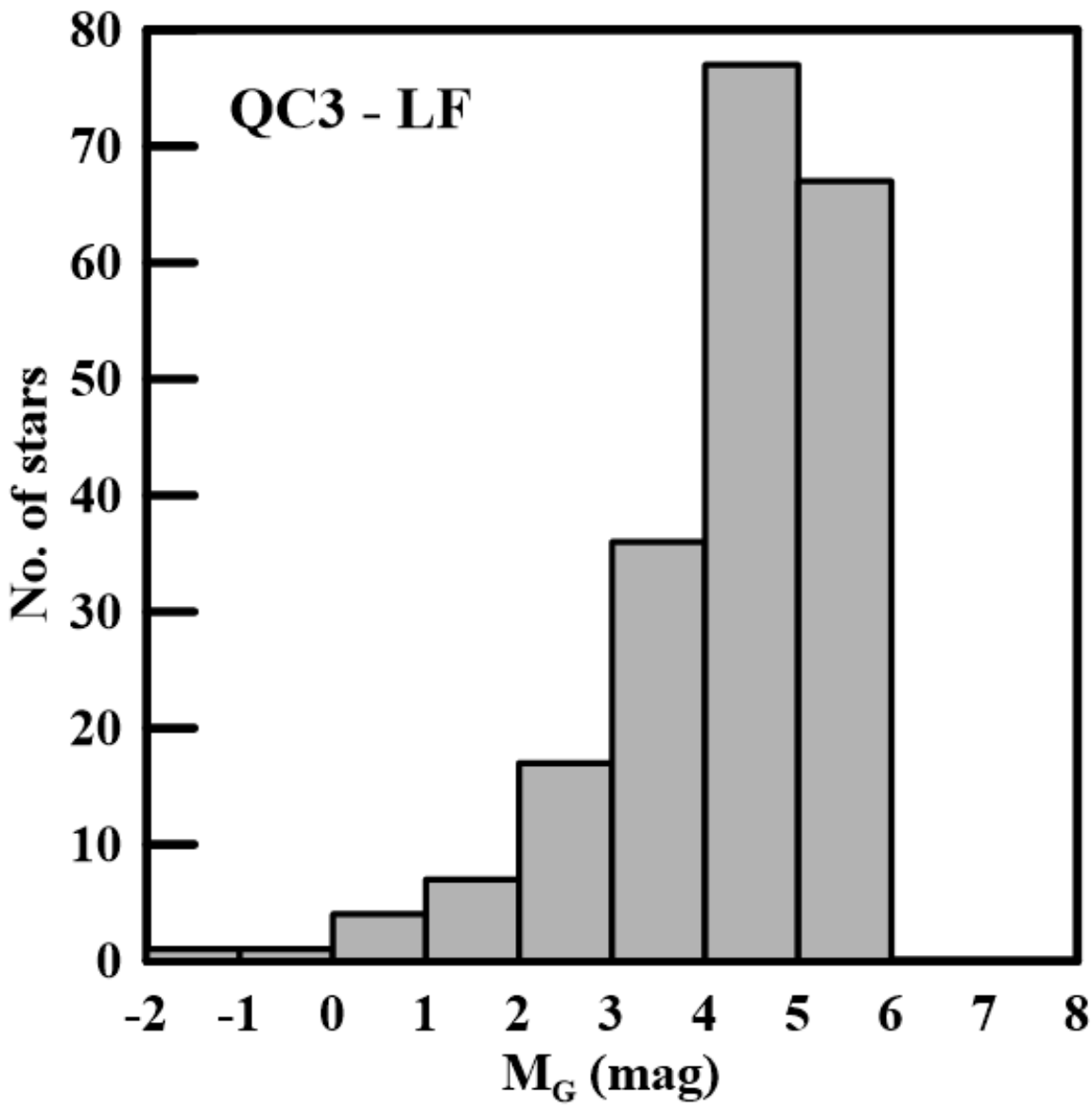}
%\hfill
  \includegraphics[width = 4.25cm]{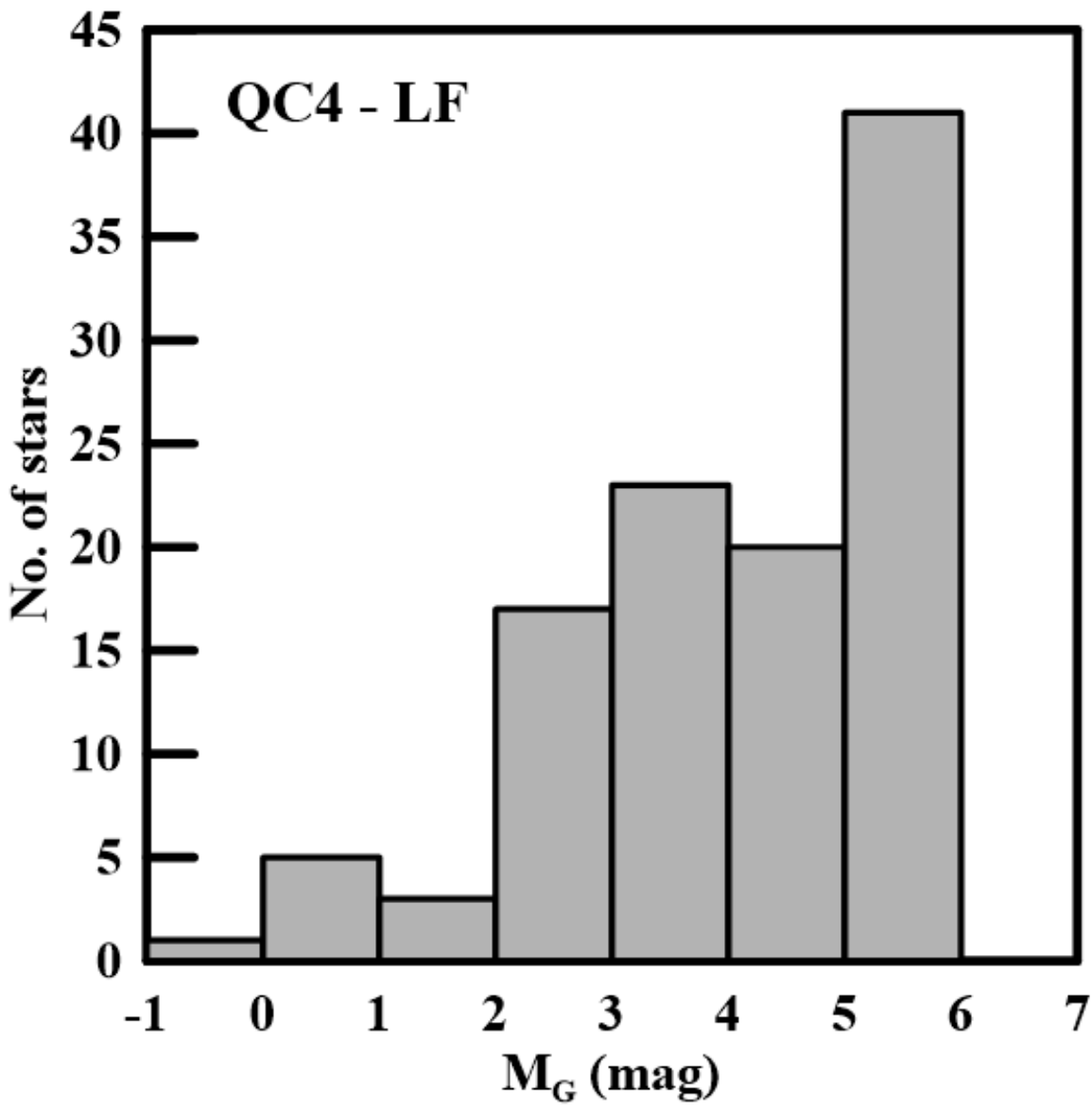}
%\end{minipage}%
\vspace{1cm}%\begin{minipage}{0.9\textwidth}
  \includegraphics[width = 4.25cm]{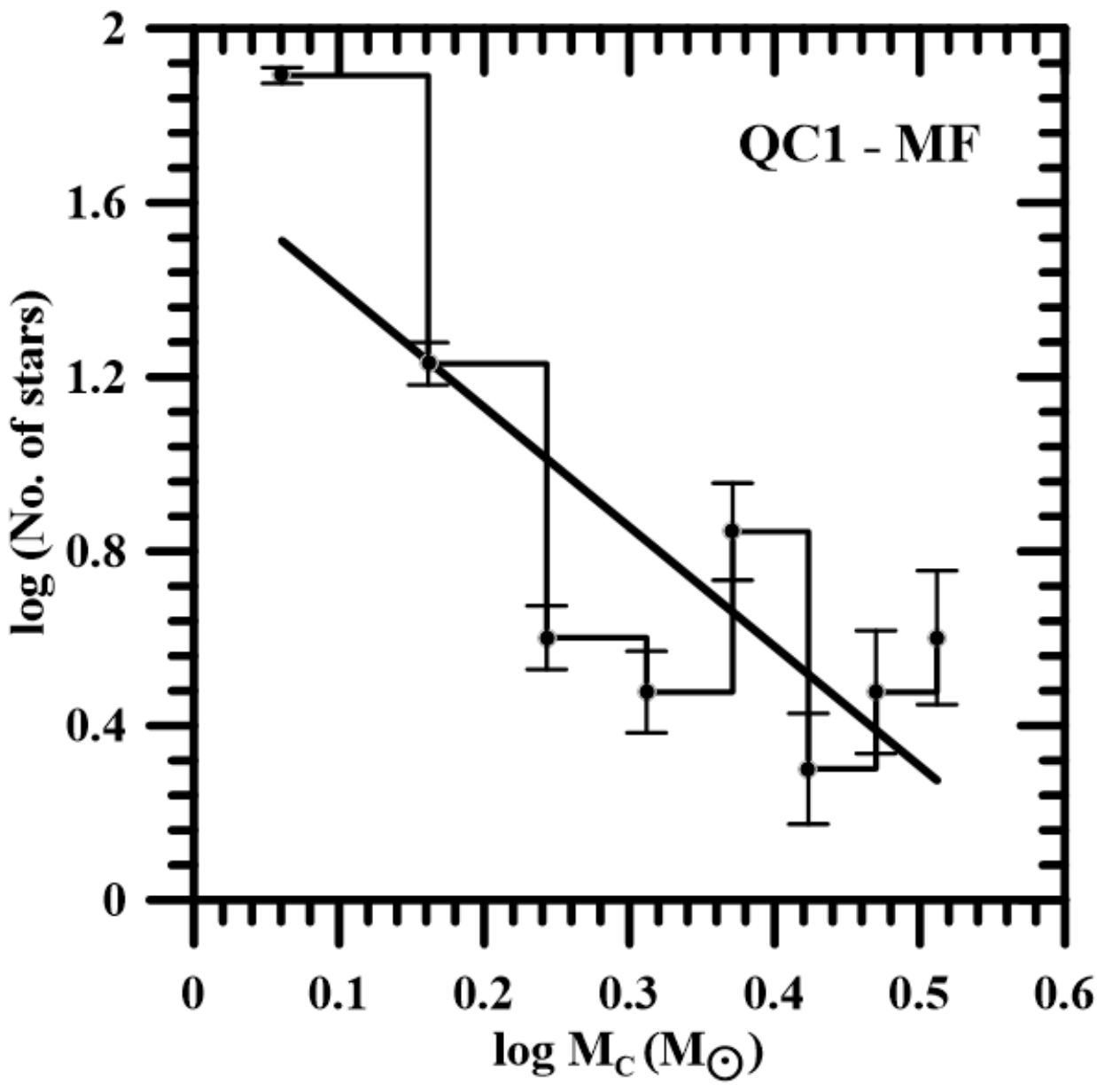}
  \includegraphics[width = 4.25cm]{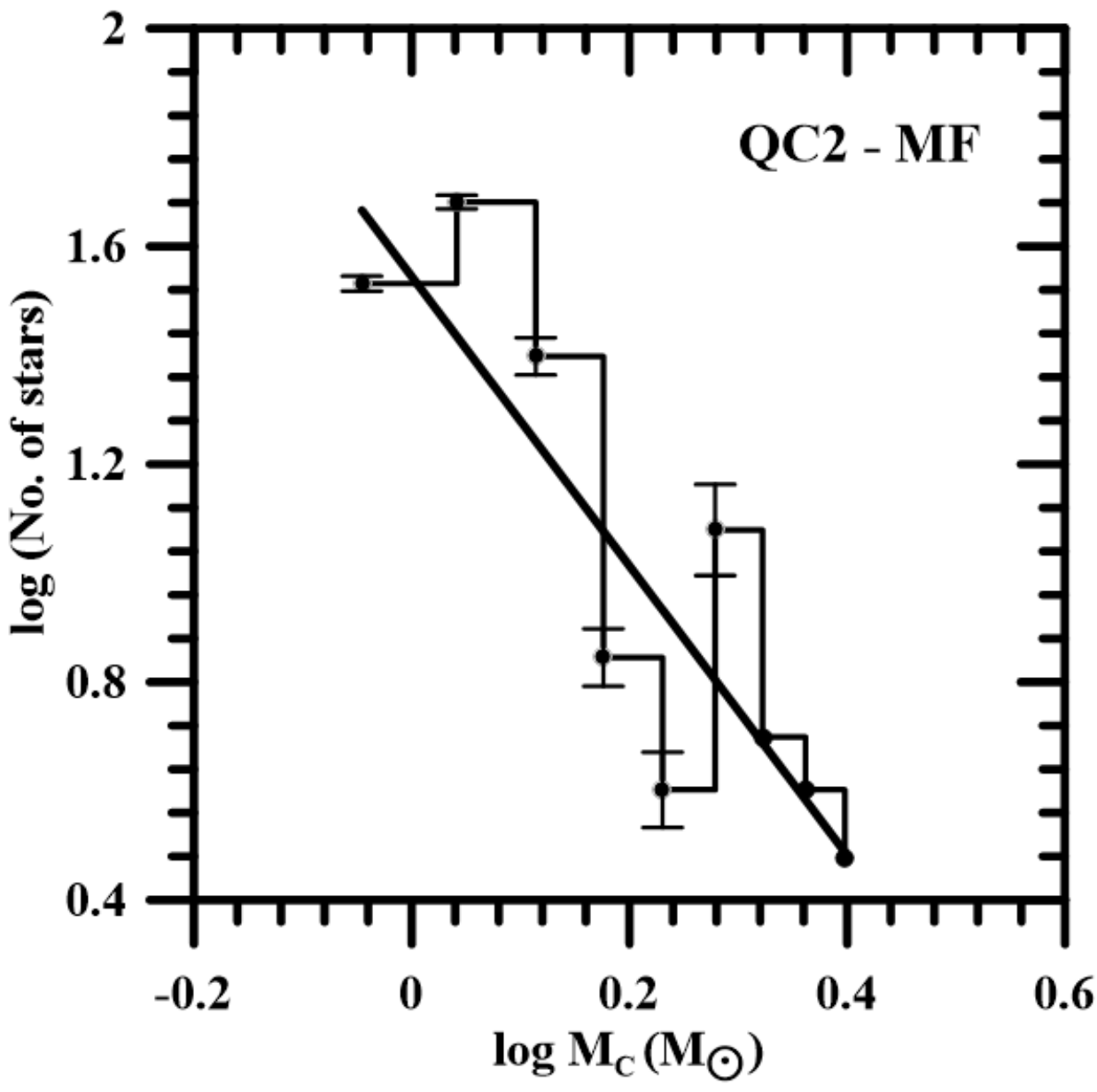}
%\vspace{0.5cm}
  \includegraphics[width = 4.25cm]{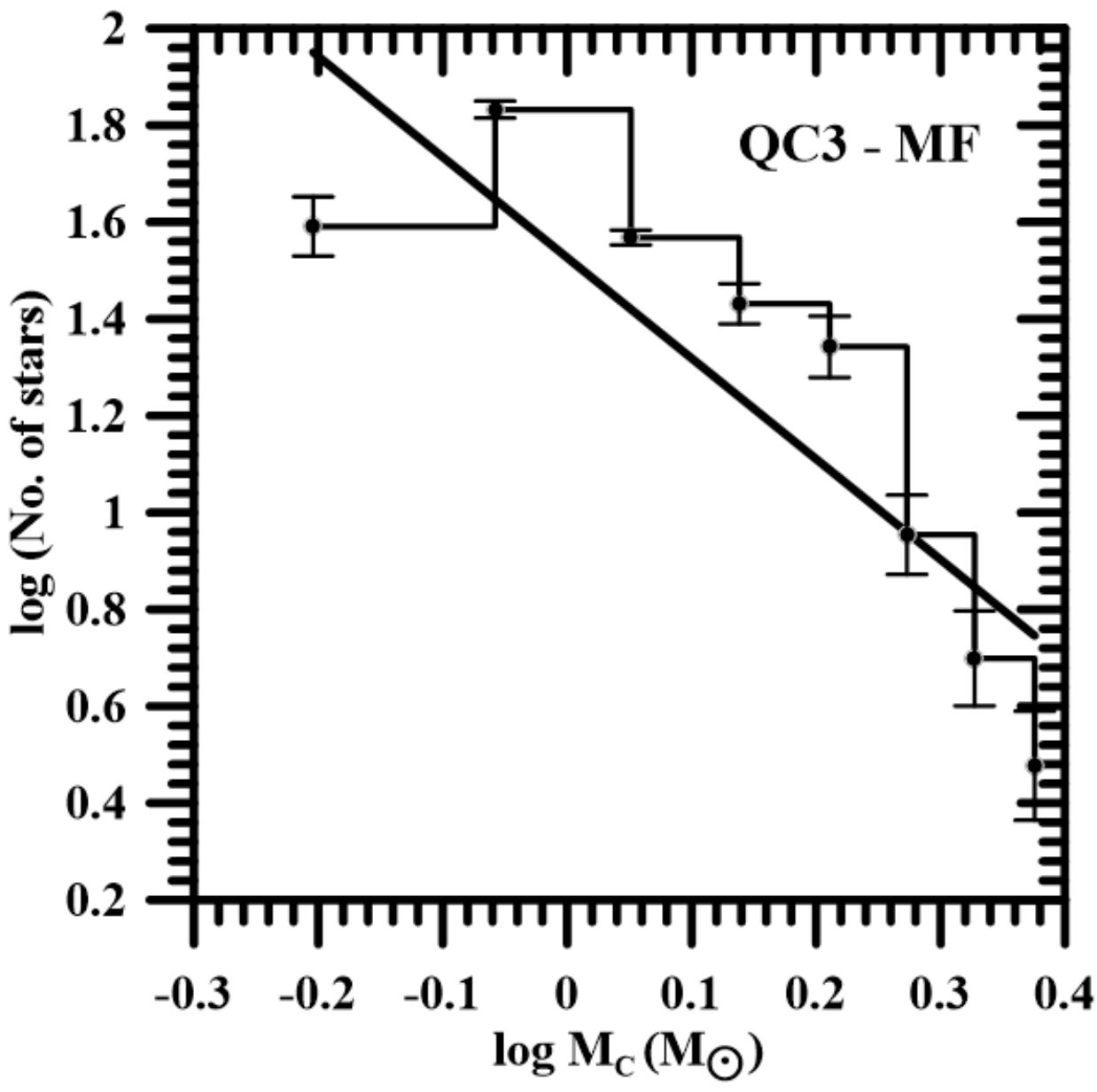}
  \includegraphics[width = 4.25cm]{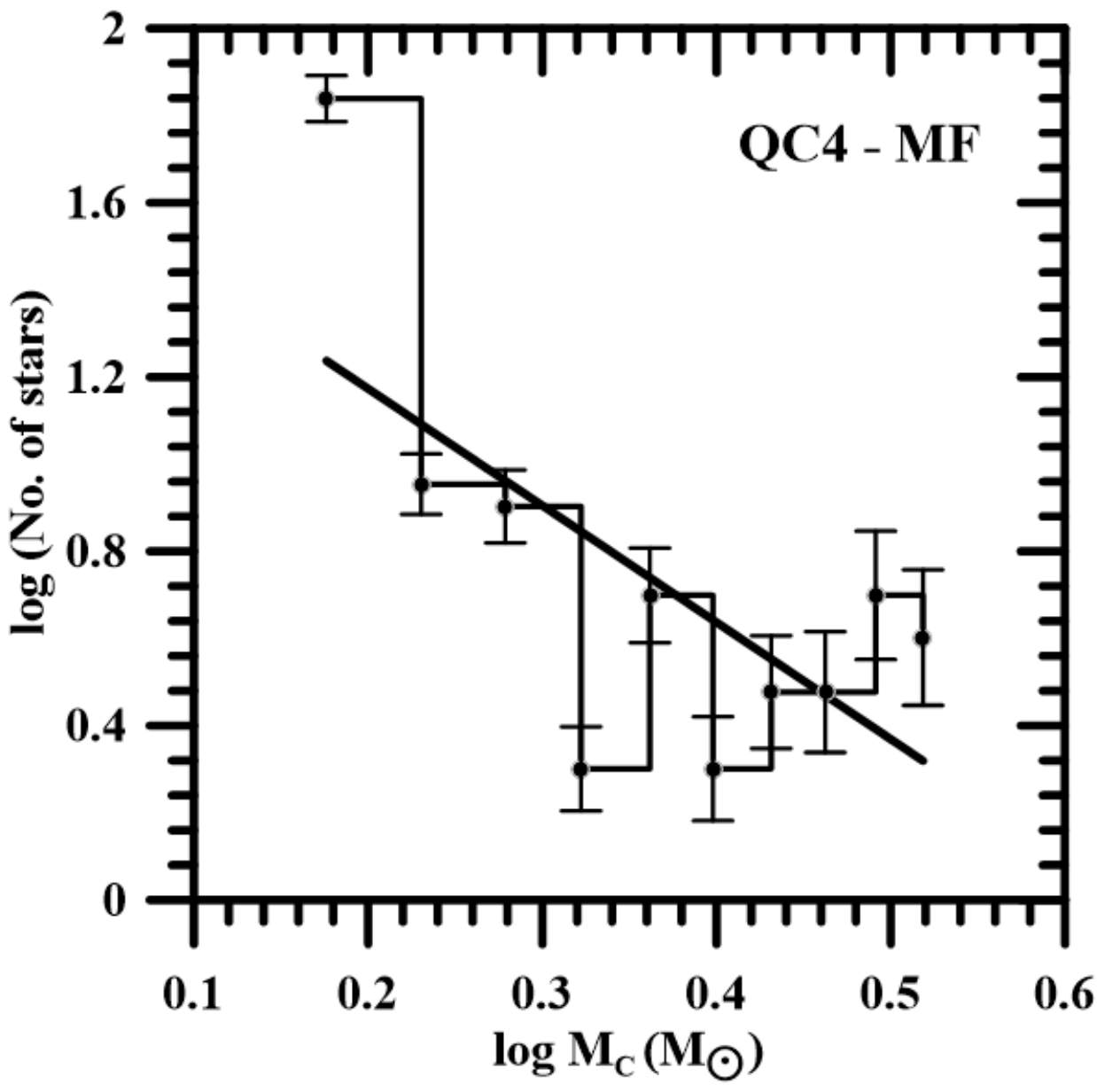}
\caption {Upper panel: the true luminosity function (LF) for each of the studied clusters. Lower panel: The mass function (MF) for the cluster members fitted by \citet{Salpeter55} power-law (solid black line) to compute the slope $\Gamma$ in Eq. \ref{eq:mf}.}
\label{fig:mlf}
\end{minipage}
\end{center}
%\end{frame}
\end{figure*}

\citet{Salpeter55} described the IMF as a power law that relates the total stellar number density ($dN$) distributed over a mass scale in a mass bin ($dM$) with central mass (M) as ($dN/dM$=$M^{-\alpha}$) and found that $\alpha$ = 2.35 for massive stars than our Sun. 
The expansion of Salpeter's power law, implies that the number of stars in each mass range (bin) decreases rapidly with increasing stellar masses. The MF slope can be derived by using Eq. \ref{eq:mf} \citep{Bisht20}.
\begin{equation}
\label{eq:mf}
log~\bigg(\frac{dN}{dM_{G}}\bigg)~=-(1+\Gamma)~log(M_G) ~ + ~\text{constant}
\end{equation}
where $\Gamma$ (= $\alpha$ - 1) is a dimensionless parameter refers to the slope of the straight line that represents the MF (see lower panel of Fig. \ref{fig:mlf}) which equals to 1.35 for \citet{Salpeter55}. %This value supports our findings for the four clusters in which w
Our computed slops 1.08 $\le \Gamma \le$ 1.74 obtained by applying the least-square fitting to our MF data are in line with the findings of \citet{Salpeter55} for QC2 and QC3 but have larger values for QC1 and QC4.

The second-order polynomial function, in Eq. \ref{eq:mlf}, relates the cluster absolute magnitude (M$_G$) and its collective masses (total mass, M$_c$) that can be obtained by fitting the adopted isochrones for our CMDs \citep{Bressan12} to the clusters' MLR. This fitting enables us to infer the mass for individual member stars 
\begin{equation}
\label{eq:mlf}
M_{C}= a_0 ~ + ~a_1 M_G ~ + ~a_2 M_G^2
\end{equation}
where $a_0$, $a_1$ and $a_2$ are three characteristic constants obtained from the fitting of the isochrones to the MF of each cluster. 
From LF of each cluster, we can estimate the cluster absolute magnitudes M$_G$ like: 4.33 (QC1), 3.80 (QC2), 4.25 (QC3), and 4.10 (QC4). While from MLR we can concluded for each cluster both the total mass M$_C$ and the average mass $\bar{M_C}$ in Solar mass unit (M$_{\odot}$), i.e. (M$_C$, $\bar{M_C}$) = (158, 1.34; QC1), (177, 1.25; QC2), (232, 1.10; QC3), and (182, 1.65; QC4). These parameters are computed for the first time in the present work and are summarized in Table \ref{tab:LF}.% = 1.65)   , M$_C$ = 158, $\bar{M_C}$ = 1.34) for QC1; (M$_G$ = 3.8, M$_C$ = 177, $\bar{M_C}$ = 1.25) for QC2; 
%(M$_G$ = 4.25, M$_C$ = 232, $\bar{M_C}$ = 1.1) for QC3; and (M$_G$ = 4.1, M$_C$ = 182, $\bar{M_C}$ = 1.65) for QC4.
%From MLR fitting of each cluster we can estimate the cluster absolute magnitude M$_G$ (in mag), total mass M$_C$ and the individual stellar masses $\bar{M_C}$ (in M$_{\odot}$) as follows: 
%(M$_G$ = 4.33, M$_C$ = 158, $\bar{M_C}$ = 1.34) for QC1; (M$_G$ = 3.8, M$_C$ = 177, $\bar{M_C}$ = 1.25) for QC2; 
%(M$_G$ = 4.25, M$_C$ = 232, $\bar{M_C}$ = 1.1) for QC3; and (M$_G$ = 4.1, M$_C$ = 182, $\bar{M_C}$ = 1.65) for QC4.
%[0.18 $\le M_G~ \text{(mag)} \le$ 5.87 and \zmat{135 $\le M_C ~\text{(M$_{\odot}$)} \le$ 401], }
%[0.19 $\le M_G ~\text{(mag)} \le$ 5.57 and \zmat{129 $\le M_C ~\text{(M$_{\odot}$)} \le$ 365], }
%[-1.04 $\le M_G ~\text{(mag)} \le$ 5.62 and \zmat{126 $\le M_C ~\text{(M$_{\odot}$)} \le$ 475],} 
% -0.30 $\le M_G ~\text{(mag)} \le$ 5.87 and \zmat{162 $\le M_C ~\text{(M$_{\odot}$)} \le$ 358} 
% \zmat{The computed mean mass, $\bar{\text{M_C}}$ (M$_{\odot}$), of each cluster is 1.34, 1.25, 1.10 and 1.65 for QC1 through QC4. }
%\zmaa{Dr can you double check the value of the mean mass of each cluste!!}%, don't you think that the masses are too small and they are not the ones written in the conclusion section!!! -- please double check the values and correct when needed}. 

%These parameters are computed for the first time in the present work and are summarized in Table \ref{tab:mlr}.% summarises all the parameters obtained from the MLR fitting. 
%%%%%%%%%%%%%%%%%%%%%%%%%%%%%%%%%%%%%%%%%%%%%
% Fig. 7  %
%%%%%%%%%%%%%%%%%%%%%%%%%%%%%%%%%%%%%%%%%%%%%
%%%%%%%%%%%%%%%%%%%%%%%%%%%%%%%%%
%%%%%%%%%%%%%%%%%%%%%%%%%%%%%
% Table 5 
%%%%%%%%%%%%%%
\begin{table*}
\caption{The estimated absolute magnitudes from LFs, masses with MLR fitting, and the slopes of MFs.} %fitting of with MLR    obtained parameters from the MLR fitting.}%Our obtained astrophysical and photometric parameters of QCs cluster as compared with Qin et al. (2021).}
\centering
%\resizebox{\textwidth}{!}{
\begin{tabular}{lllll}
\topline
%Parameters & Koposov 12 & Koposov 43 & References \\
%& (FSR 802) & (FSR 848) &\\\midline
%f$_{bg}$ (stars arcmin$^{-2}$)& 0.069$\pm$0.016& 0.093$\pm$0.006&      \\
{\bf Parameters}  &	{\bf QC1} &	{\bf QC2}	&	{\bf QC3}	&	{\bf QC4}	\\   \hline
M$_G$ (mag)	&	4.33 $\pm$ 0.75	&	3.80 $\pm$ 0.23	&	4.25 $\pm$ 0.01	&	4.10 $\pm$ 0.01	\\[0.8 ex]
a$_0$	&	3.567 $\pm$ 0.050	&	2.659 $\pm$ 0.062	&	2.170 $\pm$ 0.046	&	3.047 $\pm$ 0.090	\\[0.8 ex]
a$_1$	&	-1.060 $\pm$ 0.050	&	-0.506$\pm$ 0.045	&	-0.118 $\pm$ 0.008	&	-0.650 $\pm$ 0.063	\\[0.8 ex]
a$_2$	&	0.115 $\pm$ 0.040	&	0.034 $\pm$ 0.031	&	-0.029 $\pm$ 0.003	&	0.067 $\pm$ 0.001	\\[0.8 ex]
M$_C$  (M$_{\odot}$)	&	158 $\pm$ 13	&	177 $\pm$ 13	&	232 $\pm$ 15	&	182 $\pm$ 14	\\[0.8 ex]
$\bar{M_C}$ (M$_{\odot}$)	&	1.34	&	1.25	&	1.10	&	1.65	\\[0.8 ex]
$\Gamma$	&	1.74 $\pm$ 0.02	&	1.66 $\pm$ 0.10	&	1.08 $\pm$ 0.05	&	1.68 $\pm$ 0.03	\\[0.8 ex]
\hline
\label{tab:LF}
\end{tabular}
%}
\end{table*}
%%%%%%%%%%%%%%%%%%%%%%%%%%%%%%%%%%
%%%%%%%%%%%%%%%%%%%%%%%%%%%%%%%%%%%%%%%%%

\subsection{\bf Dynamical and kinematical structure}%ructural parameters}
\label{kin}
\subsubsection{\bf The dynamical tidal radius\\}
\label{dyn}
For any cluster, there are two gravitational forces act; one towards the Galactic center and another towards the cluster center; to keep it bound. The tidal radius is the distance at which a balance between these two forces is reached. Thus, it may act as a separator between gravitationally bound and unbound stars to a cluster. \citet{Roser19} studied the effect of the gravitational massive bodies such as stellar clusters in the Galactic disk and expressed it in terms of the distance ({$x_L$; Eq. \ref{eq:xl}}) between the Lagrangian points and the Galactic center. These distances are equivalent to the tidal radius r$_t$ of the cluster (i.e. $x_L \approx$ r$_t$) and can be expressed as follows
\begin{equation}
\label{eq:xl}
x_{L}=\left[\frac{GM_C}{4A(A-B)}\right]^{1/3}=\left[\frac{GM_C}{4\Omega_{o}^2-\kappa^2}\right]^{1/3}.
\end{equation}
where M$_{C}$ is the cluster mass (in M$_{\odot}$), G is the gravitational constant (= 4.30 $\times$ 10$^{-6}$ kpc M$^{-1}_{\odot}$ km$^2$ s$^{-2}$), $\Omega_{o}$ (= A - B) is the angular velocity, and $\kappa$ (= $\sqrt{\textrm{-4B (A - B)}}$) is the epicyclic frequency at the position of the Sun; both of the $\Omega_{o}$ and $\kappa$ parameters are measured in km s$^{-1}$ kpc$^{-1}$ \citep{Roser11}. The two constants A (= 15.6 $\pm$ 1.6) and B (= -13.9 $\pm$ 1.8) are the Oort constants (in km s$^{-1}$kpc$^{-1}$) with adopted values from \citet{Nouh20}.

The calculated tidal radii (in pc) for the clusters QC1, QC2, QC3 and QC4 are 7.17 $\pm$ 1.68, 7.45 $\pm$ 1.73, 8.16 $\pm$ 1.86, and 7.52 $\pm$ 1.75, respectively. %\zmaa{\sc Dr. Can you please double check these values?!}%much lower than those obtained by \citep{Qin21} -- this have differnt units we will see the comparison when we get Qin21 data in pc instead of (')}%\sc dr can you give an explaination to this differences in light of the use of EDR3 instead of dr2?! {\bf Also, why we do not see the trend large Mc means small r$_t$?} Please also note that the values of r$_t$ in Table \ref{tab:mlr} from asteca or dynamical are not consistent with the  values on the Rdp plots - can you correct them} }%almost less than 

\subsubsection{\bf Dynamical evolution times\\}
\label{dtime}
%{\sc start}\\
Unlike compact halo counterparts (i.e., globular clusters), open clusters have a looser spatial distribution, and the interaction among the stars in an open cluster leads to the energy exchange \citep{Inagaki85, Baumgardt03}. %For a force of contraction and/or destruction, mass-segregation for massive stars are concentrated towards the cluster core than low-mass ones. 
This phenomenon has been recently reported for many open clusters (e.g. \citealt{Dib18, Bisht20, Joshi20}). 

The dynamical relaxation time (T$_{relax}$), measured in years, is the characteristic time needed for a cluster to reach equilibrium. \citet{Spitzer71} expressed this time, Eq. \ref{eq:Trelax}, in terms of both the number of member stars of the cluster (N) and the cluster diameter (D)
\begin{equation}
\label{eq:Trelax}
T_{relax} ~=~ \frac{8.9\times10^5~N^{1/2}~ R_h^{3/2}}{(\bar {M_C})^{1/2}~log(0.4~N)}
\end{equation}
where R$_h$ (in pc) is the radius containing half of the cluster mass and $\bar{M_C}$ is the average mass of all cluster members in Solar masses. \citet{Lada03} assumed that the cluster diameter is twice its limiting radius; D $\sim$ 2 r$_{cl}$. 

Estimating T$_{relax}$ enables us to estimate the evaporation time, $\tau_{ev}$ which is the time needed to eject all member stars due to internal stellar encounters \citep{Adams01} and it is estimated to be about 10$^2$ T$_{relax}$. Low-mass stars continue to escape from the cluster, at low speeds through the Lagrange points \citep{Kupper08}. 

For a cluster to remain bound, the escaping velocity (V$_{esc}$) of rapid gas removal from the cluster has to satisfy Eq. \ref{eq:vesc} \citep{Fich91, Fukushige00}.
\begin{equation}
\label{eq:vesc}
V_{esc} = R_{gc}~ \sqrt{2~ G~ M_{C}~ /~ 3 r_{t}^3}
\end{equation}
Therefore, abound group will emerge only if the star-formation efficiency (SFE) is greater than 50\% \citep{Wilking83}.

The dynamical state of QCs clusters can be described by computing the dynamical evolution parameter, $\tau$ (= age / T$_{relax}$). The cluster is relax (i.e. no dynamical effects) if $\tau >>$ 1, otherwise the cluster suffers dynamical interactions. Our calculations showed that the dynamical parameter for the four clusters, apart from QC1, is larger than 1 which means that they are dynamically relaxed. We found that $\tau(QC1) \sim$ 0.82. 
The result of QC1 supports our finding (in Table \ref{tab:4}) that this cluster is the youngest (log(age) $\sim$ 6.9) among the other four open clusters. Moreover, the %Another explaination besides its age could be that its 
stellar content of QC1 could be dominated by low-mass instead of massive star members and hence it will have, on average, a larger random velocities that generally leads to more interactions with the surrounding space %hence larger volumes than clusters with massive stars, and therefore, these clusters are capable of interacting with their surrounding during their relaxation time 
\citep{Mathieu86}. %(T$_{relax}$) low-mass stars , bigger volume than the high mass does, .}%\zmaa{is this also relaxed??!!}. 
All of the calculations for the dynamical study are listed are Table \ref{tab:mlr}. 

\subsubsection{\bf Ellipsoidal motion and the kinematical structure \\}
\label{struc}
To highlight the gravitationally bound system of the stellar groups in a limited volume of space within the Galactic system characterized by the parallelism and equality of their motions, we studied the velocity ellipsoid parameters VEPs and those kinematics using a computational algorithm developed by \citet{Elsanhoury18} and \citet{Bisht20}.
%; Elsanhoury 2021; Bisht et al. 2021). 

For any cluster members with coordinates ($\alpha, \delta$) at distance (d) and have a proper motion components ($\mu_{\alpha}\cos\delta, \mu_\delta$) and radial velocity (V$_r$), the space velocity vectors (V$_x$, V$_y$, V$_z$) are given by 
\begin{equation}
\label{eq:Vs}
%\resizebox{\textwidth}{!}{
\begin{split}
V_x & = -4.74~d~\mu_\alpha \cos\delta \sin\alpha~-~4.74~d~\mu_\delta\sin\delta \cos\alpha \\
&~+~V_r \cos\delta \cos\alpha, \\ %\\%vspace{1cm}\\
%\end{split}
%\end{equation}
%\begin{equation}
%\label{eq:Vs-y}
%\begin{split}
V_y & = +4.74~d~\mu_\alpha \cos\delta \cos\alpha~-~4.74~d~\mu_\delta\sin\delta \sin\alpha \\
&~+~V_r \cos\delta \sin\alpha, \\% \\
%\end{split}
%\end{equation}
%\begin{equation}
%\label{eq:Vs-z}
%\begin{split}
V_z & = +4.74~d~\mu_\delta\cos\delta~+~V_r\sin\delta.%\\
\end{split}
\end{equation}
%where V$_r$ (in km/s) is the average radial velocity (the velocity along the line-of-sight) is one of the available measurements on EDR3. 
where the average radial velocity (V$_r$; in km/s) for the four clusters (taken from the Gaia EDR3 database) is: -13.79 $\pm$ 1.86 (QC1), -17.75 $\pm$ 3.21 (QC2), -34.67 $\pm$ 2.94 (QC3), and -22.82 $\pm$ 2.39 (QC4). %\zma{The negative sign indicates the direction of the velocity vector.}

\begin{figure*}
\begin{center}
%% trim left bottom right top
  \includegraphics[width = 5.75cm]{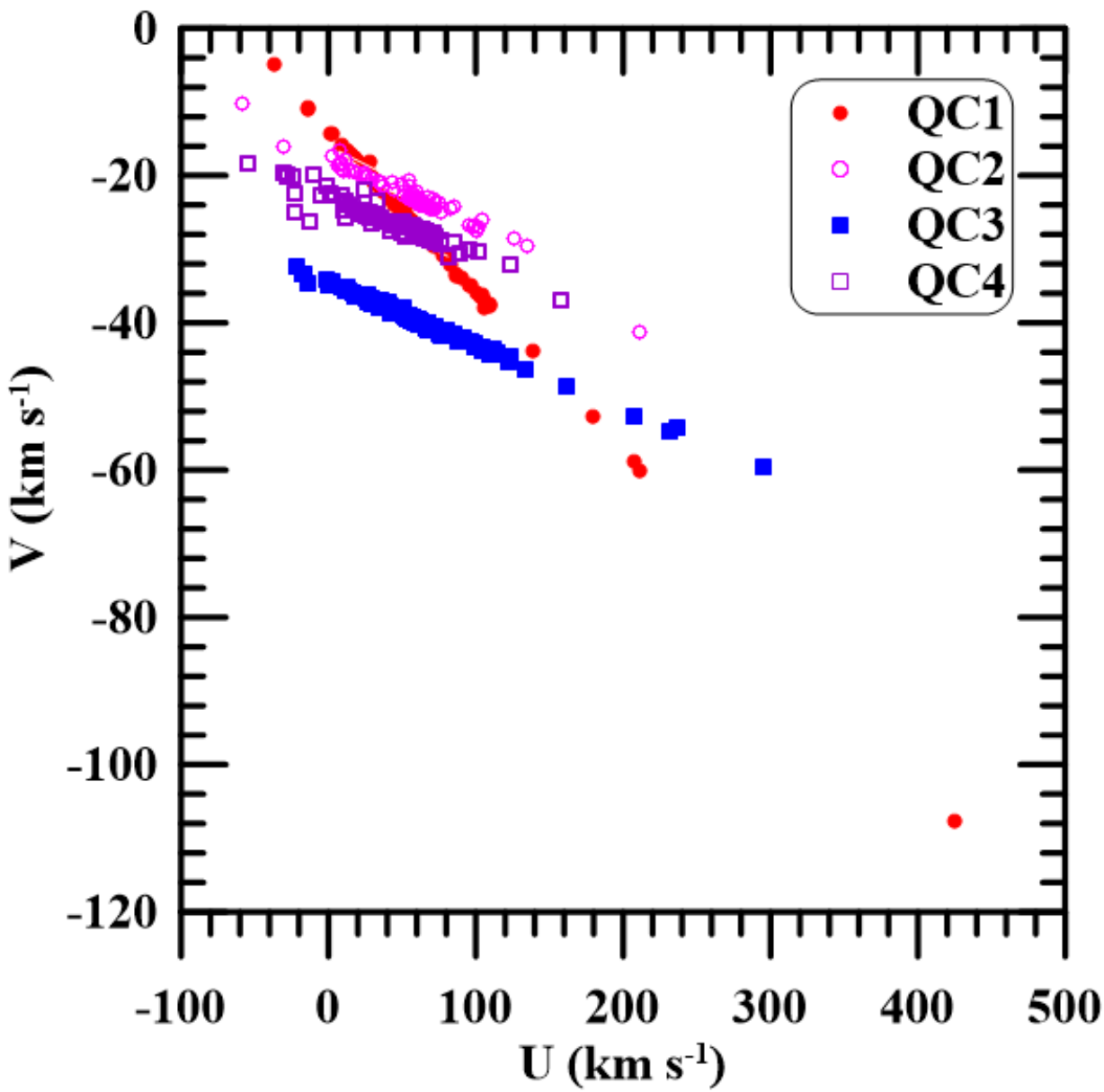}
  \includegraphics[width = 5.75cm]{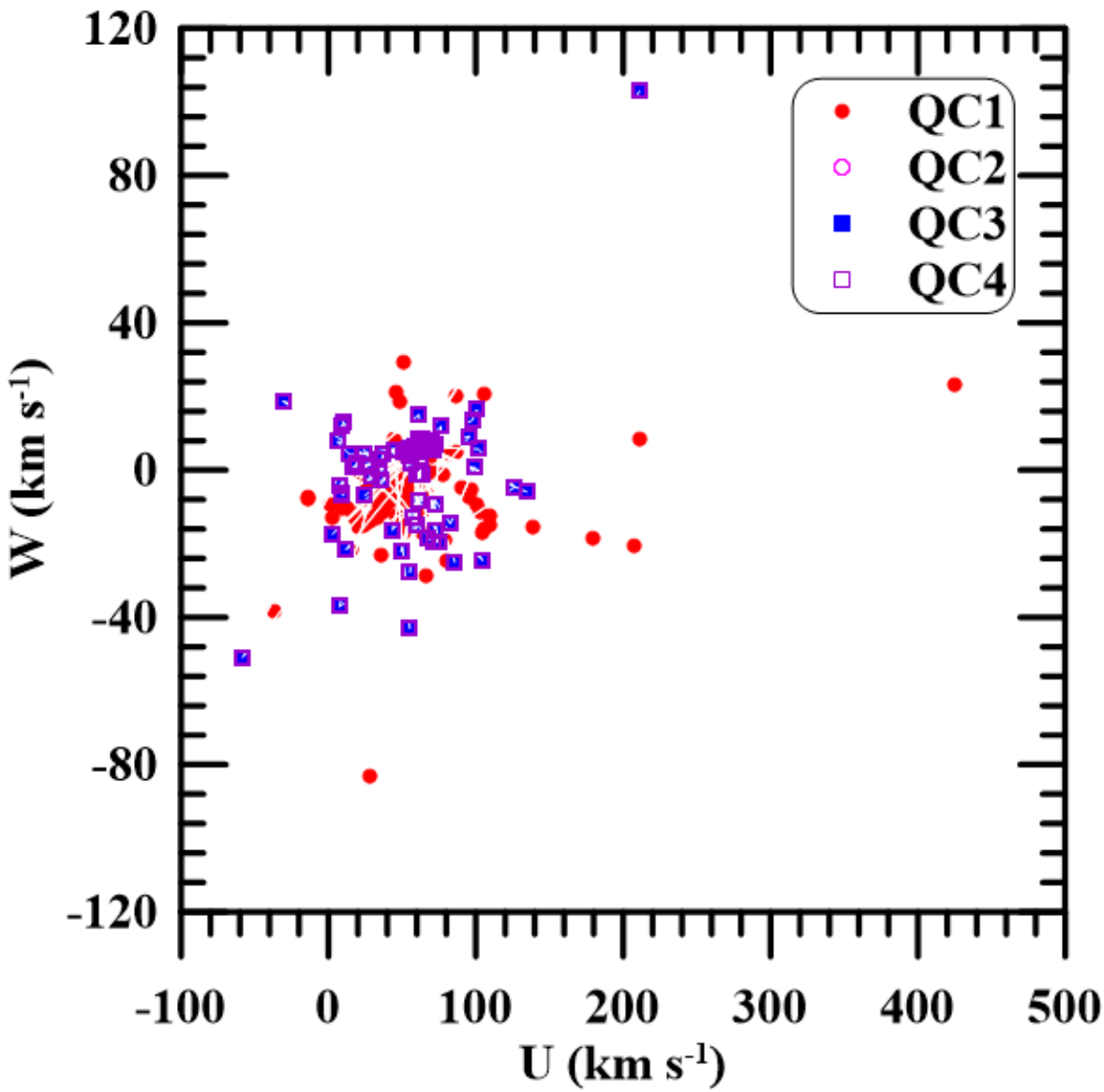}
  \includegraphics[width = 5.75cm]{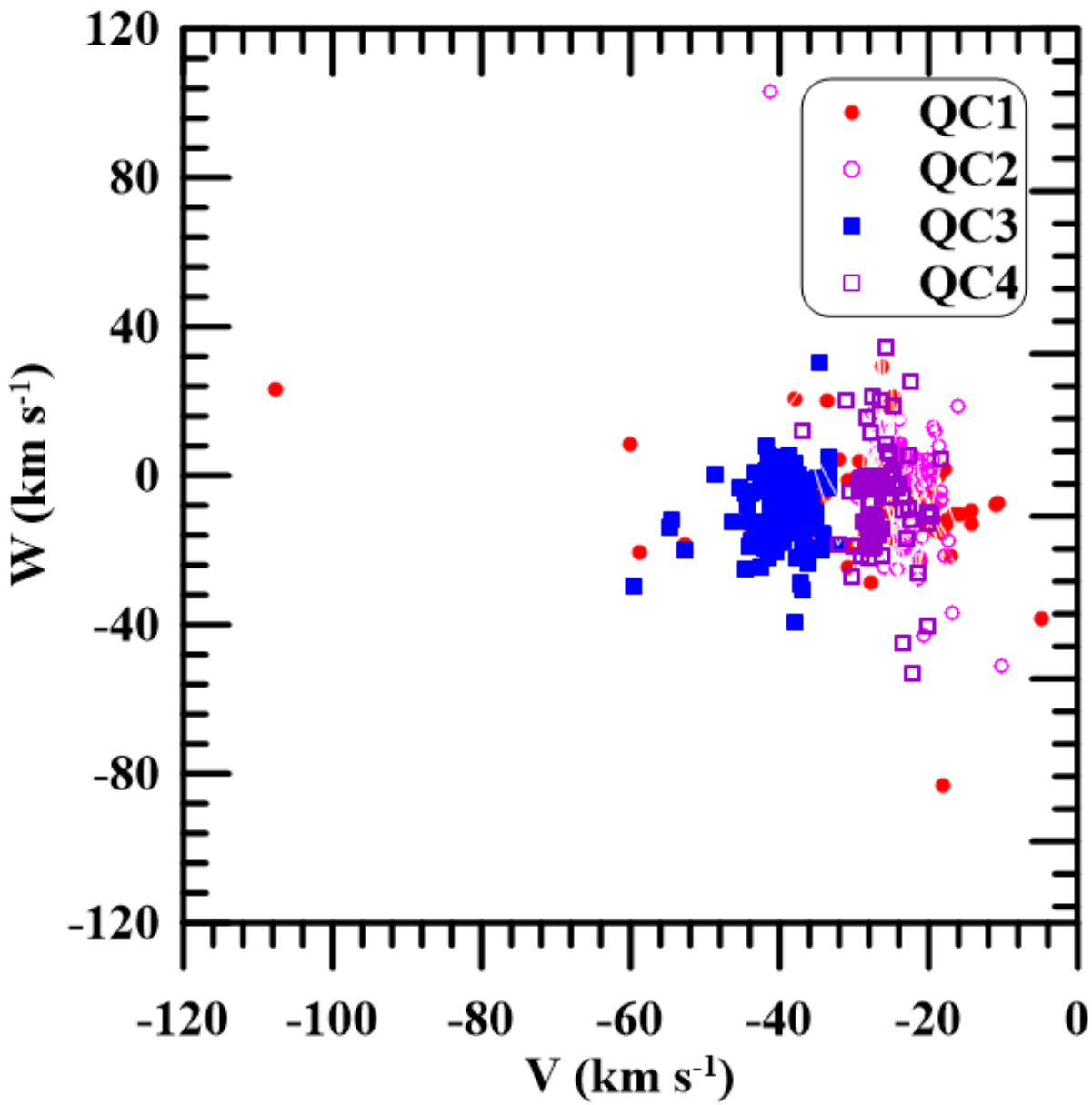}
\caption {The distribution of the spatial space velocity components along the Galactic coordinates of the star members of the studied clusters; see Fig. key.}
\label{fig:8}
\end{center}
\end{figure*}

\begin{figure*}
\begin{center}
%% trim left bottom right top
\includegraphics[width = 12cm]{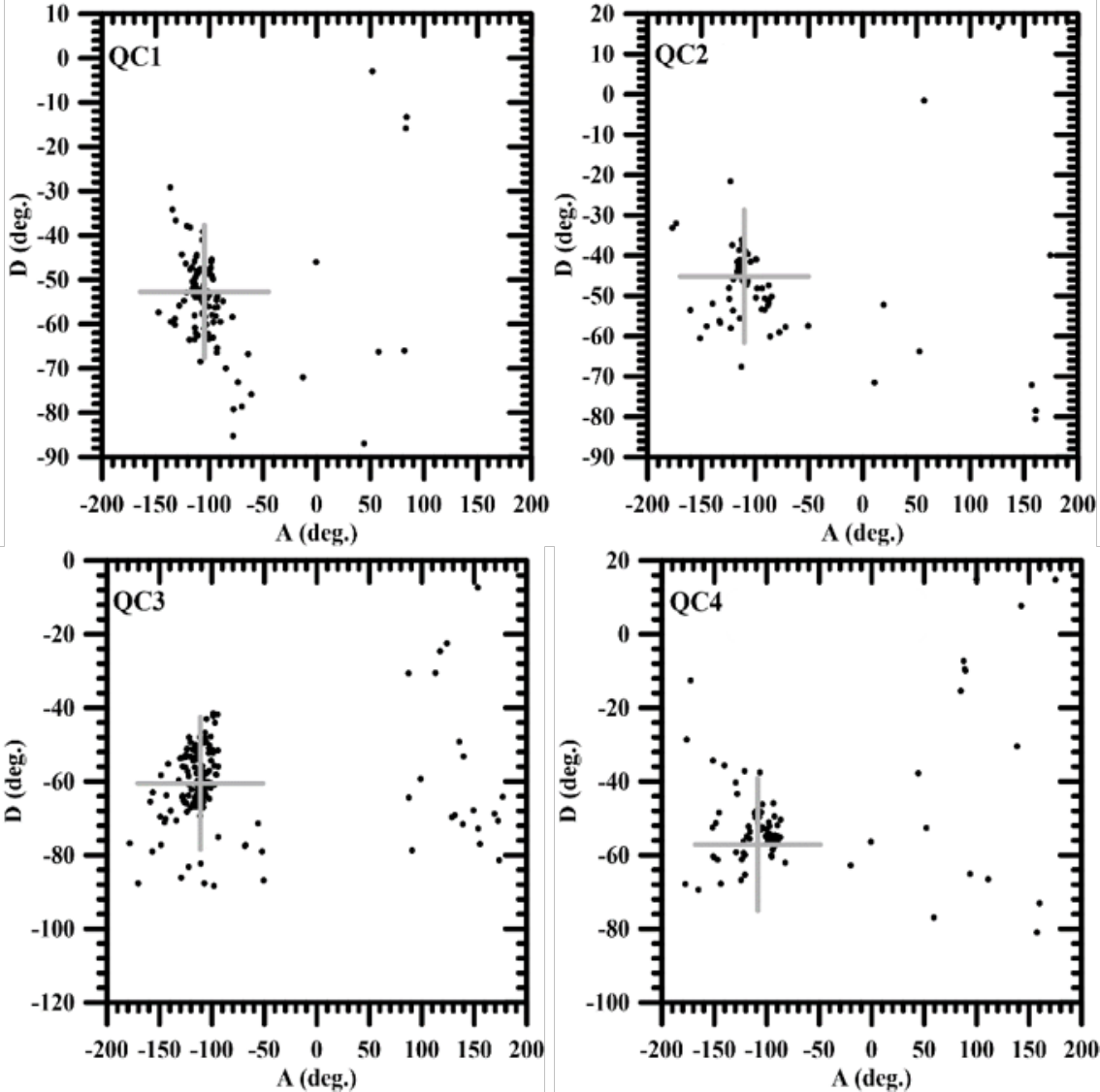}
\caption {The AD -- diagrams for the clusters QC1, QC2, QC3, and QC4 with the cross mark denotes the location of the apex point (A, D).}
\label{fig:9}
\end{center}
\end{figure*}

%\zmaa{need to restructure this part\\}\zma{say; 
The Galactic spatial velocity distribution of the stellar members of the four studied clusters is shown in Fig. \ref{fig:8}. 
The spatial velocity components (U, V, W) in the Galactic coordinates, Eq.s \ref{eq:uvw}, were derived in light of the calculated space velocity components, given in Eq.s \ref{eq:Vs}, by using an equatorial-Galactic transformation matrix based on 
%the ear-infrared 2MASS point source catalog and 
the SPECFIND v2.0 catalog of radio continuum spectra; see Eq. 14 in \citet{Liu11}. %Eq. \ref{eq:uvw} is a simplithe expression of the Galactic coordinates %NIR of 2MASS\footnote{2MASS refers to Two-Micron All-Sky Survey} \citep{Skrutskie06} and the radio observation data.\zmaa{Dr., are the radio data also from 2MASS?}
\begin{equation}
\label{eq:uvw}
\begin{split}
%U ~=~-0.05 V_{x}~-~0.87 V_{y}~-~0.49 V_{z},\\
%V ~=~+0.48 V_{x}~-~0.45 V_{y}~ +~ 0.75 V_{z},\\ 
%W ~=~-0.87 V_{x}~-~0.20 V_{y} ~+~0.45 V_{z},%\\
U &=-0.0518807421 V_{x}-0.8722226427 V_{y} \\
&-0.4863497200 V_{z},\\ %\\
%\end{split}
%\end{equation}
%\begin{equation}
%\label{eq:V}
%\begin{split}
V &=+0.4846922369 V_{x}-0.4477920852 V_{y} \\
&+0.7513692061 V_{z},\\ %\\
%\end{split}
%\end{equation}
%\begin{equation}
%\label{eq:U}
%\begin{split}
W &=-0.8731447899 V_{x}-0.1967483417 V_{y} \\
&+0.4459913295 V_{z},%\\
\end{split}
\end{equation}
%%%%%%%%%%%%%%%%%%%%%%%%
% Fig. 8 %                                                 
%%%%%%%%%%%%%%%%%%%%%%%%%%%%%%%%%%%%%%%%%%%%%%%%%%%%%%%%%%%%%%%%%%%%%%%%%
%%%%%%%%%%%%%%%%%%%%%%%%%%%%%%%%%
The apex position (A, D) is a convergent point into which member stars of the cluster will be coherently directed to. This point represents the intersection of the stellar spatial velocity vectors with the celestial sphere. \citet{Chupina01, Chupina06} introduced a method to determine the apex components in the equatorial coordinates by knowing the average space velocity vectors as expressed in Eq.s \ref{eq:Vs}. %-x} to \ref{eq:Vs-z}
The apex coordinates can be determined as follows
\begin{equation}
\label{eq:ad}
\begin{split}
A &= \tan^{-1} \Bigg(\frac{\bar V_y}{\bar V_x}\Bigg) \\
%\end{equation}
%\begin{equation}
D &= \tan^{-1} \Bigg(\frac{\bar V_z}{\sqrt {\bar V_x^2 + \bar V_{y}^{2}}}\Bigg)
\end{split}
\end{equation}
The cross mark in Fig. \ref{fig:9} illustrates the apex position for each of the four open clusters and list their coordinates in Table \ref{tab:mlr}.
%%%%%%%%%%%
% Fig. 9                                                  %
%%%%%%%%%%%%%%%%%%%%%%%%%%%%%%%%%%%%%%%%%%%%%%%%%%%%%%%%%%%%%%%%%%%%%%%%%
%%%%%%%%%%%%%%%%%%%%%%%%%%%%%%%%%
\subsubsection{\bf Other kinematic parameters}
\begin{figure*}
\begin{center}
%% trim left bottom right top
  \includegraphics[width = 6.85cm]{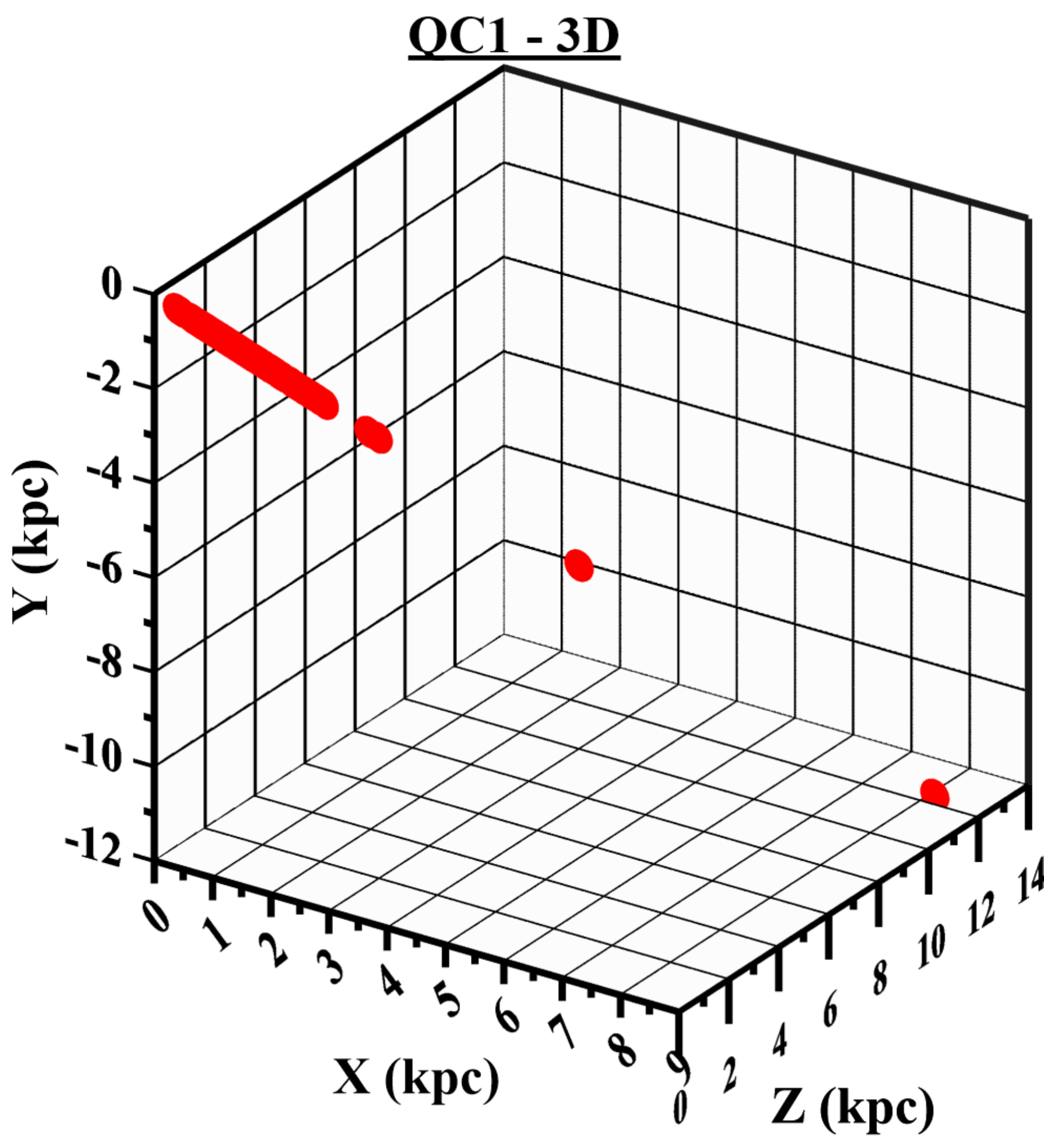}
%\hfill
  \includegraphics[width = 6.85cm]{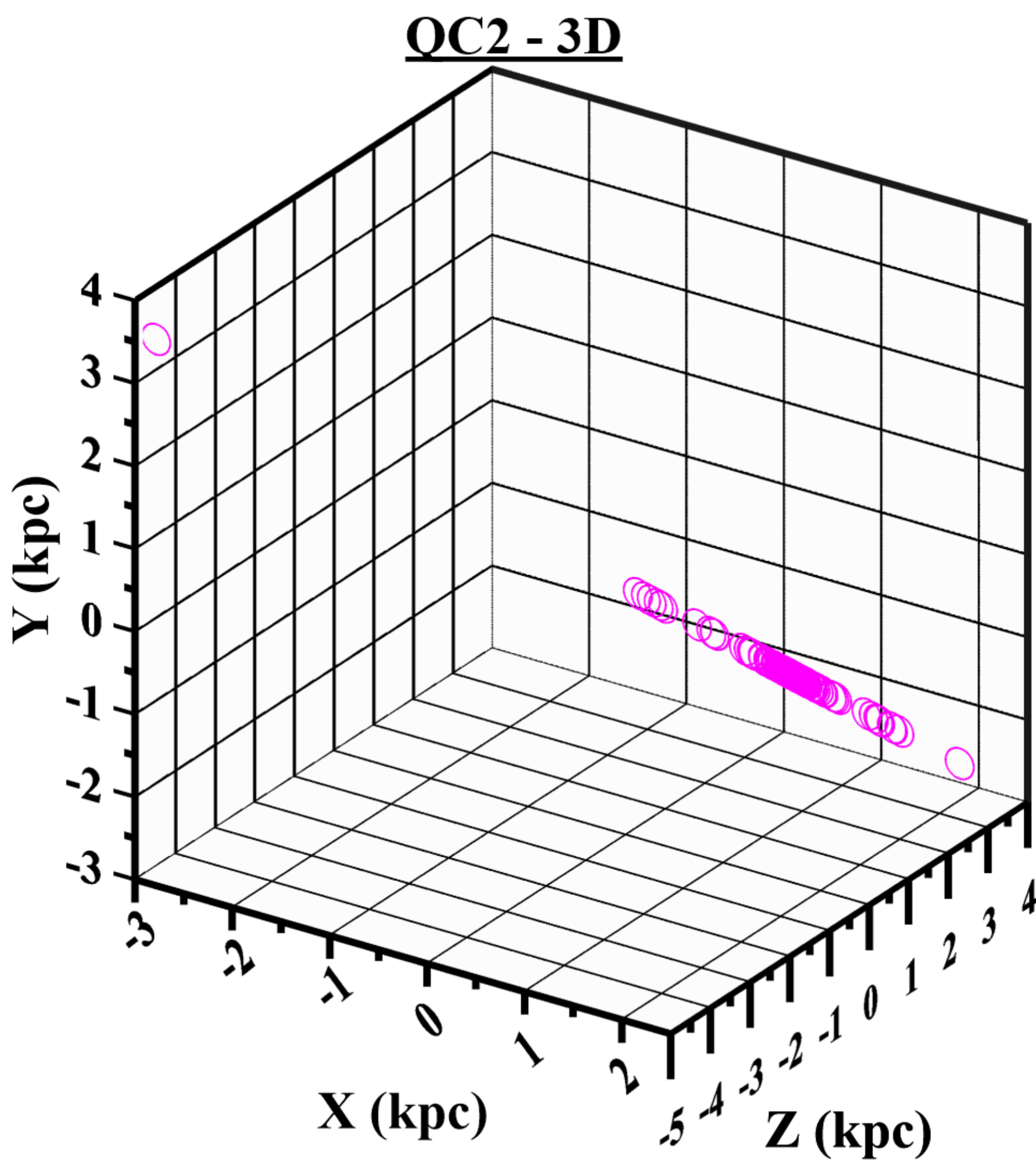}
\vspace{0.5cm}
  \includegraphics[width = 6.85cm]{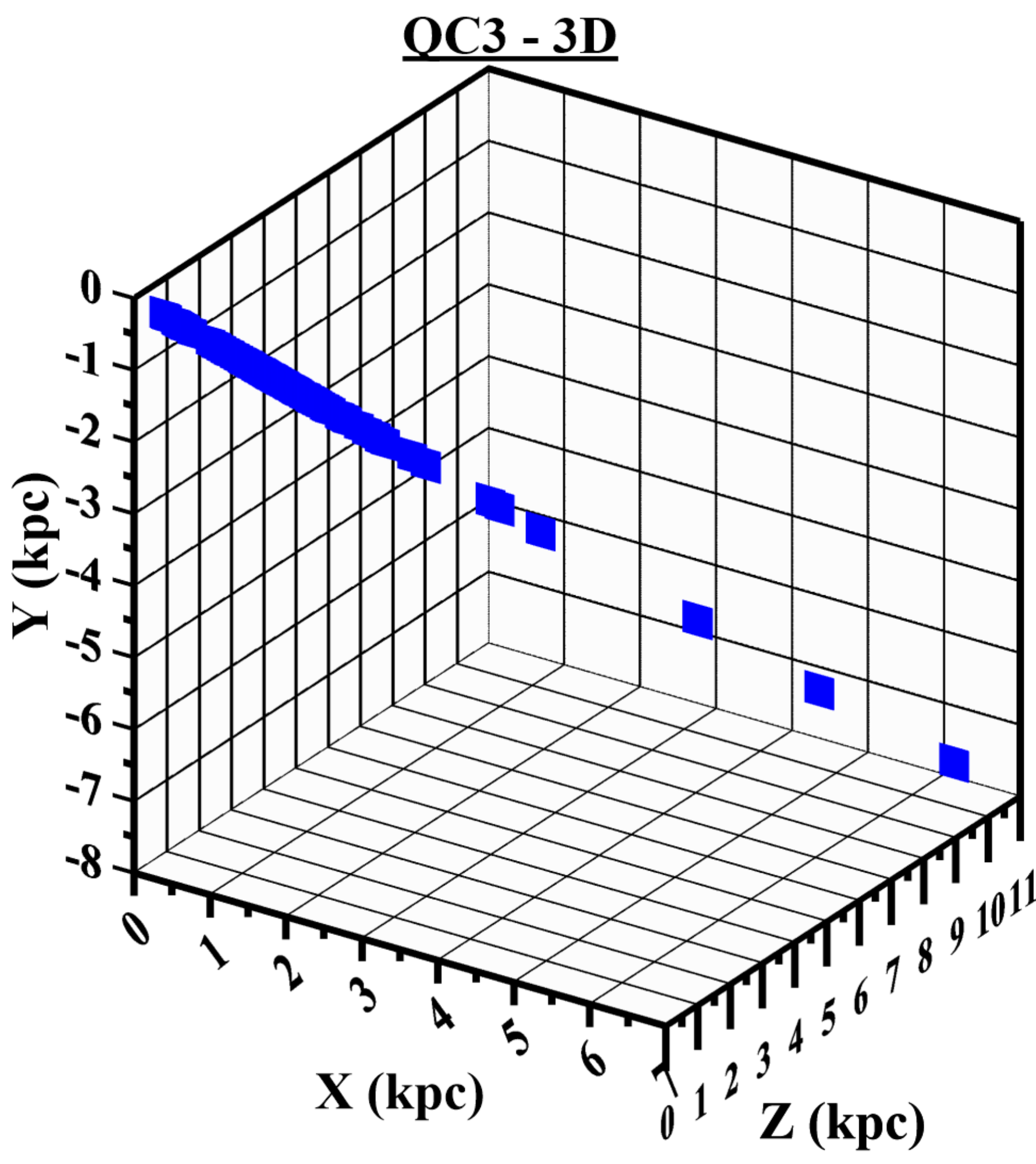}
  \includegraphics[width = 6.85cm]{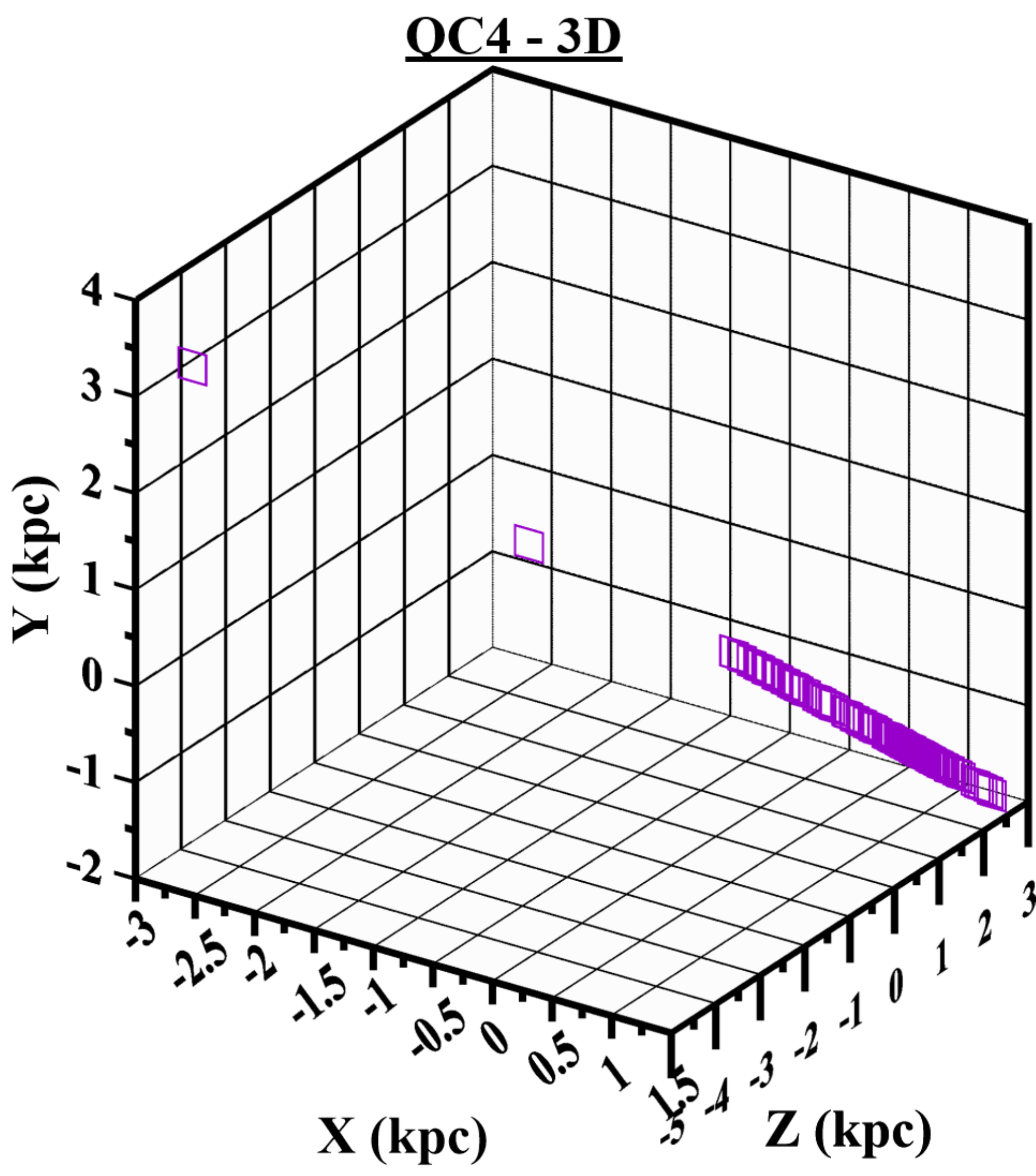}
\caption {The 3D spatial morphology plots in heliocentric Cartesian coordinates (x, y, z) of the four studied clusters.}%; QC1 (upper left corner), QC2 (upper right corner), QC3 (lower left corner), and QC4 (lower right corner). }
\label{fig:10}
\end{center}
\end{figure*}

%\zmaa{This part needs to be enhanced with some discussion that shows the importance of these parameters and the meaning of them. I think the morphology of the cluster has to be discussed in more details given the meaning of the shape and the importance of knowing this morphology is really important to strengthen the paper. \zma{I did not modify this part that much as I did not get the point beyond conducting it}.}
\begin{enumerate}[label=(\alph*)]
{\bf \item {The cluster center}}% of mass}}

%\zmaa{\sc start}\\
For a cluster consists of N members each has a location ($\alpha, \delta$) at a distance d, the position of the cluster center (x$_{c}$, y$_{c}$, z$_{c}$) can be calculated by finding the center of mass of the N member stars in equatorial coordinates following Eq.s \ref{eq:Xc}. %with respect to that of the Galaxy is given by the simple formulae in Eq. \ref{eq:Xc}. 
\begin{equation}
\label{eq:Xc}
\begin{split}
x_c~&=~\frac{1}{N}~\sum\limits_{i=1}^{N}~d_i~\cos\alpha_i~\cos\delta_i, \\%\Bigg{/}N,\\
%\end{equation}
%\begin{equation}
y_c~&=\frac{1}{N}~\sum\limits_{i=1}^{N}d_i\sin\alpha_i\cos\delta_i, \\%\right]\Bigg{/}N,\\
%\end{equation}
%\begin{equation}
z_c~&=~\frac{1}{N}~\sum\limits_{i=1}^{N}d_i\sin\delta_i.%\right]\Bigg{/}N.
\end{split}
\end{equation}
%\\
These coordinates are expressed in units of parsecs because they represent the distance between the cluster center and the observer.%at of the Galaxy. 
{\bf \item {The Solar elements}}

If the mean spatial velocity components for a cluster in the Galactic coordinates are ($\bar{U}, \bar{V}, \bar{W}$), then we may deduce the Solar space velocity components (U$_{\odot}$, V$_{\odot}$, W$_{\odot}$) via the following relation 
$$ U_{\odot}~=~-\bar{U}, \quad V_{\odot}~=~-\bar{V}, \quad \text{and} \quad W_{\odot}~=~-\bar{W}$$
from which we can determine the absolute value of the Solar space velocity with respect to the studied objects as follows%our groups under study.
\begin{equation}
\label{eq:s-gal}
S_{\odot}~=~\sqrt{(\bar{U})^2~+~(\bar{V})^2~+~(\bar{W})^2}
\end{equation}
and then we may estimate the location of the Solar apex (l$_A$, b$_A$) in the Galactic coordinates to be 
\begin{equation}
\label{eq:ad-gal}
\begin{split}
l_{A}&=tan^{-1}\Bigg(\frac{-\bar{V}}{\bar{U}}\Bigg), \qquad \text{and} \\
%\end{equation}
%\begin{equation}
b_{A}&=sin^{-1}\Bigg(\frac{-\bar{W}}{S_{\odot}}\Bigg).
\end{split}
\end{equation}

Eq.s \ref{eq:s-gal} and \ref{eq:ad-gal} can be expressed in terms of the Cartesian coordinates, (x, y, z)% ($X_{\odot}^{\bullet}, Y_{\odot}^{\bullet}, Z_{\odot}^{\bullet}$)
, assuming that the center of the axes is located at the Sun. Given the following relation
%supposed to be centred at the SUn, the Solar space velocity components the Now consider the position along x, y, and z-axes in the coordinate system whose centered at the Sun, then the Sun's velocities with respect to this same group and referred to the same axes are given as; (
$$ X_{\odot}^{\bullet}~=~ -\bar{V_{x}}, \quad Y_{\odot}^{\bullet}~=~-\bar {V_y}, \quad \text{and} \quad Z_{\odot}^{\bullet}~ =~-\bar{V_z} $$ 
where ($\bar{V_{x}}, \bar{V_{y}}, \bar{V_{z}}$) are the mean space velocity components of the cluster that can be determined from Eq.s \ref{eq:Vs}. 
Hence, the Solar space velocity is %we have obtained the Solar elements with radial velocities considered as;
\begin{equation}
\label{eq:s-car}
S_{\odot}~=~\sqrt{(X_{\odot}^{\bullet})^2~+~(Y_{\odot}^{\bullet})^2~+~(Z_{\odot}^{\bullet})^2},
\end{equation}
and therefore, the position of the Solar apex point ($\alpha_A, \delta_A$) in the equatorial coordinates is
\begin{equation}
\label{eq:ad-car}
\begin{split}
\alpha_{A}&~=~tan^{-1}~\Bigg(\frac{Y_{\odot}^{\bullet}}{X_{\odot}^{\bullet}}\Bigg), \qquad \text{and}\\
%\end{equation}
%\begin{equation}
\delta_{A}&~=~tan^{-1}~\Bigg(\frac{Z_{\odot}^{\bullet}}{\sqrt{(X_{\odot}^{\bullet})^2~+~(Y_{\odot}^{\bullet})^2}}\Bigg).
\end{split}
\end{equation}
%where ($l$$_{A}, $\alpha$$_{A}) is the Galactic; longitude and right ascension of the Solar apex and ($b$$_{A}, $\delta$$_{A}) are the Galactic; latitude and declination of the Solar apex, where (S$_{${\odot}$}) is considered as the absolute value of the Sun's velocity relative to our groups under investigations. \\

{\bf \item {The cluster 3D-morphology}}

The shapes of young star clusters must reflect the conditions in the parental molecular clouds and during the cluster formation process \citep{Chen04}. The median age for clusters associated with clouds is 4 Myr, whereas it is 50 Myr for clusters that are sufficiently separated from a molecular cloud to be considered unassociated. After $\sim$ 6 Myr, the majority of the star clusters lose association with their molecular gas \citep{Grasha19}.

We analyzed the 3D spatial position (X, Y, Z) of the member stars in the heliocentric Cartesian coordinates (x, y, z) by knowing the estimated distance to the cluster, d
\begin{equation}
\label{3d-mor}
\begin{split}
X~&=~d~\cos\delta~\cos\alpha,\\
%\end{equation}
%\begin{equation}
Y~&=~d~\cos{\delta}~\sin{\alpha},\\
%\end{equation}
%\begin{equation}
Z~&=~d~\sin{\delta}.
\end{split}
\end{equation}
%\\
%\item a durian
The 3D morphology for the four clusters is plotted in Fig. \ref{fig:10}. It is noticeable that cluster members show an elongated expansion in their cluster region. % the stars of those are located and expand through separate elongate regions in space. 
This expansion may be regarded as fast gas expulsion and virilization \citep{Pang21} and hence we may %. Morphologically and as suggested in Figure 10, we can 
conclude that the birthplaces of those member stars are in the same region of the disk. This result, in particular for QC1, supports our calculations for the parameters $\delta_c$ and C that unveiled that the cluster is not compact but slightly scattered.
\end{enumerate}
%%%%%%%%%%%%%%%%%%%%%%%%%%%%%%%%%%%%%%%%%%%%%%%%%%%%%%%%%%%%%%%%%%%%%%%%%

%%%%%%%%%%%
% Fig. 10                                                  %
%%%%%%%%%%%%%%%%%%%%%%%%%%%%%%%%%%%%%%%%%%%%%%%%%%%%%%%%%%%%%%%%%%%%%%%%%
%%%%%%%%%%%%%%%%%%%%%%%%%%%%%%%%%
%%%%%%%%%%%%%%%%%%%%%%%%%%%%%
% Table 6 
%%%%%%%%%%%%%%
\begin{table*}
\caption{Our computed dynamical evolution and kinematical parameters for the four clusters.} %of QCs cluster as compared with \citet{Qin21}.}
\centering
%\resizebox{\textwidth}{!}{
\begin{tabular}{lllll}
\topline
%Parameters & Koposov 12 & Koposov 43 & References \\
%& (FSR 802) & (FSR 848) &\\\midline
%f$_{bg}$ (stars arcmin$^{-2}$)& 0.069$\pm$0.016& 0.093$\pm$0.006&      \\
{\bf Parameters}  &	{\bf QC1} &	{\bf QC2}	&	{\bf QC3}	&	{\bf QC4}	\\   \hline
T$_{relax}$ (Myr)  &11.80 $\pm$ 1.72	&	19.17 $\pm$ 4.38	&	31.40 $\pm$ 5.61	&	17.48 $\pm$ 4.18	\\[0.8 ex]
$\tau_{ev}$ (Myr) &	1180 $\pm$ 43.00	&	1917 $\pm$ 44.00	&	3140 $\pm$ 56.00	&	1748 $\pm$ 42.00	\\[0.8 ex]
$\tau$	         &	0.82	&	17.44	&	22.97	&	13.32	\\[0.8 ex]
V$_{esc}$ (km/s)	&	281 $\pm$ 16.76	&	290 $\pm$ 17.03	&	289 $\pm$ 17.00	&	288 $\pm$ 16.97	\\[0.8 ex]
$\bar{V}_X$ (km/s)&-9.99 $\pm$ 1.58	&	-16.23 $\pm$ 4.03	&	-13.32 $\pm$ 1.83	&	-9.23 $\pm$ 1.52	\\[0.8 ex]
$\bar{V}_Y$ (km/s)&-36.50 $\pm$ 6.04	&	-40.70 $\pm$ 6.38	&	-31.66 $\pm$ 5.63	&	-27.71 $\pm$ 5.26	\\[0.8 ex]
$\bar{V}_Z$  (km/s)&-50.32 $\pm$ 7.09	&	-44.58 $\pm$ 6.68	&	-62.97 $\pm$ 7.94	&	-45.96 $\pm$ 6.78	\\[0.8 ex]
A	 &-105$^o$.32 $\pm$ 0$^o$.10&-111$^o$.74 $\pm$ 0$^o$.09	&-112$^o$.82 $\pm$ 0$^o$.10	&	-108$^o$.41 $\pm$ 0$^o$.10\\[0.8 ex]
D	&-53$^o$.05 $\pm$ 0$^o$.14&	-45$^o$.50 $\pm$ 0$^o$.15&	-61$^o$.39 $\pm$ 0$^o$.13	&	-57$^o$.56 $\pm$ 0$^o$.13\\[0.8 ex]
$\bar{U}$ (km/s)&56.83 $\pm$ 7.54	&	58.02 $\pm$ 7.62	&	58.94 $\pm$ 7.68	&	46.99 $\pm$ 6.85	\\[0.8 ex]
$\bar{V}$ (km/s)&-26.31 $\pm$ 5.13	&	-23.14 $\pm$ 4.81	&	-39.60 $\pm$ 6.30	&	-26.59 $\pm$ 5.16	\\[0.8 ex]
$\bar{W}$(km/s)	&-6.53 $\pm$ 0.40	&	2.29 $\pm$ 0.66	&	-10.23 $\pm$ 3.20	&	-6.99 $\pm$ 2.64	\\[0.8 ex]
x$_c$ (pc)	&1018 $\pm$ 31.91	&	964 $\pm$ 31.05	&	1204 $\pm$ 34.70	&	829 $\pm$ 28.79	\\[0.8 ex]
y$_c$ (pc)	&-1459 $\pm$ 38.20	&	-1174 $\pm$ 34.26	&	-1413 $\pm$ 37.59	&	-1019 $\pm$ 31.92	\\[0.8 ex]
z$_c$ (pc)	&1492 $\pm$ 38.63	&	1599 $\pm$ 39.99	&	1932 $\pm$ 43.95	&	1400 $\pm$ 37.42	\\[0.8 ex]
S$_{\odot}$ (km/s)&62.96 $\pm$ 7.94	&	62.51 $\pm$ 7.91	&	71.74 $\pm$ 8.47	&	54.45 $\pm$ 7.38	\\[0.8 ex]
(l$_A$, b$_A$)$_o$ &(24.84, 5.96)	& (21.74, -2.11)	&	(33.90, 8.20)	&	(29.50, 7.37)	\\[0.8 ex]
($\alpha_A, \delta_A$)$_o$ & (74.69,53.05)	&	(68.27, 45.50)	&	(67.19, 61.39)	&	(71.59, 57.56)	\\[0.8 ex]
\hline
\label{tab:mlr}
\end{tabular}
%}
\end{table*}
%%%%%%%%%%%%%%%%%%%%%%%%%%%%%%%%%%

%%%%%%%%%%%%%%%%%%%%%%%%%%%%%%%%%%%%%%
\section{Conclusion}
\label{conc}
We presented the first complete comprehensive astrometric, photometric and kinematical study of the newly discovered open clusters namely; QC1, QC2, QC3, and QC4 using the most recent data from Gaia EDR3 with the aid of ASteCA code. We derived most of the fundamental astrophysical and dynamical parameters of these clusters after estimating their most probable members (118; QC1, 142; QC2, 210; QC3, and 110; QC4). Moreover, we computed some astrometric parameters for the first time such as the density contrast, $\delta_c$, and the concentration, C, parameters. %carried out with a computational routine devoted to Mathematica software. 
We summarize the main results and conclusions as follows:
\begin{enumerate}
\item The new positions of the clusters' centers are in agreement in the RA direction and slightly different in the Dec. direction with those obtained by \citet{Qin21}. % re-estimated, and the obtained results have slightly different from those obtained with Qin et al. (2021).

\item Employing the ASteCA code, for each cluster, we determined its radial density profile (RDP) from which we inferred the cluster internal spatial structure, the number of most probable member candidates, metallicity, log (age), reddening, distance modulus, and all of the astrophysical and photometric parameters of the cluster.

\item For the estimated MF and LF of those clusters, we constructed the MLR of individual member stars and estimated the total mass of each cluster to be 158 M$_{\odot}$ (QC1), 177 M$_{\odot}$ (QC2), 232 M$_{\odot}$ (QC3), and 182 M$_{\odot}$ (QC4). In addition, the slopes of the MF are in good agreement with \citet{Salpeter55}.%\zmaa{\bf \sc Dr these values are different from those written in section \ref{mlf} - please match}% (1955) for masses .

\item For all clusters, the space velocity components (U, V, W) relative to the Galactic coordinates and ($V_x, V_y, V_z$) with respect to the Cartesian coordinates were computed, and the position of their corresponding apex coordinates (A, D) were also determined. Moreover, the Solar elements for each cluster were also determined.

\item The dynamical evolution parameters showed that the four clusters are dynamically relaxed, exception is QC1 which is dynamically active with $\tau \sim$ 0.82. In addition, the 3D morphology of the clusters showed that they are extended along the plane of the Galactic disc which is supported by our calculations of the contract parameter $\delta_c$. %\zmaa{ Dr please check $\tau$ for QC1 is less than 1}. %where the dynamical evolution parameter is more than one. On the other hand, we have found that their 3D morphology is expanded with elongation through the same region of the Galactic disc.
\end{enumerate}

We conclude that EDR3 might improved the overall determined parameters of the clusters, in particular those obtained from the astrometric and photometric observations, by reducing their uncertainties. This would lead to a more accurate characteristics of these new open clusters. However, more studies on these clusters are necessary to validate the current findings and to support our conclusions.

\section*{Acknowledgements}
The authors are deeply thankful to the referee for his/her valuable and constructive comments that improved the original manuscript. %quality of the paper.

This work presents results from the European Space Agency (ESA) space mission Gaia. Gaia$\_$data are being processed by the Gaia$\_$Data Processing and Analysis Consortium (DPAC). Funding for the DPAC is provided by national institutions, in particular the institutions participating in the Gaia Multi-Lateral Agreement (MLA). The Gaia$\_$mission website is https://www.cosmos.esa.int/gaia. The Gaia$\_$archive website is https://archives.esac.esa.int/gaia.
%Acknowledgements here.
%\vspace{-1em}
%%use \balance somewhere in the left column of the last page to balance the two columns in the end page
%%%%%%%%%%%%%%%%%%%%%%%%%%%%%%%%%%%%%%%%%%%%%%%%%%%%
%%References section
%%%%%%%%%%%%%%%%%
% References
%%%%%%%%%%%%%%%%%%
%\bibliographystyle{mnras} 
\bibliography{QCRef}
%\bibliographystyle{apj}
%%%%%%%%%%%%%%%%%%
%\begin{theunbibliography}{}
%\bibitem{latexcompanion}
%Adams, F. C., \& Myers, P., 2001, ApJ., 533, 744.
%\bibitem{latexcompanion}
%Zwicky, F. 1957, Morphological Astronomy, Springer-Verlag, Berlin, p. 258
%\end{theunbibliography}

%}]

\end{document}